\documentclass{article}
\usepackage{apacite}

\usepackage{PRIMEarxiv}

\usepackage[utf8]{inputenc} 
\usepackage[T1]{fontenc}    
\usepackage{hyperref}       
\usepackage{url}            
\usepackage{booktabs}       
\usepackage{amsfonts}       
\usepackage{nicefrac}       
\usepackage{lipsum}
\usepackage{fancyhdr}  
\usepackage[dvipdfmx]{graphicx}       
\graphicspath{{media/}}     
\usepackage{natbib}
\usepackage{bm}
\usepackage{amsmath}
\usepackage{amssymb}
\usepackage{enumitem}

\usepackage{algorithmic}
\usepackage[ruled]{algorithm2e}

\usepackage{float}

\usepackage{subcaption}

\usepackage{multirow}
\usepackage{color}

\title{Bayesian Geographically Weighted Regression \\ using Fused Lasso Prior
}

\author{
  Toshiki Sakai \\
  Graduate School of Culture and Information Science, Doshisha University\\
  \texttt{toshikisakai0711@gmail.com} \\
   \And
  Jun Tsuchida \\
  Faculty of Data Science, Kyoto Women's University\\
  \AND
  Hiroshi Yadohisa \\
  Faculty of Culture and Information Science, Doshisha University\\
}

\begin{document}
\maketitle
\begin{abstract}
A main purpose of spatial data analysis is to predict the objective variable for the unobserved locations.
Although Geographically Weighted Regression (GWR) is often used for this purpose, estimation instability proves to be an issue.
To address this issue, Bayesian Geographically Weighted Regression (BGWR) has been proposed.
In BGWR, by setting the same prior distribution for all locations,
the coefficients' estimation stability is improved.
However, when observation locations' density is spatially different, these methods do not sufficiently consider the similarity of coefficients among locations. Moreover, the prediction accuracy of these methods becomes worse.
To solve these issues, we propose Bayesian Geographically Weighted Sparse Regression (BGWSR) that uses Bayesian Fused Lasso for the prior distribution of the BGWR coefficients.
Constraining the parameters to have the same values at adjacent locations is expected to improve the prediction accuracy at locations with a low number of adjacent locations. 
Furthermore, from the predictive distribution, it is also possible to evaluate the uncertainty of the predicted value of the objective variable.
By examining numerical studies, we confirmed that BGWSR has better prediction performance than the existing methods (GWR and BGWR) when the density of observation locations is spatial difference.
Finally, the BGWSR is applied to land price data in Tokyo.
Thus, the results suggest that BGWSR has better prediction performance and smaller uncertainty than existing methods.
\end{abstract}

\section{Introduction}
Spatial data denote data that comprise individual values and the spatial location information where they were observed.
The application of spatial data, such as mapping to understand visual features and creating hazard maps for disasters~\citep{fischer2010}, is used in various research fields.
When we obtain spatial data, we sometimes set objective variables and covariates for the variables obtained.
The purposes of analyzing such spatial data include visualization of spatial data, classification and division of space, and prediction of the value of the objective variable (e.g., \citealp{jena2023}; \citealp{wang2023}). The prediction of the objective variable's value is important in analyzing spatial data because there are many unobserved locations for the objective variables in many cases.

There are two main ways to predict the objective variable.
The first applies to the spatial data containing only the objective variable.
The methods used in this case include Kriging~\citep{krige1951}, inverse distance weighting~\citep{bailey1995}, and spline~\citep{cline1973}.
For example, Kriging predicts the value of the objective variable at the prediction location using a weighted average of the available objective variable data.
The second applies to the spatial data containing covariates in addition to the objective variable.
The methods used in this case include Kriging with an external drift~\citep{ahmed1987}.
One of the most frequently used methods is Geographically Weighted Regression (GWR)~\citep{brunsdon1996, fotheringham2003}.
GWR is based on linear regression and assumes a multiple regression model with different coefficients for each location.
Further, it assumes that the values of the coefficients are similar for locations close to each other---the method assumes spatial autocorrelation in the coefficients.
In \cite{yu2007}, GWR is used to investigate the relationship among housing prices, floor space, and age of the house in Milwaukee, Wisconsin, in the United States.
In \cite{koh2020}, GWR is used to investigate the relationship among
Nitrate concentrations, land use, and precipitation in South Korea.

Notably, when applying GWR to real data, the estimation of coefficients becomes unstable when the number of observation locations is small~\citep{lesage2004, subedi2018} because of the small number of observations used to estimate the coefficients.
In extreme cases, for estimation coefficients, we could use only one or two locations, and the solution of the weighted least squared method may not be uniquely determined.
To solve this issue, Bayesian Geographically Weighted Regression (BGWR) has been proposed~\citep{lesage2004}.
BGWR is a method for estimating parameters of GWR within the framework of Bayesian statistics.
Specifically, parameters such as coefficients are assumed to follow a certain probability distribution, and the estimator is the expected value of the posterior distribution calculated by using the data.
Setting a common prior distribution for all locations allows for a stable estimation of coefficients when the number of locations is small, as information from the prior distribution becomes more weighted in the estimation.
Additionally, BGWR allows us to evaluate the instability of the objective variable's predicted value because the predicted value's credibility is calculated by the predictive distribution.

However, these methods sometimes estimate considerably different coefficients for adjacent locations.
For example, two close yet adjacent locations do not share the same location for estimation.
In such a situation, the similarity of the coefficients at adjacent locations, as assumed by spatial autocorrelation, cannot be fully captured because the coefficients are estimated using completely different data~\citep{fotheringham2017}.

To resolve this issue, we propose a novel method by using the idea of Fused Lasso~\citep{tibshirani2005}.
Fused Lasso penalizes the absolute value of the difference between the coefficients at adjacent locations, making it easier to obtain the same estimated value of the coefficient at adjacent locations.
For example, \cite{li2019} and \cite{zhong2023} use Fused Lasso penalties to consider the similarity of the coefficient of regression and show improved prediction performance compared to the GWR.
Hence, the proposed method---Bayesian Geographically Weighted Sparse Regression (BGWSR)---combines BGWR and Bayesian Fused Lasso~\cite {kyung2010} and is expected to obtain similar estimates of coefficients between adjacent locations. This is because the Laplace distribution is set for the difference in coefficients between adjacent locations, and the difference between adjacent locations is easily estimated as $0$.
Moreover, as BGWSR can consider the similarity between adjacent locations, it is expected to improve the prediction accuracy of coefficients and objective variables compared to existing methods.

In BGWR and GWR, when observation locations are obtained from a spatially heterogeneous environment (i.e., there are areas whose density is sparse and dense),
the estimation of coefficients and predictions for the observed locations may be less accurate.
For example, if adjacent locations are defined as extremely close in distance, the number of observation locations used to estimate the coefficients will not be sufficient in areas where the density of observation locations is sparse, and the resulting estimation will be unstable.
Therefore, for stable estimation, it is necessary to have excessive information in the prior distribution.
Nevertheless, when adjacent locations are defined as distant locations, the estimated values of the coefficients may be the same for most locations because the same data are shared by many locations and the penalty of Fused Lasso.
Furthermore, in areas with dense observation locations, it is necessary to define adjacent locations as those close to each other to consider rapid changes adequately.
By introducing a prior distribution for setting the width of each location to be an adjacent location, it is expected that the appropriate adjacent location will be determined for all locations.
This allows for a trade-off between the breadth of the definition of the adjacency points and the strength of the to-be-determined Fused Lasso penalty so that the estimation performance is better for each location. Therefore, introducing this prior is expected to improve the prediction accuracy of coefficients and objective variables even when locations are spatially heterogeneous.

This paper is organized as follows. In Section \ref{chap2:prop}, we introduce the proposed method and other predictive algorithms.
In Section \ref{chap3:numerical studies}, we evaluate the numerical performance of the proposed methods together with some existing methods through numerical studies.
In Section \ref{chap4:realdata}, we demonstrate the proposed method through spatial regression modeling of the land prices in the Tokyo metropolitan area.
Finally, Section \ref{chap5:summary} summarizes our main conclusions and identifies areas where further research is necessary.
\section{Bayesian Geographically Weighted Sparse Regression (BGWSR)\label{chap2:prop}}
\subsection{Model}
This section describes the model for Bayesian estimation of the BGWSR.
Let $D \subset \mathbb{R}^{2}$ be the region to be analyzed, and let $\bm{s}_i = (s_{1i},\,s_{2i})^{\prime} \in D$ be the $i$-th location.
Suppose a set of spatial data ${(x(\bm{s}_{i}), \, y(\bm{s}_{i}))}$ is observed locations $\bm{s}_{1}, \, \ldots, \, \bm{s}_{n} \in D$, \, where the objective variable $y(\bm{s}_{i})$ is assumed to be spatially correlated, and $\bm{x}(\bm{s}_{i}) = (x_{1}(\bm{s}_{i}), \, \ldots , \, x_{p}(\bm{s}_{i}))^{\prime}$ is the $p$-dimensional vector of covariates at the location $\bm{s}_{i}$.
\begin{align*}
    y(\bm{s}_{i}) = \bm{x}(\bm{s}_{i})^{\prime} \bm{\beta}(\bm{s}_{i}) + \varepsilon(\bm{s}_{i}) .
\end{align*}
We assume that $\bm{\beta}(\bm{s}_{i})$ is the coefficient and that the error terms $\varepsilon(\bm{s}_{i})$ follow a Normal distribution ${\rm{N}}(0, \, \sigma^{2}) (\sigma^{2}>0)$ independent of each other.

To estimate the coefficients, we consider that the coefficients are described by $n$-dimensional vectors.
Let $\bm{\beta}_{k}$ be a vector of coefficients whose $i$-th elements are the coefficient $\beta_{k}(\bm{s}_{i}) (k=1, \, 2, \, \ldots, \, p)$,
and be defined as $\bm{y}_{k}=\bm{X}_{k}\bm{\beta}_{k}+\bm{\varepsilon}_{k}$, where $\bm{\varepsilon}_{k} \sim {\rm{MN}}(\bm{0}_{n}, \, \sigma^{2} \bm{I}_{n}/p)$, and $\bm{I}_{n}$is $n$-dimensional identity matrix.
${\rm{MN}}(\bm{\mu}, \, \bm{\Sigma})$ denotes the Multivariate Normal distribution of the mean vector $\bm{\mu}$ and covariance matrix $\bm{\Sigma}$.
The $\bm{X}_{k}$ is an $n$-dimensional diagonal matrix and $\bm{X}_{k}= {\rm{diag}}\{x_{k}(\bm{s}_{1}), \, x_{k}(\bm{s}_{2}), \, \ldots, \, x_{k}(\bm{s}_{n}))\}\in \mathbb{R}^{n \times n}$.
Using these terms, the $\bm{y}=(y(\bm{s}_{i}))$ is represented as $\bm{y}=\sum_{k=1}^{p}\bm{y}_{k}=\sum_{k=1}^{p}\bm{X}_{k}\bm{\beta}_{k}+\bm{\varepsilon}_{k}$.
The model for estimating the parameters at location $\bm{s}_{i}$ is given as follows based on BGWR~\citep{lesage2004}.
\begin{align}
    \label{model:BGWSR}
    \bm{W}(\bm{s}_{i})\bm{y}_{k} = \bm{W}(\bm{s}_{i})\bm{X}_{k} \bm{\beta}_{k} + \bm{\varepsilon}_{k}.
\end{align}
$\bm{W}(\bm{s}_{i})$ is an $n$-dimensional diagonal matrix, and $\bm{W}(\bm{s}_{i})={\rm{diag}}\{w_{1}(\bm{s}_{i}), \, w_{2}(\bm{s}_{i}), \, \ldots, \, w_{n}(\bm{s}_{i})\}$ and $w_{j}(\bm{s}_{i})\,(\geq 0; j=1, \, 2, \, \ldots, \, n)$ are weights determined by the distance between two locations $\bm{s}_{i}$ and $\bm{s}_{j}$.

For handling the spatial autocorrelation of coefficients, the Fused Lasso Prior is selected as the prior distribution of the coefficients in BGWSR.
The following distribution is set as the prior distribution of the coefficient $\bm{\beta}_{k}\,(k=1, \, 2, \, \ldots, \, p)$ for the $k$-th covariates.
{\small{
\begin{align}
    \begin{split}
        \bm{\beta}_{k}|\sigma^{2}, \, S, \, N, \, C, \, \lambda_{k, \, 1}, \, \lambda_{k, \, 2}
        \sim (\sigma^{2})^{-\frac{2n+\sum_{i=1}^{n}n_{i}}{4}} &\prod_{i=1}^{n} {\rm{Laplace}} \left (\frac{\beta_{k}(\bm{s}_{i})}{\sqrt{\sigma^{2}}} \Big{|} 0,\lambda_{k, \, 1} \right) \\
        &\quad \times \prod_{(i,j) \in \bm{C}}{\rm{Laplace}} \left( \frac{\beta_{k}(\bm{s}_{i})-\beta_{k}(\bm{s}_{j})}{\sqrt{\sigma^{2}}}\Big{|}0, \, \lambda_{k, \, 2} \right),
    \end{split}
\label{beta_prior}
\end{align}
}}
where ${\rm{Laplace}}(x|\mu, \, b)$ denotes the Laplace distribution of mean $\mu$ and scale parameter $b$ with $x$ as a random variable.
$S$ is the set of the observed locations and $S=\{\bm{s}_{1}, \, \bm{s}_{2}, \, \ldots, \, \bm{s}_{n}\} \subset D$.
$N$ is the set of the number of adjacent locations $n_{i}$ at location $\bm{s}_{i}$, where $N=\{n_{1}, \, n_{2}, \, \ldots, \, n_{n}\}$.
$C$ is a set of pairs of adjacent locations, and $(i, \, j) \in C$ indicates that two locations $\bm{s}_{i}, \, \bm{s}_{j}$ are spatially close to each other.
$\lambda_{k,1}, \, \lambda_{k,2}$ are parameters that determine the strength of the shrinkage.
For example, the following is obtained.
\begin{align*}
    C = \{(i,j)\mid w_j(\bm{s}_i)>0 \}.
\end{align*}

The first term on the right-hand side of the equation (\ref{beta_prior}) is a Laplace distribution with mean $0$ for the coefficient at each location based on the idea of Bayesian Lasso ~\citep{park2008}.
The role of this term is to shrink the coefficient to $0$ regardless of location.
As it tends to shrink to $0$ of the estimated value of the coefficient at a location with a small size adjacent set to $0$, the estimation is more numerically stable than GWR.
The second term on the right side of the equation (\ref{beta_prior}) is based on the idea of Bayesian Fused Lasso~\citep{kyung2010}, setting a Laplace distribution with mean $0$ for the difference of coefficients at adjacent locations.
This allows us to estimate similar coefficients for adjacent locations.

As the coefficients $\bm{\beta}_{k}$ have a Laplace distribution for the prior distribution, the full conditional distribution for Gibbs sampling cannot be derived explicitly without modification.
We apply parameter augmentation ~\citep{andrews1974}.
For calculating the posterior distribution, the prior distribution of the coefficient $\bm{\beta}_{k}$ can be rewritten using the parameter $\tau_{k, \, i}^{2}, \, \omega_{k, \, i, \, j}^{2}\,(i=1, \, 2, \, \ldots, \, n; j=1, \, 2, \, \ldots, \, n, \, i\neq j; k=1, \, 2, \, \ldots, \, p)$ as follows, where $\tau_{k,i}^{2} > 0, \, \omega_{k, \, i, \, j}^{2} > 0$.
\begin{align*}
    \bm{\beta}_{k}|\sigma^2,T_{k}, \, \Omega_{k} &\sim {\rm{MN}}(\bm{0}_{n}, \, \sigma^{2} \bm{\Sigma}_{k}^{-1}), \, \\
    T_{k} | \lambda_{k, \, 1}^{2} &\sim \prod_{i=1}^{n} \frac{\lambda_{k, \, 1}^{2}}{2} {\rm{exp}}\left\{ -\frac{\lambda_{k, \, 1}}{2}\tau_{k,i}^{2}\right\}, \, \notag\\
    \Omega_{k} | \lambda_{k, \, 2}^{2} &\sim \prod_{(i, \, j) \in \bm{C}} \frac{\lambda_{k, \, 2}^{2}}{2} {\rm{exp}}\left\{ -\frac{\lambda_{k, \, 2}}{2}\omega_{k,i,j}^{2} \right\}. \notag
\end{align*}
Here, $T_{k} = \{ \tau_{k,1}^{2}, \, \tau_{k,2}^{2}, \, \ldots,\tau_{k,n}^{2} \}, \, \Omega_{k} = \{ \omega_{k,1,2}^{2}, \, \omega_{k,1,3}^{2}, \, \ldots, \, \omega_{k,n-1,n}^{2} \}$ and $\omega_{k, \, i, \, j}^{2} = \omega_{k, \, j, \, i}^{2} \quad (k = 1, \, 2, \, \ldots, \, p)$.
Let $\sigma^{2} \bm{\Sigma}_{k}^{-1}$ be the covariance matrix of the coefficients defined as
\begin{align*}
    (\bm{\Sigma}_{k}^{-1})_{ij} = 
    \left\{
        \begin{array}{ll}
            \frac{1}{\tau_{k,i}^{2}} + \sum_{(i,\ell) \in C_{i}} \frac{1}{\omega_{k, \, i,\ell}^{2}} & (i=j) \\
            -\frac{1}{\omega_{k,i,j}^{2}} & (i \neq j, \, (i, \, j) \in C) \\
            0 & ({\rm{otherwise}}),
        \end{array}
    \right. 
\end{align*}
where $C_{i}\subset C$ is a set comprising only pairs of locations adjacent to $\bm{s}_{i}$.

Based on \cite{kyung2010}, the models for the other parameters are set as
\begin{align*}
    \lambda_{k, \, 1}^{2} | r_{1}, \, q_{1} \sim {\rm{Gamma}}(r_{1}, \, q_{1}),
    \quad \lambda_{k, \, 2}^{2} | r_{2}, \, q_{2} \sim {\rm{Gamma}}(r_{2}, \, q_{2}), \\
    \sigma^{2} | r, \, q \sim {\rm{IGamma}}\left(\frac{r}{2}, \, \frac{q}{2}\right),
    \quad w_{j}(\bm{s}_{i}) | h = f(d_{ij}|h),
\end{align*}
where ${\rm{Gamma}}(a, \, b), \, {\rm{IGamma}}(a, \, b)$ denote the gamma distribution and Inverse Gamma distribution with shape parameter $a$ and reciprocal scale parameter $b$. Meanwhile $r, \, q, \, r_{1}, \, q_{1}, \, r_{2}, \, q_{2}$ are hyperparameters determined by the analyst.
The function $f(d_{ij}|h)$ determines the weights of two locations $\bm{s}_{i}, \, \bm{s}_{j}$ with the distance $d_{ij}$ as an argument.
For example, the following equation is used.
\begin{align*}
    f(d_{ij}|h) = 
        \begin{cases}
            1 - \left( \frac{d_{ij}}{h} \right)^{2} & \,\left(\left\|\boldsymbol{s}_i-\boldsymbol{s}_j\right\|<h\right) \\
            0 & \text { (otherwise) }
        \end{cases}
    .
\end{align*}

$h$ is a parameter that determines the degree of spatial correlation.
It may be determined by the analyst based on domain knowledge, or it can be incorporated as a model.
For example, based on ~\cite{boehm2015, ma2021}, we have
\begin{align}
\label{func:h}
    h | a \sim {\rm{U}}(0, \, a),
\end{align}
where ${\rm{U}}(0, \, a)$ represents a Uniform distribution with lower limit $0$ and upper limit $a$, where $a$ is the hyperparameter determined by the analyst.
For example, the maximum distance between locations in a region is set as $a$.
Sampling methods of the posterior distribution of $h$ include the Metropolis–Hastings Algorithm (MH method)~\citep{ma2021}.
The method of determining the optimal adjacency for each location is described in \ref{chap:adjacent}.

\subsection{Parameter estimation}
The probability density function of the simultaneous posterior distributions of the parameters for BGWSR has a complex form.
Therefore, Gibbs sampling and MH method are used to estimate the posterior distribution of the parameters.
For $\bm{\beta}_{k}, \, \sigma^{2}, \, \lambda_{k, \, 1}^{2}, \, \lambda_{k, \, 2}^{2}$, Gibbs sampling is used because the probability density function of the conditional distribution can be derived explicitly.
For the parameter $h$, we use the MH method based on \cite{ma2021}.

First, We show the conditional distribution for Gibbs sampling.
The full conditional distribution of the coefficients $\bm{\beta}_{k}$ is given as follows,
{\small{
\begin{align*}
    \begin{split}
        \bm{\beta}_{k}|&\bm{X}_{k}, \, \bm{y}_{k}, \, \bm{\Sigma}_{k}^{-1}, \, W, \, \sigma^{2}, \, T_{k}, \, \Omega_{k} \\
        &\sim {\rm{MN}}\left(\left(\bm{X}_{k} \sum_{i=1}^{n}(\bm{W}(\bm{s}_{i}))^{2} \bm{X}_{k} + \bm{\Sigma}_{k}^{-1}\right)^{-1} \bm{X}_{k} \sum_{i=1}^{n}(\bm{W}(\bm{s}_{i}))^{2} \bm{y}_{k},\, \sigma^{2}\left(\bm{X}_{k} \sum_{i=1}^{n}\bm{W}(\bm{s}_{i}) ^{2}\bm{X}_{k} + \bm{\Sigma}_{k}^{-1}\right)\right),
    \end{split}
\end{align*}
}}
where $W$ denotes the set of weight matrices for all locations, and $W = \{ \bm{W}(\bm{s}_{1}), \, \bm{W}(\bm{s}_{2}), \, \ldots, \, \bm{W}(\bm{s}_{n}) \}$.
The conditional distribution of the parameters $T_{k}, \, \Omega_{k}$ added by parameter augmentation is as follows.
\begin{align*}
    \frac{1}{\tau_{k,i}^{2}} \Big{|} \beta_{k}(\bm{s}_{i}), \, \sigma^{2}, \, \lambda_{k, \, 1}
        &\sim {\rm{IGauss}}\left( \sqrt{\frac{\lambda_{k, \, 1}^{2} \sigma^{2}}{\beta_{k}(\bm{s}_{i})^{2}}}, \, \lambda_{k, \, 1}^{2} \right), \\
    \frac{1}{\omega_{k,i,j}^{2}}\Big{|}\beta_{k}(\bm{s}_{i}), \, \beta_{k}(\bm{s}_{j}), \, \sigma^{2}, \, \lambda_{k, \, 2}
        &\sim {\rm{IGauss}}\left( \sqrt{\frac{\lambda_{k, \, 2}^{2} \sigma^{2}}{(\beta_{k}(\bm{s}_{i})-\beta_{k}(\bm{s}_{j}))^{2}}}, \, \lambda_{k, \, 2}^{2} \right),
\end{align*}
where ${\rm{IGauss}}(\mu, \, \lambda)$ denotes the Inverse Gaussian distribution with mean $\mu$ and shape parameter $\lambda$.

The full conditional distribution of the parameters $\sigma^{2}, \, \lambda_{k, \, 1}, \, \lambda_{k, \, 2}$ is given below.
We discuss the parameter $h$ in the next section.
\begin{align*}
    &\sigma^{2} | \bm{W}(\bm{s}_{i})\bm{y}_{k}, \, \bm{W}(\bm{s}_{i})\bm{X}_{k}, \, \bm{\beta}_{k} \sim {\rm{IGamma}}\left(r_{*}, \, q_{*}\right), \\
            & \quad \quad r_{*} = r+\frac{pn^{2}}{2}, \\
            & \quad \quad q_{*} = q + \frac{\sum_{i=1}^{n} \sum_{k=1}^{p} \|\bm{W}(\bm{s}_{i})\bm{y}_{k}-\bm{W}(\bm{s}_{i})\bm{X}_{k}\bm{\beta}_{k}\|^{2}}{2},\\
    &\lambda_{k, \, 1}^{2} | \bm{y}_{k}, \, \bm{X}_{k}, \, \bm{\beta}_{k}, \, \sigma^{2}, \, T_{k}, \, \Omega_{k}
    \sim {\rm{Gamma}}(r_{1}^{*}, \, q_{1}^{*}), \\
            & \quad \quad r_{1}^{*} = r_{1} + n, \\
            & \quad \quad q_{1}^{*} = q_{1} + \frac{1}{2}\sum_{i=1}^{n}\tau_{k,i}^{2},\\
    &\lambda_{k, \, 2}^{2} | \bm{y}_{k}, \, \bm{X}_{k}, \, \bm{\beta}_{k}, \, \sigma^{2}, \, T_{k}, \, \Omega_{k}
    \sim {\rm{Gamma}}(r_{2}^{*}, \, q_{2}^{*}), \\
            & \quad \quad r_{2}^{*} = r_{2} + \frac{1}{2}\sum_{i=1}^{n}n_{i}, \\
            & \quad \quad q_{2}^{*} = q_{2} + \frac{1}{2}\sum_{(i, \, j) \in C} \omega_{k, \, i, \, j}^{2}.
\end{align*}
\subsection{Adjacency estimation}\label{chap:adjacent}
In BGWSR, BGWR, and GWR, the parameter $h$, which determines the range of locations to be adjacent, is the same for all locations.
Therefore, these methods implicitly assume that the observed locations are obtained uniformly within the region.
However, actual data may include dense areas with many locations and sparse areas with few locations.
In an area with sparse observation locations, numerous adjacent locations are required to have adequate data for the coefficient estimation.
Nevertheless, in an area with dense locations, it is necessary to reduce the range of neighboring locations to capture local changes in the coefficients successfully.
Therefore, we consider the range of adjacent locations for each location.
Specifically, the MH method is used for sampling, allowing the parameters responsible for determining the range of adjacency to vary from one location to another.
The sampling scheme of $h$ in Equation (\ref{func:h}) is the special case of this sampling scheme.

We set the prior distribution of the parameter $h(\bm{s}_{i})$ that determines the adjacent location at location $\bm{s}_{i}$ as
\begin{align*}
    h(\bm{s}_{i})|a \stackrel{\mathrm{i.i.d.}}{\sim} {\rm{U}}(0, \, a),
\end{align*}
where $h(\bm{s}_{i})$ is a parameter that determines the adjacency range at location $\bm{s}_{i}$.
By setting $h(\bm{s}_{i})$ as a different parameter for each location, we can expect to estimate a smaller adjacency range in areas with dense observation locations and large local changes in the coefficients.
Subsequently, in areas with sparse observation locations, a large adjacency range is expected to be estimated.

We use the MH method to sample the parameter $h(\bm{s}_{i})$.
The sampling procedure for $h(\bm{s}_{i})$ at location $\bm{s}_{i}$ using the MH method is as follows.
\begin{enumerate}
    \item We use the Normal distribution as the proposal distribution.
    A new $h(\bm{s}_{i})$ candidate $h_{\rm{prop}}(\bm{s}_{i})$ is sampled using the proposal distribution as follows.
    \begin{align*}
        h_{\rm{prop}}(\bm{s}_{i})|\sigma_{h}^{2} \sim {\rm{N}}(h_{\rm{cur}}(\bm{s}_{i}), \sigma_{h}^{2}).
    \end{align*}
    In the above, $h_{\rm{cur}}(\bm{s}_{i})$ is the sampled $h(\bm{s}_{i})$ before the update, and $\sigma_{h}^{2}$ is the hyperparameter representing the variance of the proposal distribution.
    \item Decide whether to accept $h_{\rm{prop}}(\bm{s}_{i})$ based on the posterior probability. 
    Specifically, calculate $\alpha$ as defined in the following equation.
    \begin{align*}
        \alpha &= \min \left( 1, \, \frac{p(h_{\rm{prop}}(\bm{s}_{i})|{\rm{data}})}{p(h_{\rm{cur}}(\bm{s}_{i})|{\rm{data}})} \right) \\
        &= \min \left( 1, \, \frac{\prod_{k=1}^{p} p(\bm{W}_{h_{\rm{prop}}(\bm{s}_{i})}(\bm{s}_{i})\bm{y}_{k} | \bm{W}_{h_{\rm{prop}}(\bm{s}_{i})}(\bm{s}_{i})\bm{X}_{k}, \, \bm{\beta}_{k}, \, \sigma^{2})}{\prod_{k=1}^{p} p(\bm{W}_{h_{\rm{cur}}(\bm{s}_{i})}(\bm{s}_{i})\bm{y}_{k} | \bm{W}_{h_{\rm{cur}}(\bm{s}_{i})}(\bm{s}_{i})\bm{X}_{k}, \, \bm{\beta}_{k}, \, \sigma^{2})} \right) \\
        &= \min \left( 1, \, \frac{\exp \left\{ -\frac{1}{2\sigma^{2}} \|\sum_{k=1}^{p}(\bm{W}_{h_{\rm{prop}}(\bm{s}_{i})}(\bm{s}_{i})\bm{y}_{k}-\bm{W}_{h_{\rm{prop}}(\bm{s}_{i})}(\bm{s}_{i})\bm{X}_{k}\bm{\beta}_{k})\|^{2} \right\}}{\exp \left\{ -\frac{1}{2\sigma^{2}} \|\sum_{k=1}^{p}(\bm{W}_{h_{\rm{cur}}(\bm{s}_{i})}(\bm{s}_{i})\bm{y}_{k}-\bm{W}_{h_{\rm{cur}}(\bm{s}_{i})}(\bm{s}_{i})\bm{X}_{k}\bm{\beta}_{k})\|^{2} \right\}} \right).
    \end{align*}
    In the above, $\bm{W}_{h}(\bm{s}_{i})$ is the matrix $\bm{W}(\bm{s}_{i})$ representing the weights of the neighbors when $h=h$.
    \item Using $u \sim {\rm{U}}(0, \, 1)$, update $h$ as follows.
    \begin{align*}
        h_{\rm{cur}}(\bm{s}_{i}) = 
        \left\{
            \begin{array}{ll}
                h_{\rm{prop}}(\bm{s}_{i}) \quad (u<\alpha) \\
                h_{\rm{cur}}(\bm{s}_{i}) \quad ({\rm{otherwise}})
            \end{array}
        \right.
    .
    \end{align*}
\end{enumerate} 

Using these, the sampling method for BGWSR parameter estimation is represented by the following Algorithm~\ref{alg:BGWSR}.
Here $\bm{y}, \, \bm{X}, \, \bm{B}$ denote the set of objective variable vectors, the set of explanatory variable matrices, and the set of coefficient vectors, respectively, $y = \{ \bm{y}_{1}, \, \bm{y}_{2}, \, \ldots, \, \bm{y}_{p} \}, \, X = \{\bm{X}_{1}, \, \bm{X}_{2}, \, \ldots, \, \bm{X}_{p} \}, \, B = \{\bm{\beta}_{1}, \, \bm{\beta}_{2}, \, \ldots, \, \bm{\beta}_{p} \}$.
\begin{figure}[H]
\begin{algorithm}[H]
    \caption{Algorithm for BGWSR parameter sampling}
    \label{alg:BGWSR}
    \begin{algorithmic}[1]   
    \renewcommand{\algorithmicrequire}{\textbf{Input:}}
    \renewcommand{\algorithmicensure}{\textbf{Output:}}
    \REQUIRE $\bm{s}_{i}, \, X, \, y, \, r, \, q, \, r_{1}, \, q_{1}, \, r_{2}, \, q_{2}, \, \sigma_{h}^{2}, \, t_{\rm{max}} \, \, (i=1, \, 2, \, \ldots, \, n; k=1, \, 2, \, \ldots, \, p)$
    \ENSURE $\widehat{B}$
    \STATE Initialize $\bm{\beta}_{k}^{(0)}, \, T_{k}^{(0)}, \, \Omega_{k}^{(0)}, \, {\sigma^{2}}^{(0)}, \, \lambda_{k, \, 1}^{(0)}, \, \lambda_{k, \, 2}^{(0)}, \, h(\bm{s}_{i})^{(0)}\,(i=1, \, 2, \, \ldots, \, n; k=1, \, 2, \, \ldots, \, p)$. \\
    \STATE Initialize $\bm{W}^{(0)}$ using $w_{j}(\bm{s}_{i})|h(\bm{s}_{i})^{(0)}$. \\
    \STATE Initialize $B^{(0)}$ using $\bm{\beta}_{1}^{(0)}, \, \bm{\beta}_{2}^{(0)}, \, \ldots, \, \bm{\beta}_{p}^{(0)}$. \\
    \STATE Initialize ${\bm{\Sigma}_{k}^{-1}}^{(0)}$ using $T_{k}^{(0)}, \, \Omega_{k}^{(0)}$. \\
    \FOR{$t=0, \, 1, \, \ldots, \, t_{\rm{max}}$}
    \FOR{$(k=1, \, 2, \, \ldots, \, p)$}
    \STATE Sampling $\bm{\beta}_{k}^{(t+1)} \sim \pi ( \bm{\beta}_{k} | \bm{X}_{k}, \, \bm{y}_{k}, \, W^{(t)}, \, {\sigma^{2}}^{(t)}, \, {\bm{\Sigma}_{k}^{-1}}^{(t)} )$. \\
    \ENDFOR
    \STATE Update $B^{(t+1)}$ using $\bm{\beta}_{1}^{(t+1)}, \bm{\beta}_{2}^{(t+1)}, \, \ldots, \, \bm{\beta}_{p}^{(t+1)}$. \\
    \FOR{k=1, \, 2, \, \ldots, \, p}
    \STATE 
    Sampling $T_{k}^{(t+1)} \sim \pi ( T_{k} | \bm{\beta}_{k}^{(t+1)}, \, {\sigma^{2}}^{(t)}, \, \lambda_{k, \, 1}^{(t)} )$. \\
    \STATE 
    Sampling $\Omega_{k}^{(t+1)} \sim \pi ( \Omega_{k} | \bm{\beta}_{k}^{(t+1)}, \, {\sigma^{2}}^{(t)}, \, \lambda_{k, \, 2}^{(t)} )$. \\
    \STATE 
    Update ${\bm{\Sigma}_{k}^{-1}}^{(t)}$ using $T_{k}^{(t)}, \Omega_{k}^{(t)}$. \\
    \ENDFOR
    \STATE 
    Sampling ${\sigma^{2}}^{(t+1)} \sim \pi(\sigma^{2} | X, \, y, \, B^{(t+1)}, \, W^{(t)}, \, r, \, q)$. \\
    \FOR{k=1, \, 2, \, \ldots, \, p}
    \STATE 
    Sampling ${\lambda_{k, \, 1}^{2}}^{(t+1)} \sim \pi ( \lambda_{k, \, 1}^{2} | T_{k}^{(t+1)}, \, r_{1}, \, q_{1} )$. \\
    \STATE 
    Sampling ${\lambda_{k, \, 2}^{2}}^{(t+1)} \sim \pi ( \lambda_{k, \, 2}^{2} | \Omega_{k}^{(t+1)}, \, r_{2}, \, q_{2} )$. \\
    \STATE 
    Sampling $h(\bm{s}_{i})^{(t+1)} \propto \pi( h(\bm{s}_{i}) | y, \, X, \, B^{(t+1)}, \, {\sigma^{2}}^{(t+1)}, \, \sigma_{h}^{2} )$ by MH method. \\
    \ENDFOR
    \STATE 
    Update $W^{(t+1)}$ using $w_{j}(\bm{s}_{i})|h(\bm{s}_{i})^{(t+1)}$. \\
    \ENDFOR
    \end{algorithmic}
\end{algorithm}
\end{figure}
\subsection{Objective variable Prediction}
The weight parameter $w_{j}(\bm{s}_{i^{*}})$ at any prediction location $\bm{s}_{i^{*}} \in D$ is given by
\begin{align}
\label{func:w}
    w_{j}(\bm{s}_{i^{*}}) = 
    &\left\{
        \begin{array}{ll}
            1 - \left( \frac{d_{i^{*}j}}{h(\bm{s}_{j})} \right)^{2} \quad &\left(d_{i^{*}j} < h(\bm{s}_{j})\right)\\
            0 \quad &\left({\rm{otherwise}}\right)
                \end{array}
    \right.
    , \, (j = 1, \, 2, \, \ldots, \, n).
\end{align}
Using this, we predict the coefficient $\bm{\beta}(\bm{s}_{i^{*}})=(\beta_{1}(\bm{s}_{i^{*}}), \, \beta_{2}(\bm{s}_{i^{*}}), \, \ldots, \, \beta_{p}(\bm{s}_{i^{*}}))^{\prime}$ at the prediction location $\bm{s}_{i^{*}}$ by the following formula.
\begin{align}
\label{func:beta}
    \widehat{\bm{\beta}}(\bm{s}_{i^{*}}) = \sum_{j=1}^{n} w_{j}(\bm{s}_{i^{*}})\bm{\beta}(\bm{s}_{j}).
\end{align}
Using $ \widehat{\bm{\beta}}(\bm{s}_{i^{*}}), \, \bm{x}(\bm{s}_{i^{*}})$, we predict the objective variable $y(\bm{s}_{i^{*}})$ at the prediction location $\bm{s}_{i^{*}}$ with the following formula.
\begin{align}
\label{func:y}
    \widehat{y}(\bm{s}_{i^{*}}) = \bm{x}(\bm{s}_{i^{*}})^{\prime} \widehat{\bm{\beta}}(\bm{s}_{i^{*}}).
\end{align}
Therefore, the algorithm for predicting the coefficients and the objective variable at the prediction locations is given by algorithm \ref{alg:BGWSR_pred},
where $n^{*}$ is the number of locations predicted.
\begin{figure}[htb]
\begin{algorithm}[H]
    \caption{Algorithm for predicting coefficients and objective variables at the prediction location by BGWSR}
    \label{alg:BGWSR_pred}
    \begin{algorithmic}[1]   
    \renewcommand{\algorithmicrequire}{\textbf{Input:}}
    \renewcommand{\algorithmicensure}{\textbf{Output:}}
    \REQUIRE $\bm{s}_{i}, \, \bm{s}_{i^{*}}, \, \bm{X}_{k}, \, \bm{X}_{k}^{*}, \, \bm{y}_{k}, \, r, \, q, \, r_{1}, \, q_{1}, \, r_{2}, \, q_{2}, \, \sigma_{h}^{2}, \, t_{\rm{max}} \, (i=1, \, 2, \, \ldots, \, n; i^{*}=1, \, 2, \, \ldots, \, n^{*}; k=1, \, 2, \, \ldots, \, p)$
    \ENSURE $\widehat{\bm{\beta}}(\bm{s}_{i^{*}}), \, \widehat{y}(\bm{s}_{i^{*}}) \, (i^{*}=1, \, 2, \, \ldots, \, n^{*})$
    \FOR{$t=0,\,2,\,\ldots,\,t_{\rm{max}}$}
    \FOR{$i^{*}=1,2,\ldots,n^{*}$}
    \FOR{$j=1,2,\ldots,n$}
    \STATE
    Caluculate $w_{j}(\bm{s}_{i^{*}})^{(t)}$ from equation (\ref{func:w}). \\
    \ENDFOR
    \STATE
    Sampling $\widehat{\bm{\beta}}(\bm{s}_{i^{*}})^{(t)}$ from equation (\ref{func:beta}). \\
    \STATE
    Sampling $\widehat{y}(\bm{s}_{i^{*}})^{(t)}$ from equation (\ref{func:y}). \\
    \ENDFOR
    \ENDFOR
    \end{algorithmic}
\end{algorithm}
\end{figure}
\section{Numerical studies\label{chap3:numerical studies}}
In this section, we describe the details and results of numerical studies conducted to evaluate the prediction performance of the proposed method.
\subsection{Settings}
\begin{figure}[H]
\centering
\begin{minipage}[b]{0.49\columnwidth}
    \centering
	\begin{center}
		\includegraphics[scale=0.44,clip]{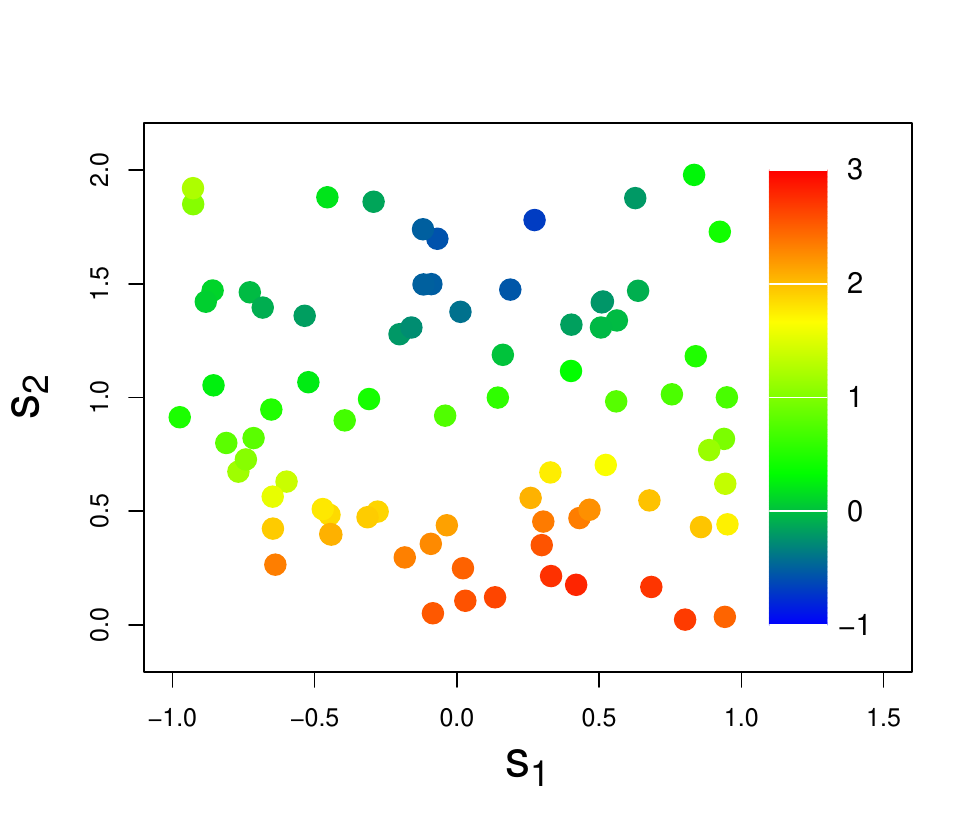}
	\end{center}
	\subcaption{Scenario \ref{a1}}
\end{minipage}
\begin{minipage}[b]{0.49\columnwidth}
	\begin{center}
		\includegraphics[scale=0.44,clip]{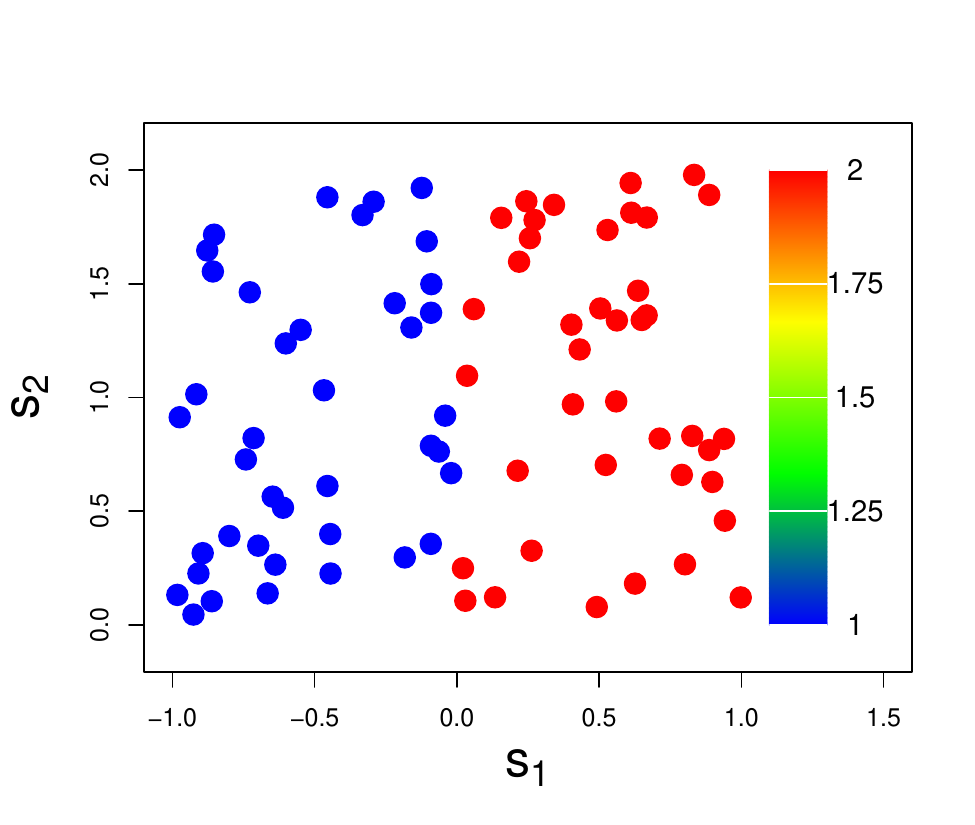}
	\end{center}
        \subcaption{Scenario \ref{a2}}
\end{minipage}
\begin{minipage}[b]{0.49\columnwidth}
	\begin{center}
		\includegraphics[scale=0.44,clip]{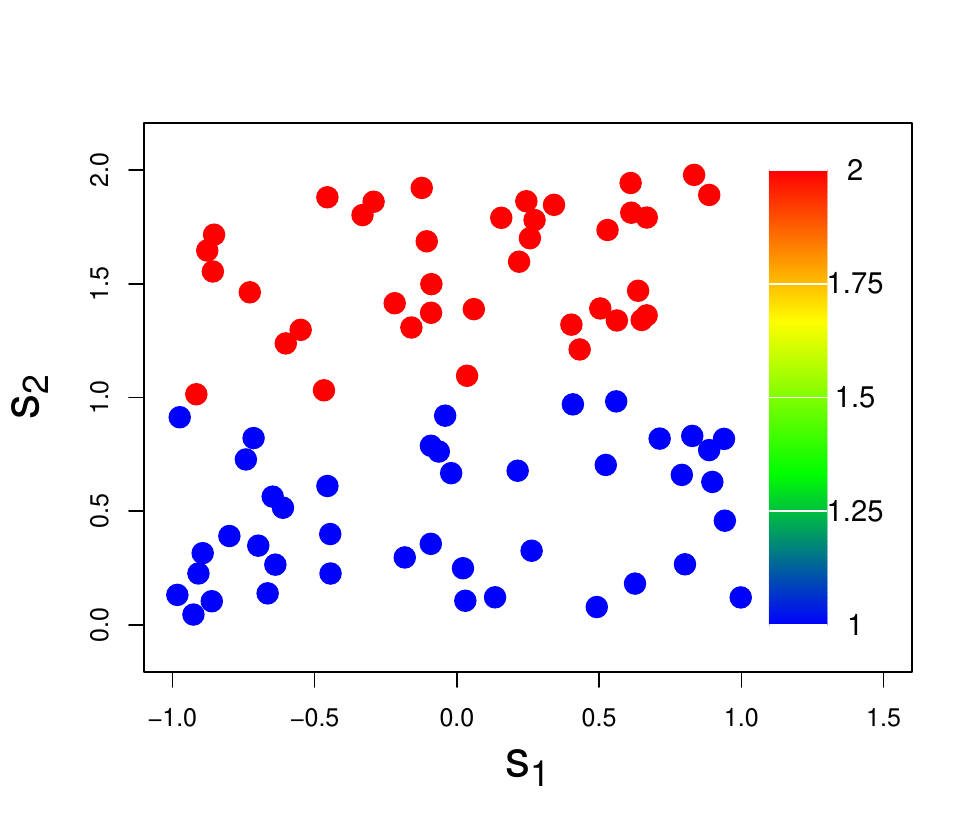}
	\end{center}
        \subcaption{Scenario \ref{a3}}
\end{minipage}
\caption{Examples of coefficients for each scenario\label{fig:senario_a}}
\end{figure}

In data generation, we first generated the location, followed by the data associated with the location.
In all scenarios, we randomly generate $n=1000$ spatial locations $\bm{s}_{1}, \, \bm{s}_{2}, \, \ldots, \, \bm{s}_{n}$ in the square domain $[-1,1] \times [0,2]$ as follows.
\begin{align*}
    \bm{s}_{i} = (s_{i1}, \, s_{i2})^{\prime}, \, s_{i1}  \stackrel{\mathrm{i.i.d.}}{\sim} {\rm{U}}(-1,1), \, s_{i2}  \stackrel{\mathrm{i.i.d.}}{\sim} {\rm{U}}(0,2).
\end{align*}
The objective variables and covariates are generated as follows, where $p=3$.
\begin{align*}
    y(\bm{s}_{i}) = \sum_{k=1}^{p} x_{k}(\bm{s}_{i}) \beta_{k}(\bm{s}_{i}) + \varepsilon(\bm{s}_{i}), &\quad i=1, \, 2, \, \ldots, \, n, \\
    x_{k}(\bm{s}_{i}) \stackrel{\mathrm{i.i.d.}}{\sim} {\rm{N}}(0, \, 1), &\quad i = 1, \, 2, \, \ldots, \, n; \,k = 1, \, 2, \, \ldots, \, p, \\
    \varepsilon(\bm{s}_{i}) \stackrel{\mathrm{i.i.d.}}{\sim} {\rm{N}}(0, \, \sigma^{2}), &\quad i = 1, \, 2, \, \ldots, \, n.
\end{align*}
The true values of the coefficients consider the following three structures of coefficients.
\begin{enumerate}[label=(\textbf{a\arabic*})]
    \item The coefficients differ \textbf{smoothly}.\label{a1}
    \item The clusters are different in the left and right regions, and the coefficients within the clusters are the same.\label{a2}
    \item The clusters are different in the upper and lower regions, and the coefficients within the clusters are the same.\label{a3}
\end{enumerate}
In \ref{a1}, the values are different for each location.
The correlation of coefficients between adjacent locations is strong, and the coefficients have similar values.
This structure of the coefficients is assumed for GWR and BGWR.
In \ref{a2} and \ref{a3}, we assume a cluster structure depending on location so that spatially close locations belong to the same cluster and have the same coefficient values.
We generated the true value of the coefficient at \ref{a1} by
\begin{align*}
    \bm{\beta}_{k} = (\beta_{k}(\bm{s}_{1}), \, \beta_{k}(\bm{s}_{2}), \, \ldots, \, \beta_{k}(\bm{s}_{n}))^{\prime} \sim {\rm{MN}}(\beta_{k}^{*}\bm{1}_{n}, \, \sigma^{2}\bm{H}), \, \quad k = 1, \, 2, \, \ldots, \, p.
\end{align*}
Here, we set the covariance matrix $\bm{H} \in \mathbb{R}^{n \times n}$ so that there is a correlation between adjacent locations,
\begin{align*}
    \bm{H} = 
    \begin{pmatrix}
        1                 & w_{2}(\bm{s}_{1}) & \cdots  & w_{n}(\bm{s}_{1}) \\
        w_{1}(\bm{s}_{2}) & 1                 & \cdots  & w_{n}(\bm{s}_{2}) \\
        \vdots            & \vdots            & \ddots  & \vdots     \\
        w_{1}(\bm{s}_{n}) & w_{2}(\bm{s}_{n}) & \cdots  & 1
    \end{pmatrix}
    ,
\end{align*}
where $w_{j}(\bm{s}_{i})=f(d_{ij}|h)=\exp\{- (d_{ij}^{2}/2h^{2})\}$.
We set the mean vector of coefficients $\bm{\beta}^{*} = (\beta_{1}^{*}, \, \beta_{2}^{*}, \, \beta_{3}^{*})^{\prime}$, the parameter of error variance $\sigma^{2}$, and bandwidth $h$ as described below.
\begin{align*}
    \bm{\beta}^{*}=(1,1,1)^{\prime}, \quad \sigma^{2} = 1^{2},\quad h = 0.5.
\end{align*}
We generated the true value of the coefficient at \ref{a2} and \ref{a3} by
\begin{align*}
    \bm{\beta}(\bm{s}_{i}) = \bm{\beta}^{(g)}, \quad \bm{s}_{i} \in C_{g}.
\end{align*}
Let $C_{g}$ denote the set of locations belonging to the $g$-th cluster ($g=1,\,2$).
We set the true values of the coefficients as follows.
\begin{align*}
    \bm{\beta}^{(1)} = (1,1,1)^{\prime}, \\
    \bm{\beta}^{(2)} = (2,2,2)^{\prime}.
\end{align*}
We set the clusters of locations based on
\begin{align*}
    \ref{a2}: \quad \bm{s}_{i} \in 
    &\left\{
        \begin{array}{ll}
            C_{1} \quad (s_{i1} \leq 0)\\
            C_{2} \quad (s_{i1} > 0),
                \end{array}
    \right. \\
    \ref{a3}: \quad \bm{s}_{i} \in 
    &\left\{
        \begin{array}{ll}
            C_{1} \quad (s_{i2} \leq 0)\\
            C_{2} \quad (s_{i2} > 0).
                \end{array}
    \right.
\end{align*}
Figure \ref{fig:senario_a} shows an example of coefficients generated in scenarios \ref{a1}, \ref{a2} and \ref{a3}.

\begin{figure}[htb]
\begin{minipage}[b]{0.49\columnwidth}
	\begin{center}
		\includegraphics[width=7cm,clip]{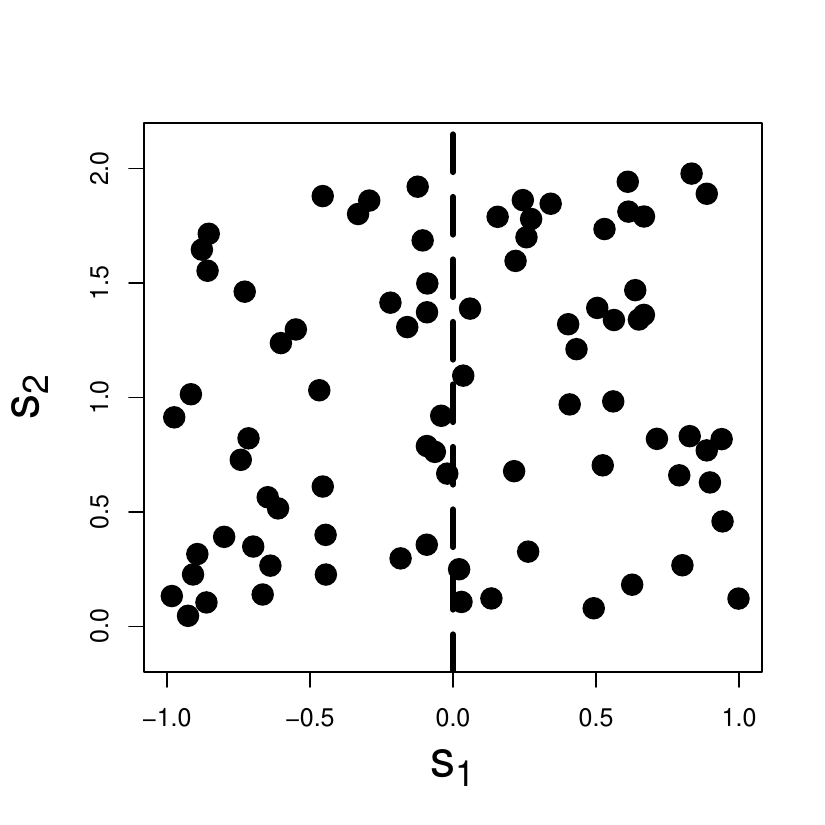}
	\end{center}
        \subcaption{Scenario \ref{b1}}
    \label{fig:r0.5}
\end{minipage}
\begin{minipage}[b]{0.49\columnwidth}
	\begin{center}
		\includegraphics[width=7cm,clip]{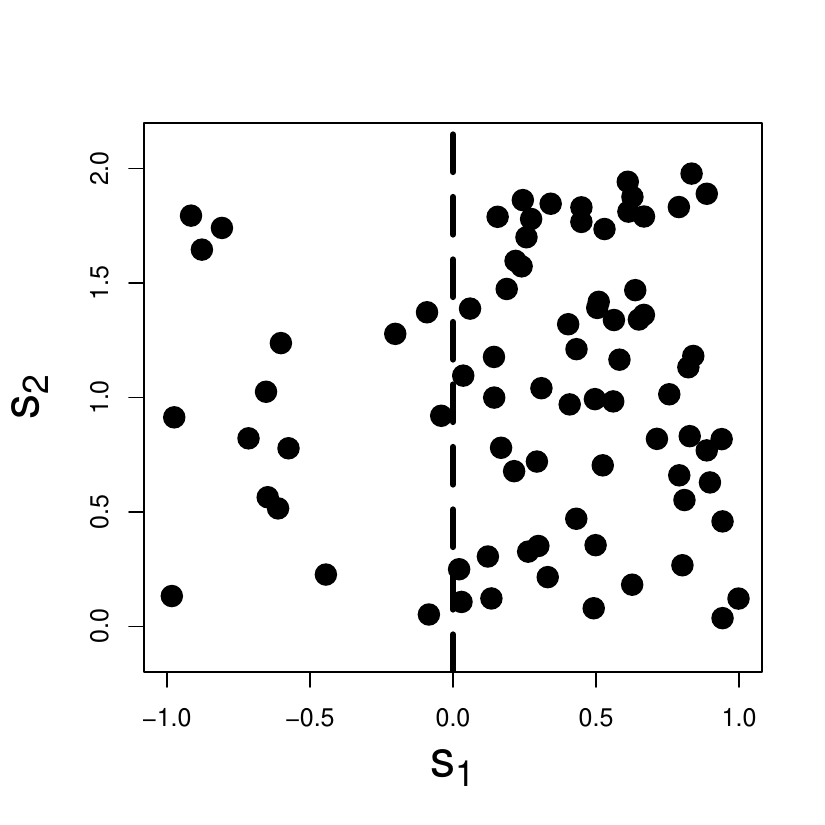}
	\end{center}
        \subcaption{Scenario \ref{b2}}
    \label{fig:r0.8}
\end{minipage}
\caption{Examples of observation locations for each scenario\label{fig:senario_b}}
\end{figure}

Next, we describe the division rule between observed and prediction locations.
In the numerical studies, we extracted the predicted and observed locations from the generated locations.
We randomly select $50$ locations for the prediction.
We set the number of observation locations to $100$ and consider the following two conditions for the selection of observation locations.
\begin{enumerate}[label=(\textbf{b\arabic*})]
    \item The observed location is \textbf{uniformly} obtained within the region.\label{b1}
    \item The observed location is \textbf{non-uniformly} obtained within the region.\label{b2}
\end{enumerate}
In \ref{b1}, we randomly select locations from the generated data.
In \ref{b2}, we extract locations from the generated data by setting the number of locations to $1:4$ on the left and right sides of the entire region.
\ref{b1} is the situation assumed by GWR and BGWR.
\ref{b2} is the situation assumed by the BGWSR, where there are not many locations near a given point in the sparse left-hand area.
Therefore, in GWR and BGWSR, the estimated coefficients are expected to differ among adjacent locations.
Figure \ref{fig:senario_b} shows an example of locations generated in the scenarios of \ref{b1} and \ref{b2}.
The dashed lines indicate the left and right sides of the target area, divided by the center $0$; in the scenario of \ref{b2}, the observation locations on the left side are sparse.

Table \ref{tab:senarios} shows the scenarios for the numerical studies.
The scenario combining setting \ref{a3} and \ref{b2} is
the same as the scenario with the rotated region in setting 3 and is omitted in these numerical studies because it does not affect the weights used to estimate coefficients.
The situation assumed for GWR and BGWR is Scenario 1; the situations assumed for BGWSR are Scenarios 2, 4, and 5.
\begin{table}[htb]
\renewcommand{\arraystretch}{1.5}
\centering
\caption{Scenarios}
\label{tab:senarios}
\begin{tabular}{c|cc}
\hline
& true values of coefficients & \begin{tabular}{c}Left-right ratio of observation locations\\(Whether there is sparse or dense)\end{tabular} \\
\hline
Scenario1 & smoothly \ref{a1} & $1:1$ \ref{b1} \\
Scenario2 & smoothly \ref{a1} & $1:4$ \ref{b2} \\
Scenario3 & left and right clusters \ref{a2} & $1:1$ \ref{b1} \\
Scenario4 & left and right clusters \ref{a2} & $1:4$ \ref{b2} \\
Scenario5 & upper and lower clusters \ref{a3} & $1:4$ \ref{b1} \\
\hline
\end{tabular}
\end{table}
\subsection{Comparative methods and Hyperparameters settings}
We use our proposed method, BGWSR, and the existing methods, GWR and BGWR, in the numerical studies.
We apply two BGWSR methods: BGWSR with Adjacency Estimation (BGWSR-AE), which determines the adjacency range differently for each location, and BGWSR, which uses a common adjacency range for all locations.
We set the hyperparameters in BGWSR and BGWSR-AE as follows:
\begin{align*}
    (r, \, q, \, r_{1}, \, q_{1}, \, r_{2}, \, q_{2}, \, a)^{\prime} = (0.1, \, 0.1, \, 0.1, \, 0.1, \, 0.1, \, 0.1, \, 3)^{\prime}.
\end{align*}
The parameters $r_{\rm{BGWR}}, \, q_{\rm{BGWR}}$ for the prior distribution of the error variance $\sigma^{2}$ in BGWR were similarly set by
\begin{align*}
    (r_{\rm{BGWR}}, \, q_{\rm{BGWR}})^{\prime} = (0.1, \, 0.1)^{\prime}.
\end{align*}
The method for determining the adjacent locations of BGWR is based on \cite{ma2021}, sampling the bandwidth $h_{\rm{BGWR}}$ using the MH method.
The prior distribution of $h_{\rm{BGWR}}$ is the following Uniform distribution.
\begin{align*}
    h_{\rm{BGWR}} \sim {\rm{U}}(0, \, 3).
\end{align*}
We selected $\{0.10, \, 0.15, \, \ldots, \, 3.00\}$ as the candidate bandwidth $h_{\rm{GWR}}$ in GWR and determined it using 5-fold cross validation.
\subsection{Performance measures}
We evaluate in the numerical studies the estimation performance of the coefficients and objective variable at the observed and prediction location.
We use the mean squared error (MSE) as the performance measure and define the MSE at the observation location as
\begin{align*}
    \frac{1}{n} \sum_{i=1}^{n} \left( \widehat{\beta}_{k}(\bm{s}_{i}) - \beta_{k}(\bm{s}_{i}) \right)^{2}, \\
    \frac{1}{n} \sum_{i=1}^{n} \left( \widehat{y}(\bm{s}_{i}) - y(\bm{s}_{i}) \right)^{2},
\end{align*}
where $\widehat{\bm{\beta}}(\bm{s}_{i})=(\widehat{\beta}_{1}, \, \widehat{\beta}_{2}, \, \ldots, \, \widehat{\beta}_{p})^{\prime}, \, \widehat{y}(\bm{s}_{i})$ are the estimated value of the coefficients and the objective variable at observed locations; $\bm{\beta}(\bm{s}_{i}), \, y(\bm{s}_{i})$ are the true value of the coefficient and the objective variables at observed locations.
The MSE at the prediction location is defined as
\begin{align*}
    \frac{1}{n^{*}} \sum_{i^{*}=1}^{n^{*}} \left( \widehat{\beta}_{k}(\bm{s}_{i^{*}}) - \beta_{k}(\bm{s}_{i^{*}}) \right)^{2}, \\
    \frac{1}{n^{*}} \sum_{i^{*}=1}^{n^{*}} \left( \widehat{y}(\bm{s}_{i^{*}}) - y(\bm{s}_{i^{*}}) \right)^{2},
\end{align*}
where $\bm{s}_{i^{*}} \in D$ is the prediction location, and 
$\widehat{\bm{\beta}}(\bm{s}_{i^{*}}), \, \widehat{y}(\bm{s}_{i ^{*}})$ are the predicted value of the coefficients and the objective variables at prediction location $\bm{s}_{i^{*}}$.
The true value of the coefficient at location $\bm{s}_{i^{*}}$ is $\bm{\beta}(\bm{s}_{i^{*}})$, and that of the objective variable is $y(\bm{s}_{i^{*}})$.

\subsection{Summary of Numerical Studies Results}\label{sec:result_numerical ex}
In the numerical studies, we generated data for each scenario $10$ times each and calculated the median of the resulting evaluation metrics for each method applied.
Table \ref{tab:res_numerical} shows the median MSE for each method.

\begin{figure}[H]
\centering
\begin{minipage}[b]{0.49\columnwidth}	
  \includegraphics[width=\linewidth,clip]{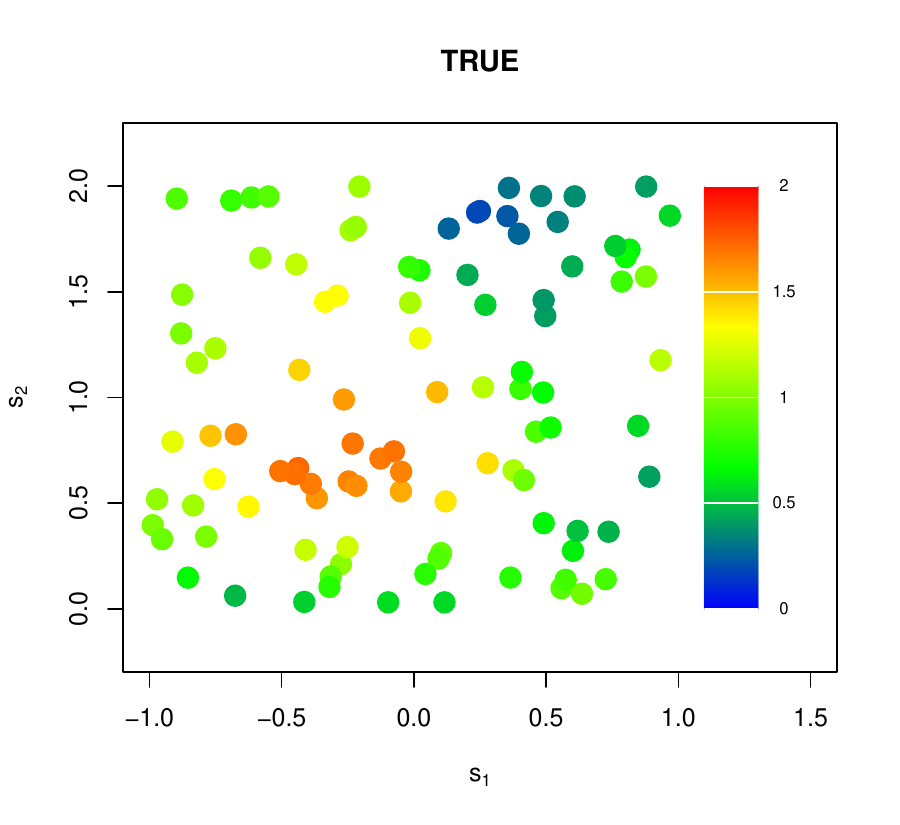}
\end{minipage}
\\
\begin{minipage}[b]{0.49\columnwidth}
	\begin{center}
		\includegraphics[width=\linewidth,clip]{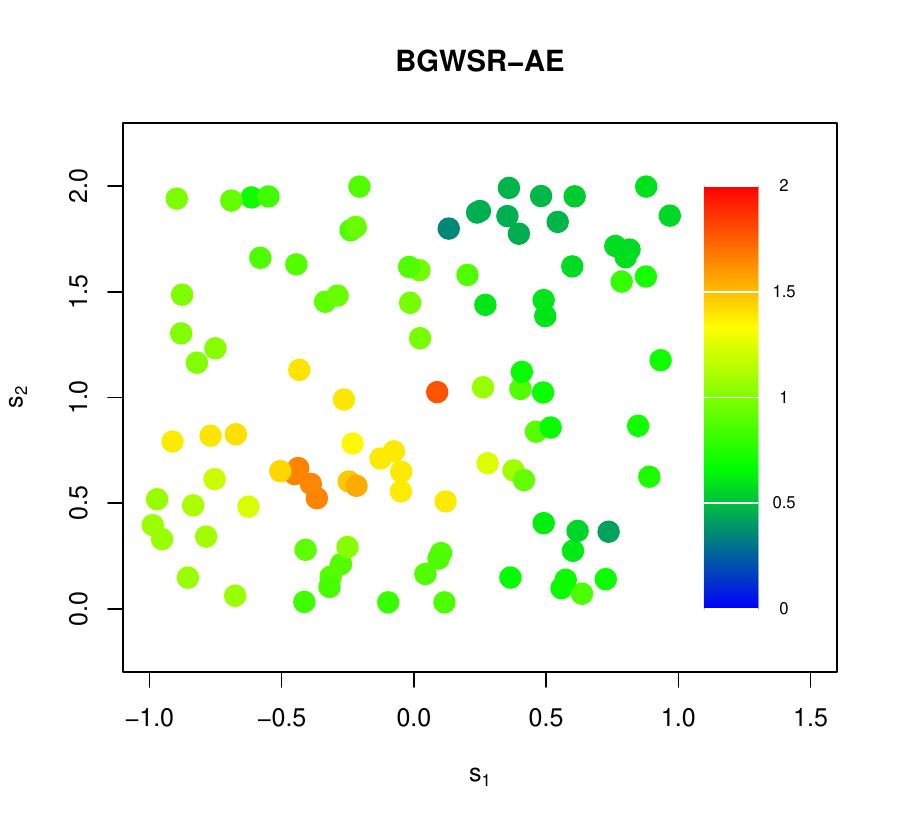}
	\end{center}
\end{minipage}
\begin{minipage}[b]{0.49\columnwidth}
	\begin{center}
		\includegraphics[width=\linewidth,clip]{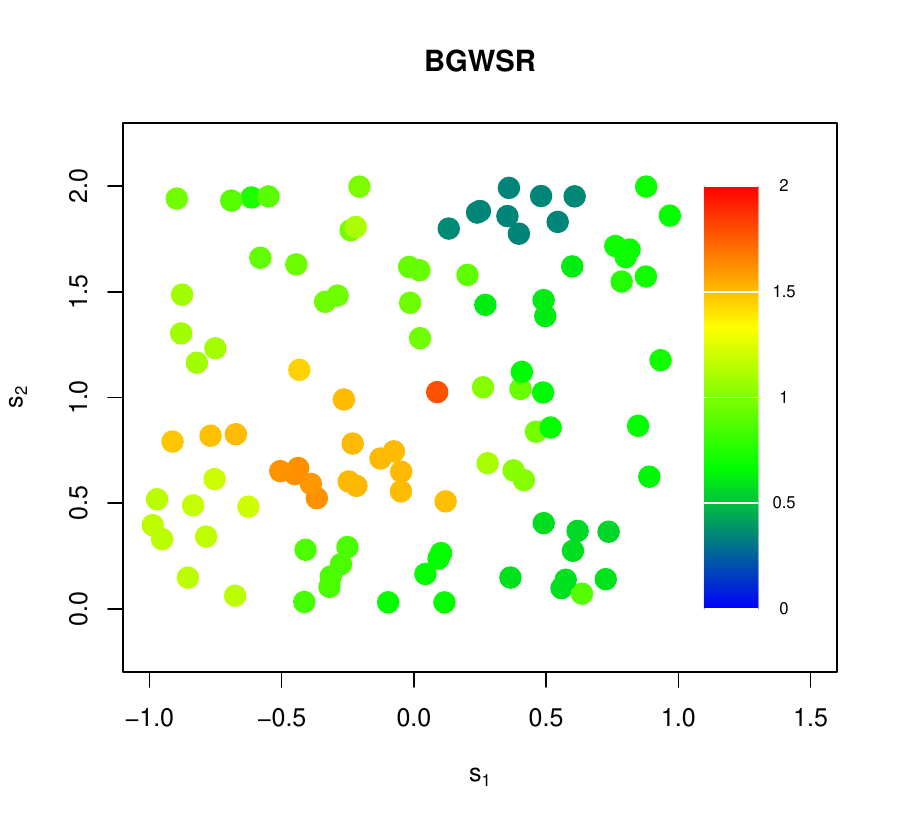}
	\end{center}
\end{minipage}
\\
\begin{minipage}[b]{0.49\columnwidth}
	\begin{center}
		\includegraphics[width=\linewidth,clip]{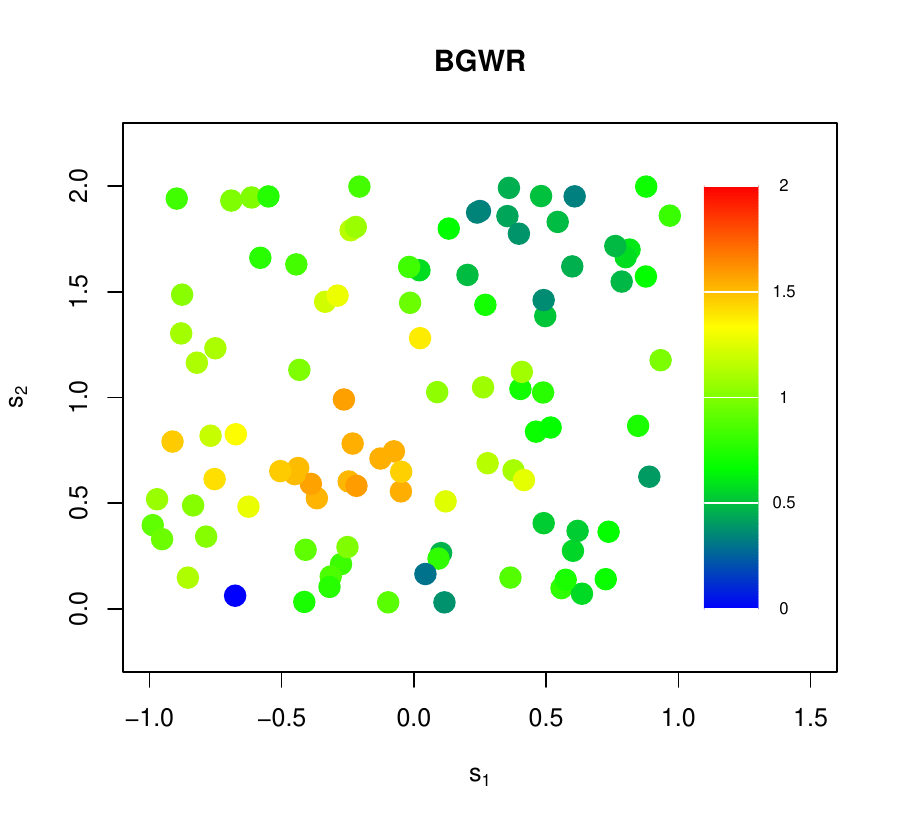}
	\end{center}
\end{minipage}
\begin{minipage}[b]{0.49\columnwidth}
	\begin{center}
		\includegraphics[width=\linewidth,clip]{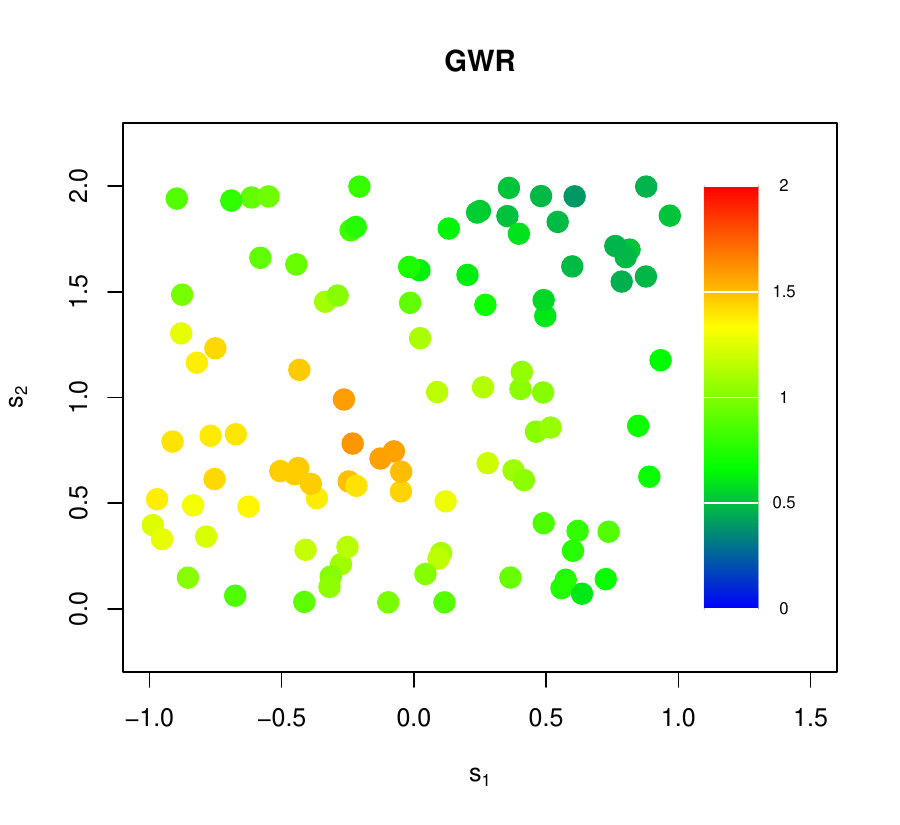}
	\end{center}
\end{minipage}
\caption{True and estimated values of coefficients for observation locations in Scenario 1\label{fig:res1}}
\end{figure}

In Scenario 1, with respect to the MSE for the observation locations, the BGWSR performed the best for the objective variables and some of the coefficients, and the BGWSR-AE performed equally well.
Similarly, for the MSE for the prediction location, the BGWSR performed best, and the BGWSR-AE performed equally well for the objective variables and most of the coefficients.
Figure \ref{fig:res1} shows an example of the true values of the coefficients and the estimates generated by each method for the location of the observation in Scenario 1.
BGWSR-AE tended to produce estimates closer to the overall average rather than abrupt changes in the coefficients. This is because the application of Bayesian Fused Lasso fails to capture the rapid changes.

\begin{figure}[H]
\centering
\begin{minipage}[b]{0.49\columnwidth}
	\begin{center}
		\includegraphics[width=\linewidth,clip]{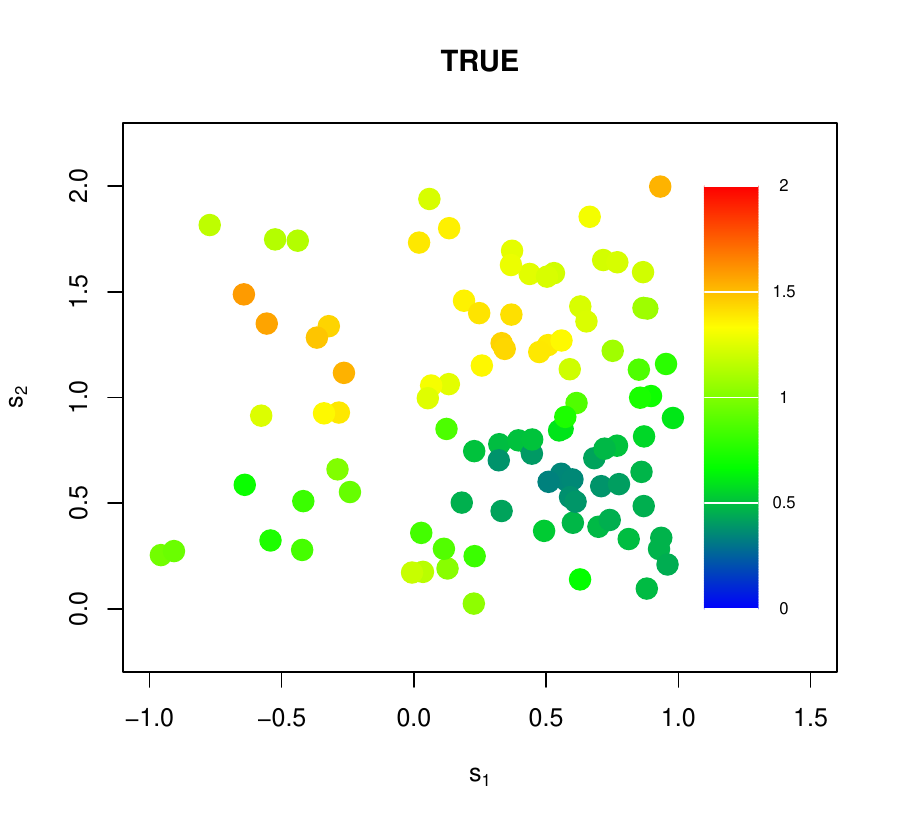}
	\end{center}
\end{minipage}
\\
\begin{minipage}[b]{0.49\columnwidth}
	\begin{center}
		\includegraphics[width=\linewidth,clip]{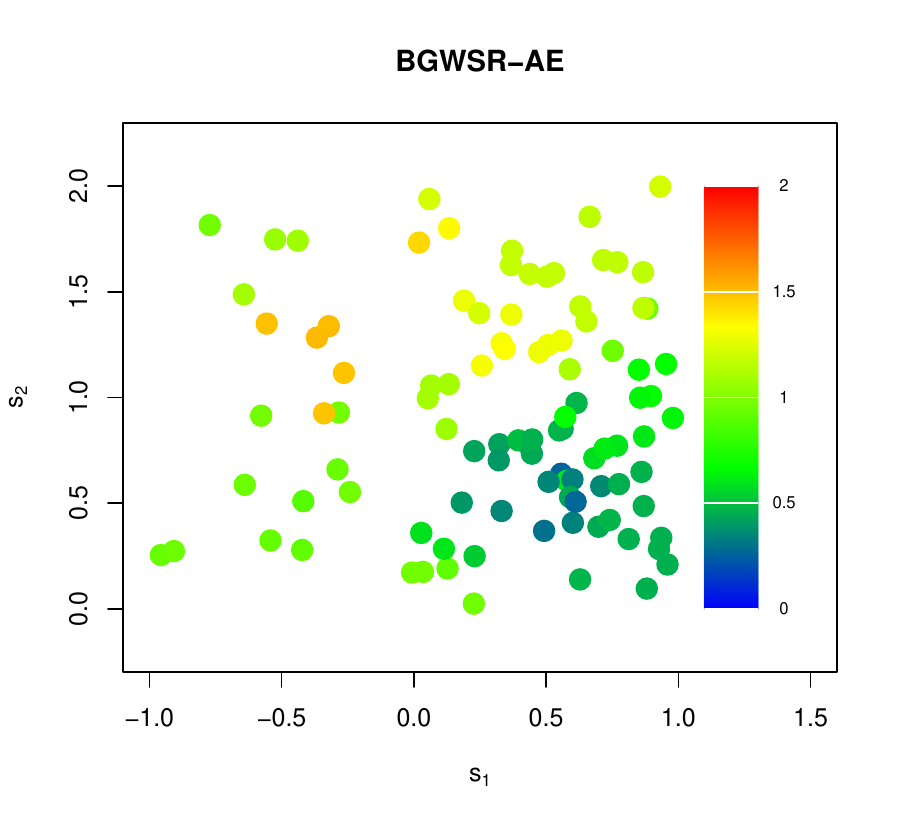}
	\end{center}
\end{minipage}
\begin{minipage}[b]{0.49\columnwidth}
	\begin{center}
		\includegraphics[width=\linewidth,clip]{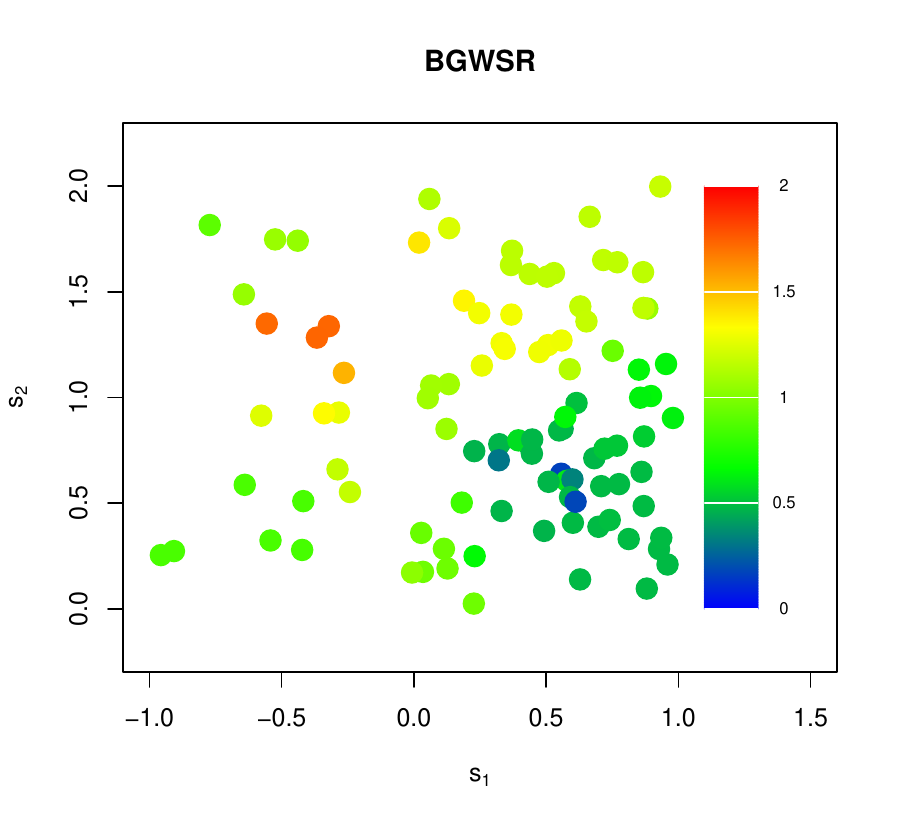}
	\end{center}
\end{minipage}
\\
\begin{minipage}[b]{0.49\columnwidth}
	\begin{center}
		\includegraphics[width=\linewidth,clip]{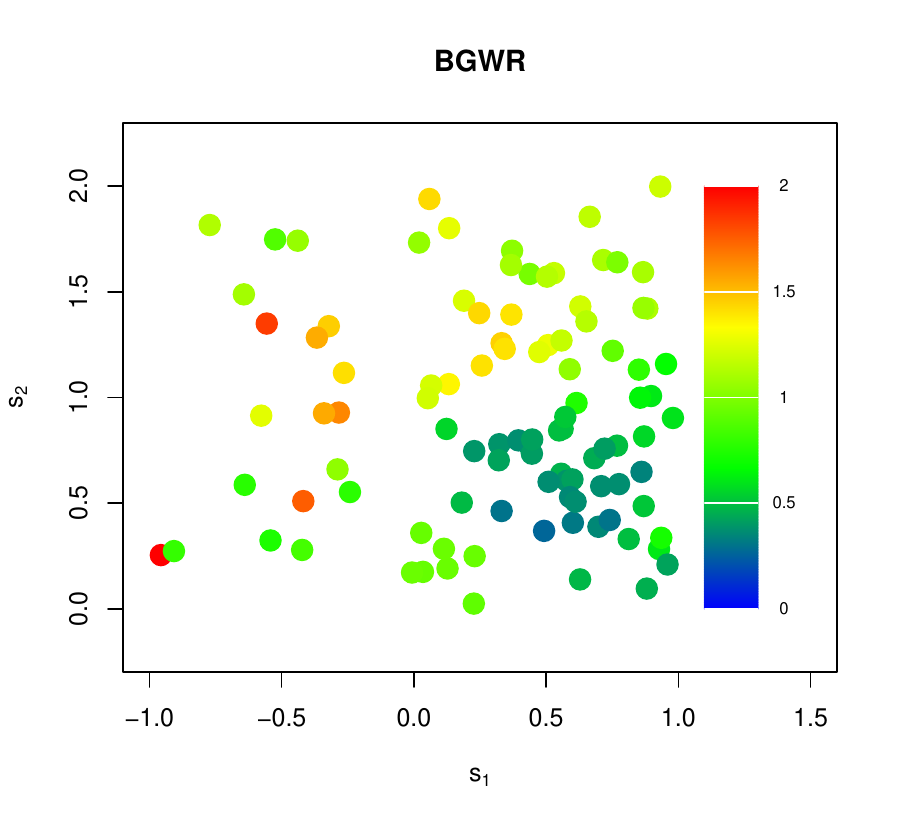}
	\end{center}
\end{minipage}
\begin{minipage}[b]{0.49\columnwidth}
	\begin{center}
		\includegraphics[width=\linewidth,clip]{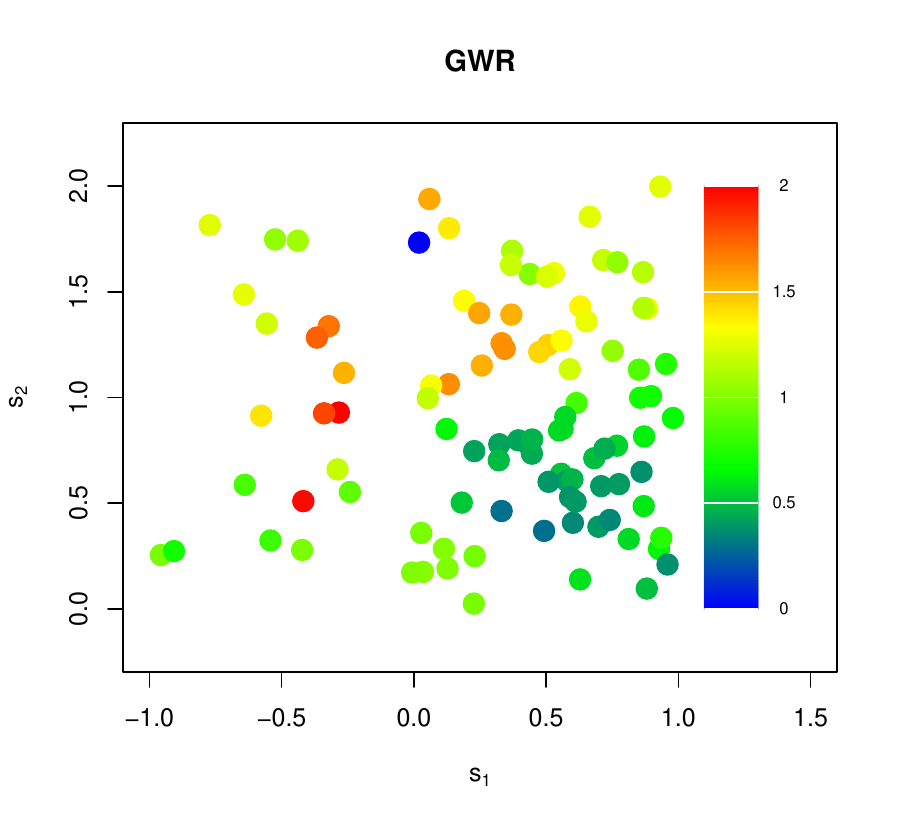}
	\end{center}
\end{minipage}
\caption{True and estimated values of coefficients for observation locations in Scenario 2\label{fig:res2}}
\end{figure}

In Scenario 2, for the MSE for the observation locations, BGWSR-AE yielded the best results for the objective variables and some coefficients.
Based on the results of the MSE for the prediction locations, BGWSR-AE and BGWSR had a high performance for the objective variables and coefficients.
In Scenario 2, BGWSR-AE and BGWSR can guarantee the similarity of locations by Bayesian Fused Lasso.
Figure \ref{fig:res2} shows an example of the true values of the coefficients and the estimated values by each method for the two locations in Scenario 2.
BGWSR-AE and BGWSR estimate values close to the true values even in regions where observed locations are sparse.
BGWR and GWR have locations with extremely large estimated values in sparse areas.
This is because the BGWR and GWR set the range of adjacent locations to be relatively small to match the dense locations.
Consequently, extreme values are estimated for locations where the data size is small.

\begin{figure}[H]
\centering
\begin{minipage}[b]{0.49\columnwidth}
	\begin{center}
		\includegraphics[width=\linewidth,clip]{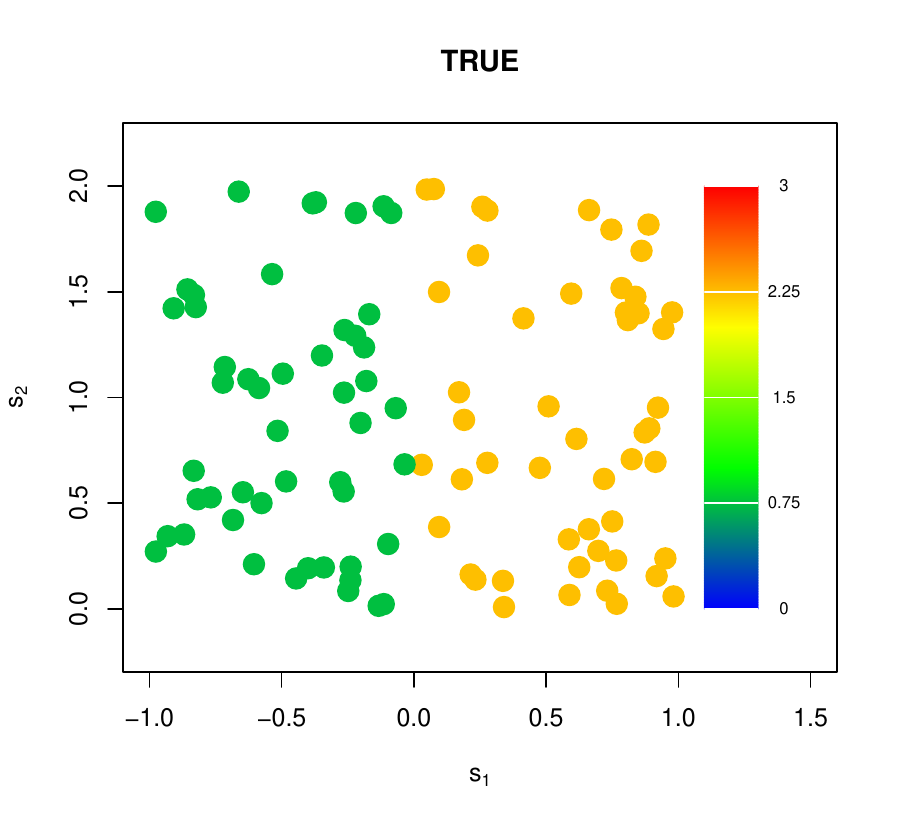}
	\end{center}
\end{minipage}
\\
\begin{minipage}[b]{0.49\columnwidth}
	\begin{center}
		\includegraphics[width=\linewidth,clip]{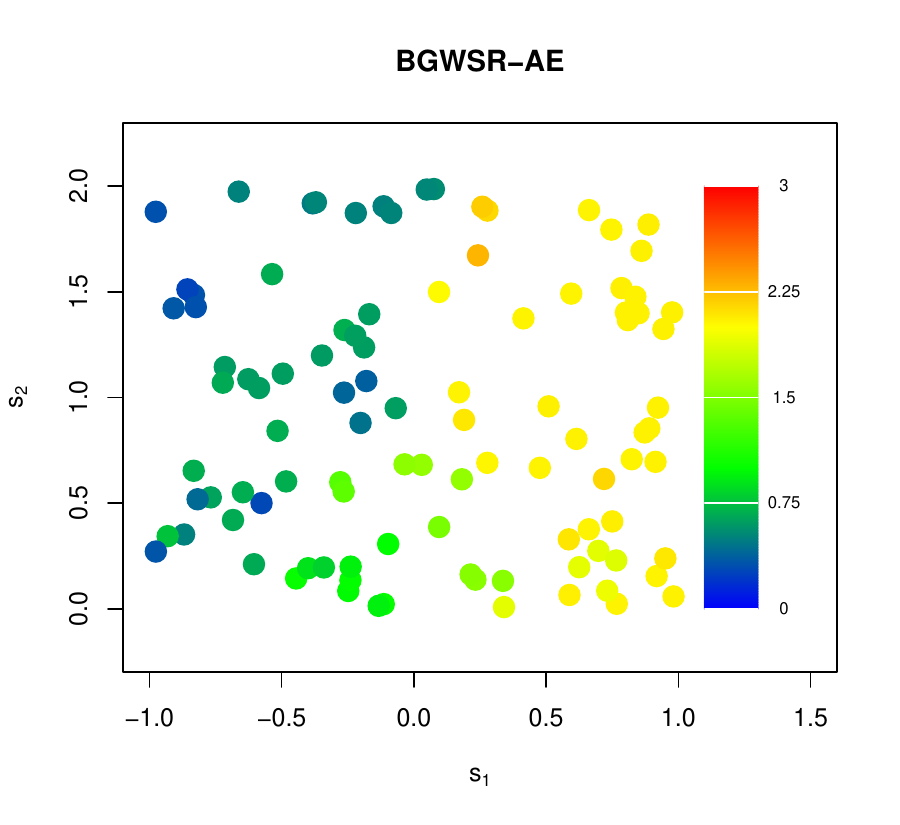}
	\end{center}
\end{minipage}
\begin{minipage}[b]{0.49\columnwidth}
	\begin{center}
		\includegraphics[width=\linewidth,clip]{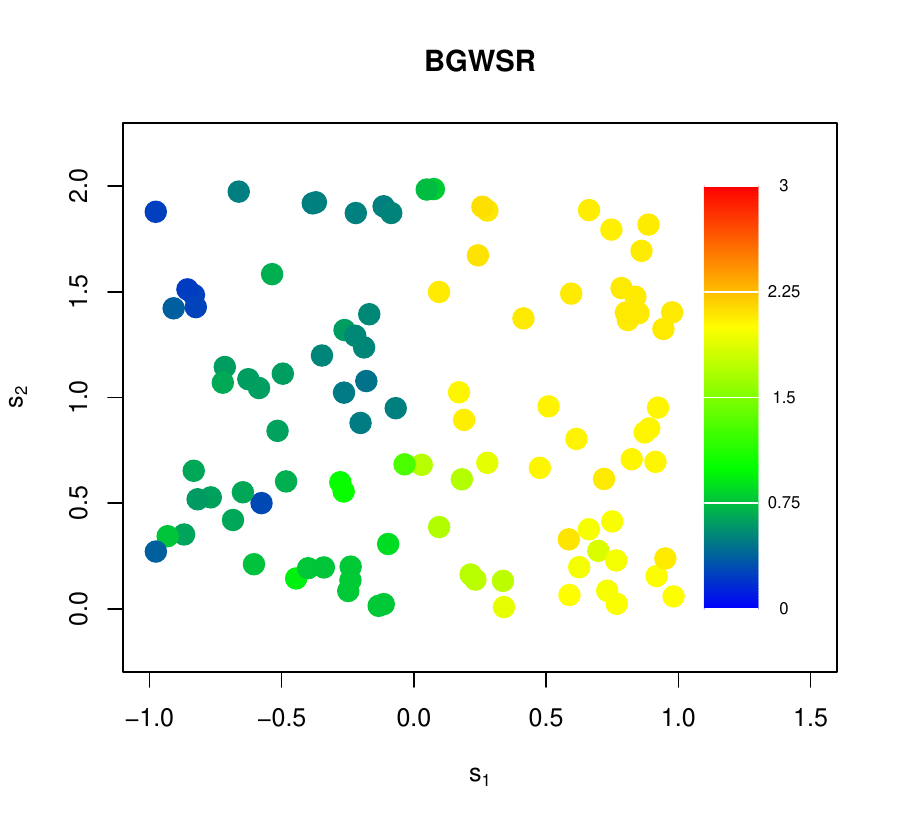}
	\end{center}
\end{minipage}
\\
\begin{minipage}[b]{0.49\columnwidth}
	\begin{center}
		\includegraphics[width=\linewidth,clip]{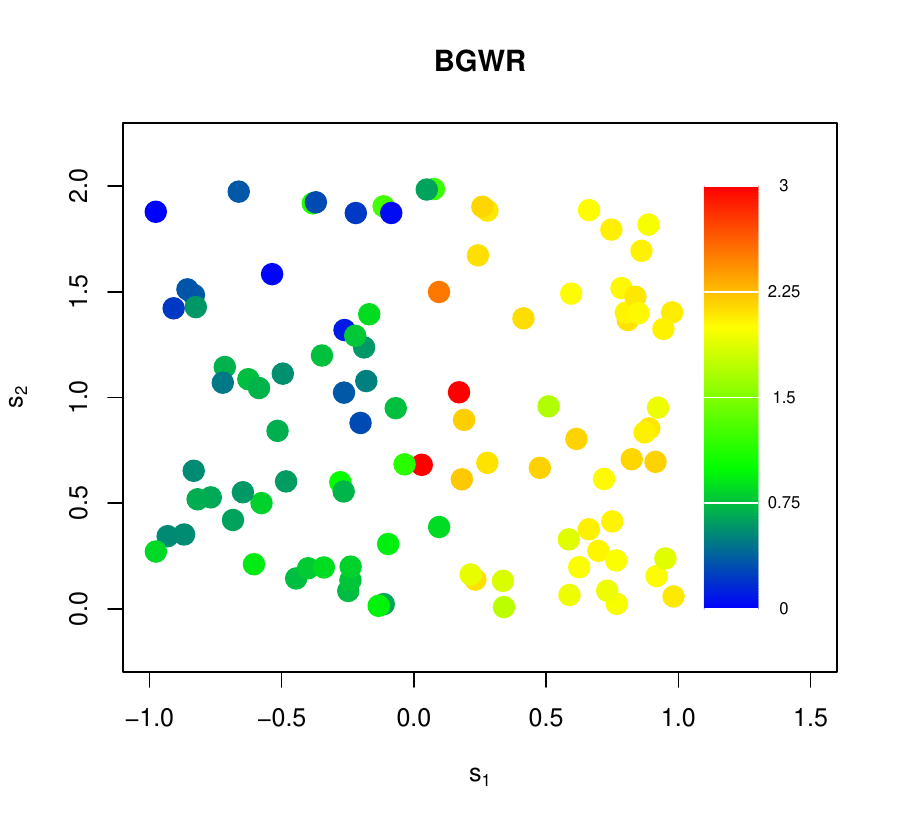}
	\end{center}
\end{minipage}
\begin{minipage}[b]{0.49\columnwidth}
	\begin{center}
		\includegraphics[width=\linewidth,clip]{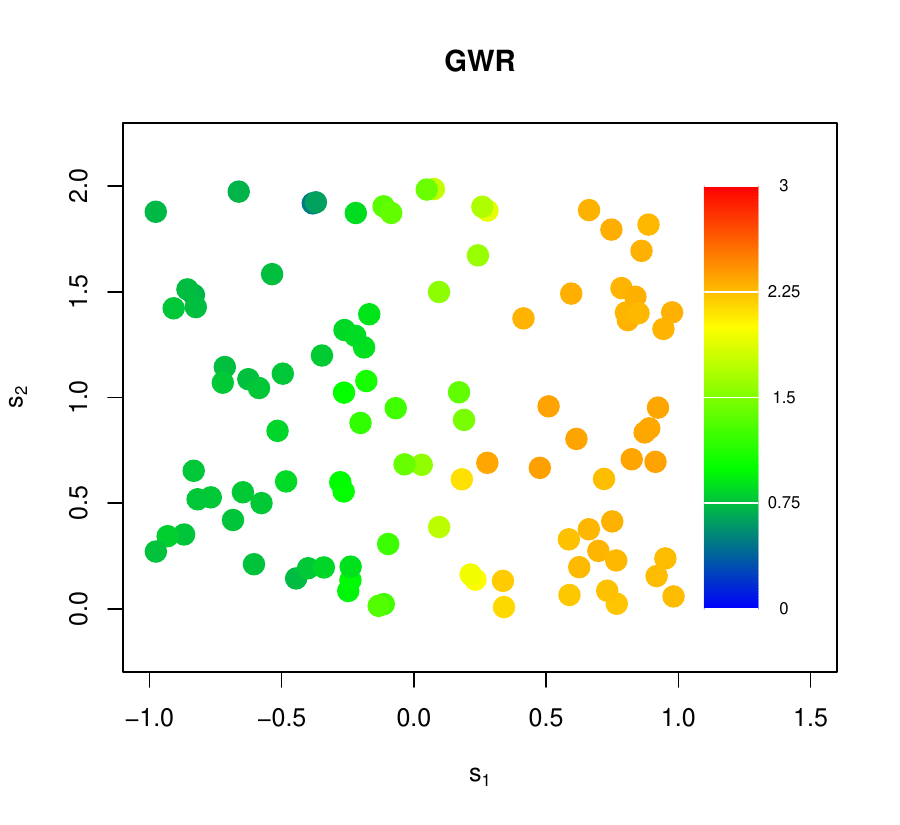}
	\end{center}
\end{minipage}
\caption{True and estimated values of coefficients for observation locations in Scenario 3\label{fig:res3}}
\end{figure}

In Scenario 3, BGWSR-AE and BGWSR exhibited high performance for objective variables and coefficients at the observed location.
According to the MSE for the prediction location, the BGWSR performed the best for the objective variables and coefficients.
Figure \ref{fig:res3} shows the true values of the coefficients for the observation locations for Scenario 3 and examples of the estimated values for each method.
Scenario 3 has a cluster structure for locations, suggesting that BGWSR-AE and BGWSR can estimate coefficients more uniformly within clusters than the other methods.
However, there are many locations where the coefficients are shrunk to $0$ as a whole compared to the true values.
Nevertheless, in the BGWR, the estimation is unstable near the edge of the cluster.
For the GWR, the estimated coefficients varied gently near the edge of the cluster and within clusters.
This suggests that the estimation performance of BGWR and GWR deteriorates near cluster boundaries.

\begin{figure}[H]
\centering
\begin{minipage}[b]{0.49\columnwidth}
	\begin{center}
		\includegraphics[width=\linewidth,clip]{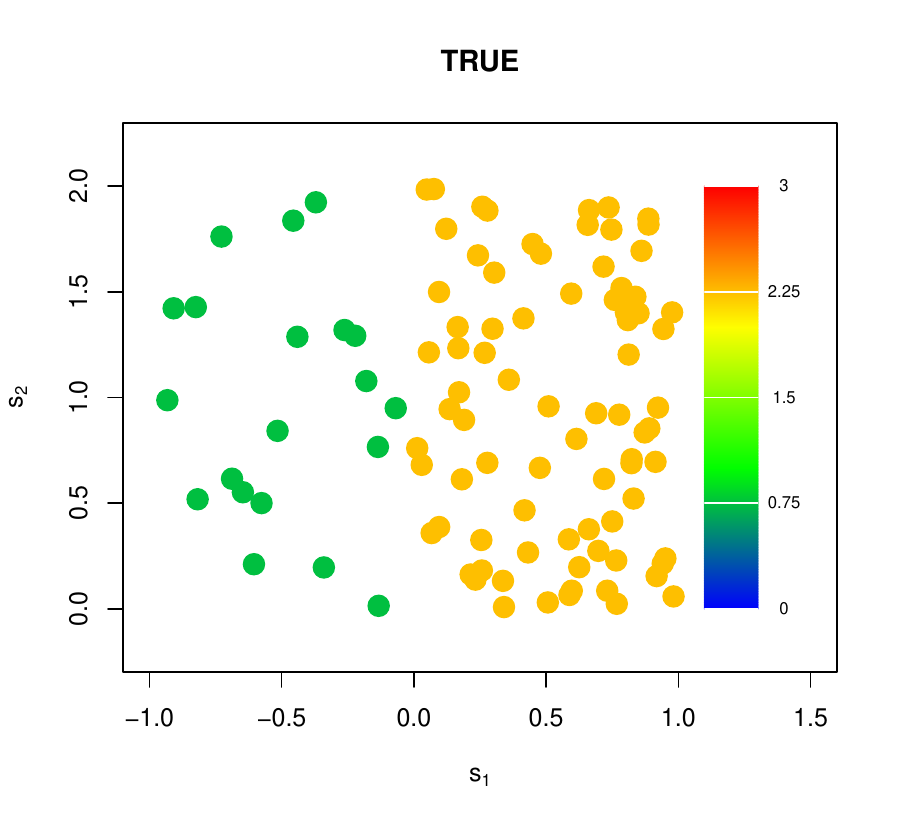}
	\end{center}
\end{minipage}
\\
\begin{minipage}[b]{0.49\columnwidth}
	\begin{center}
		\includegraphics[width=\linewidth,clip]{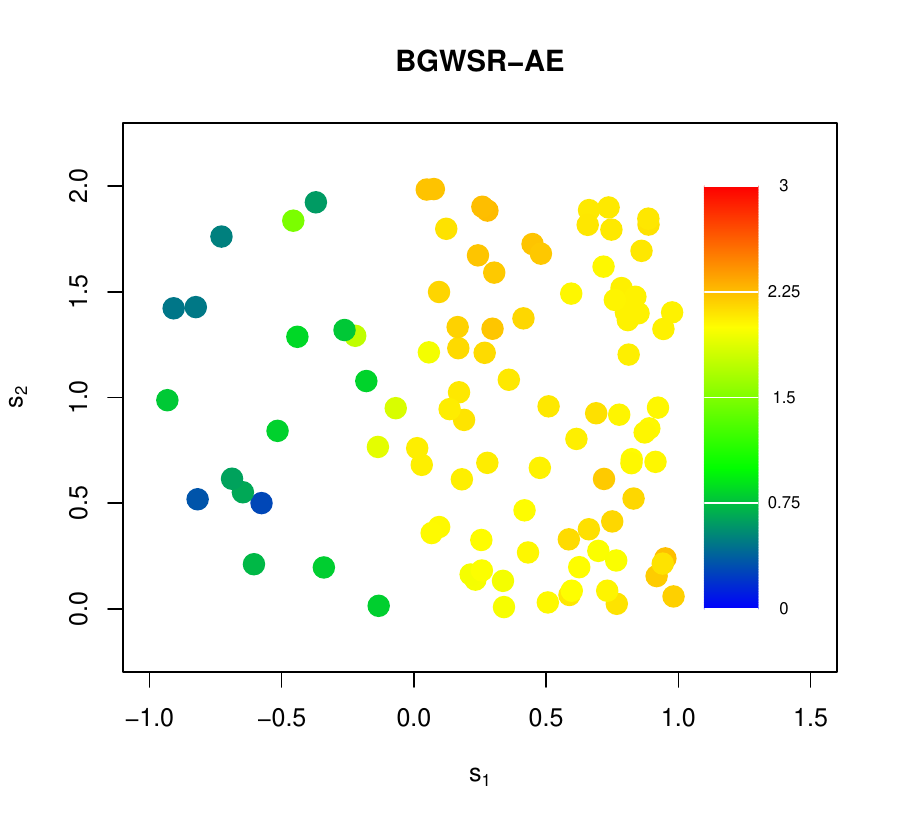}
	\end{center}
\end{minipage}
\begin{minipage}[b]{0.49\columnwidth}
	\begin{center}
		\includegraphics[width=\linewidth,clip]{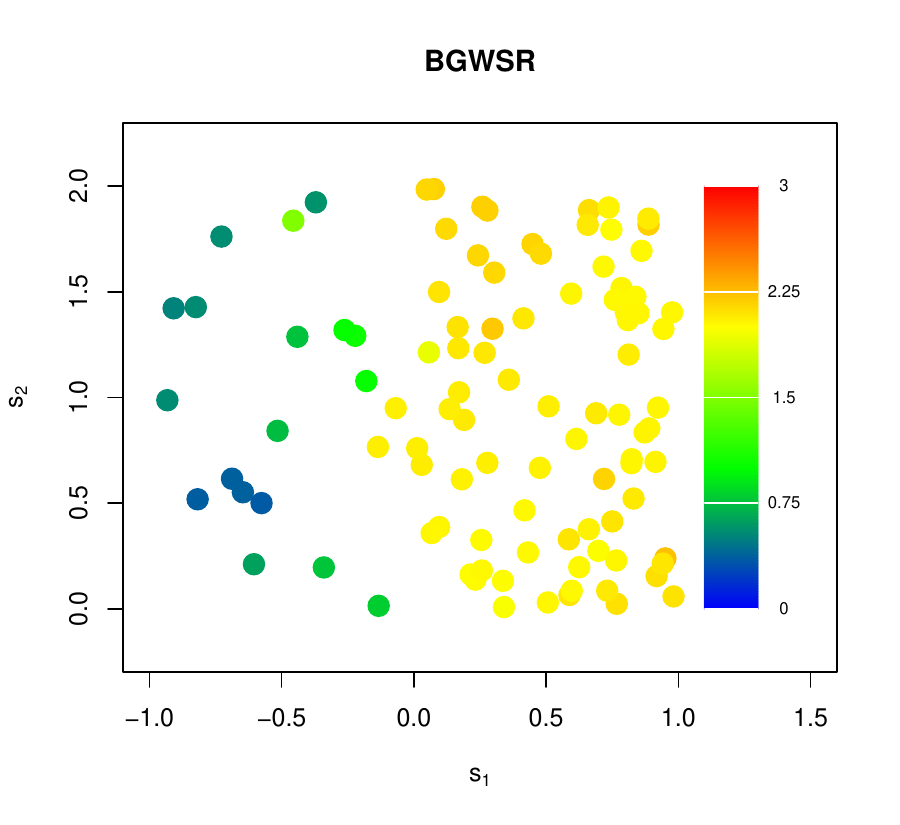}
	\end{center}
\end{minipage}
\\
\begin{minipage}[b]{0.49\columnwidth}
	\begin{center}
		\includegraphics[width=\linewidth,clip]{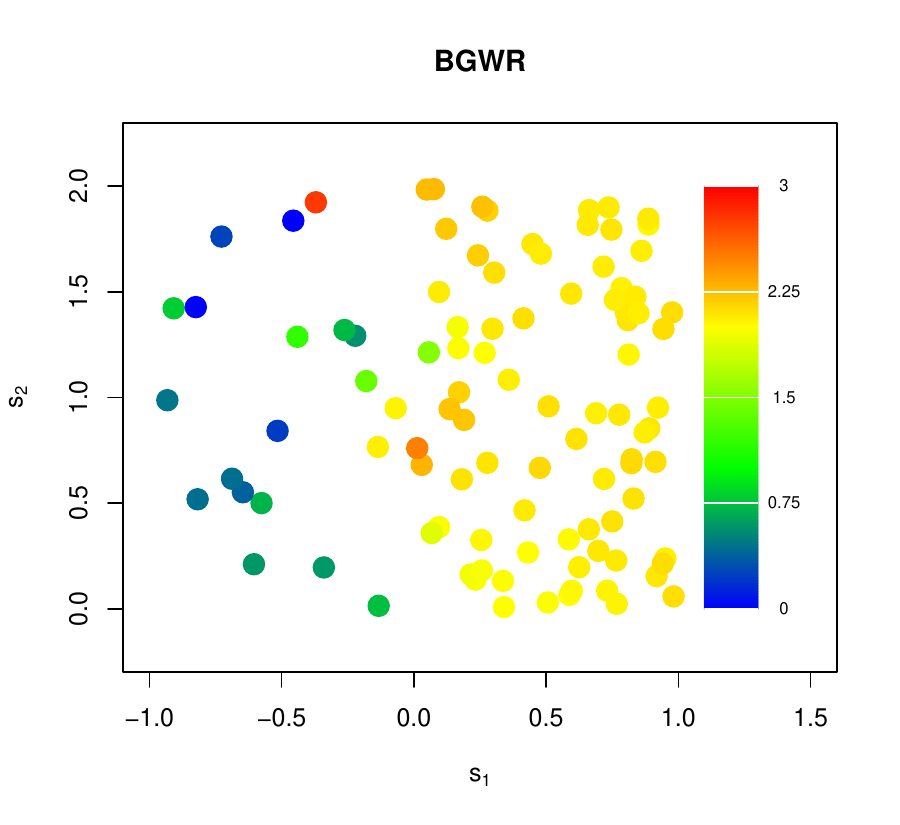}
	\end{center}
\end{minipage}
\begin{minipage}[b]{0.49\columnwidth}
	\begin{center}
		\includegraphics[width=\linewidth,clip]{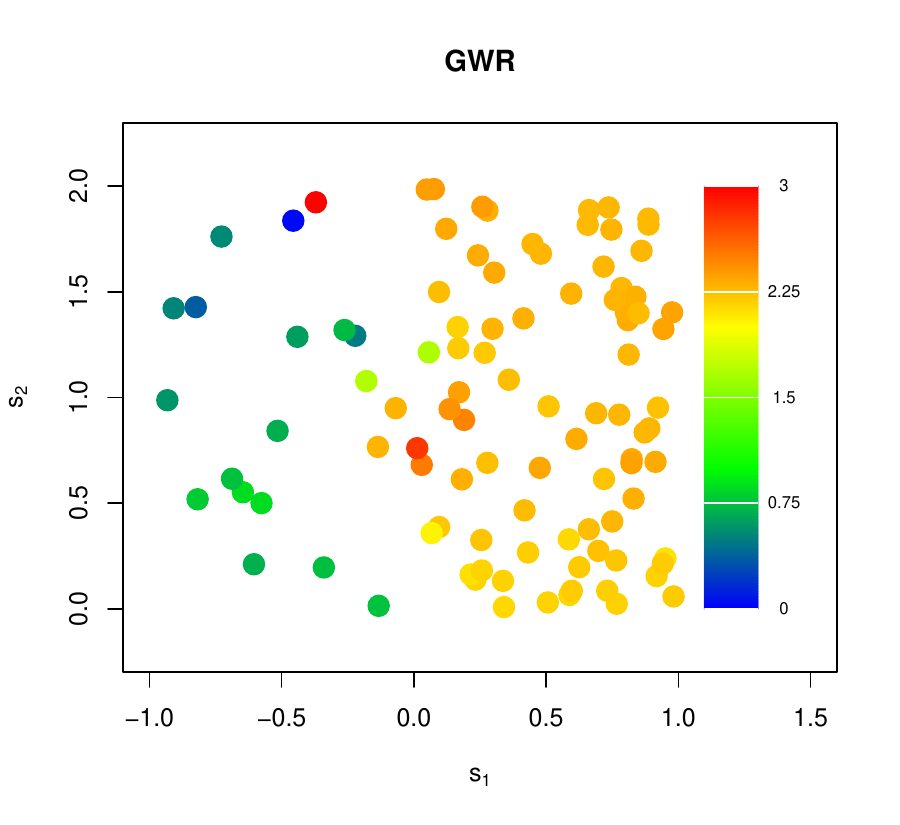}
	\end{center}
\end{minipage}
\caption{True and estimated values of coefficients for observation locations in Scenario 4\label{fig:res4}}
\end{figure}

In Scenario 4, BGWSR-AE had high performance for the objective variables and coefficients at the observation locations.
For the MSE for the prediction location, the BGWSR-AE performed the best for the objective variables and most of the coefficients, and the BGWSR performed equally well.
Figure \ref{fig:res4} shows an example of the true values of the coefficients and the estimates generated by each method in Scenario 4.
The results suggest that BGWSR-AE and BGWSR can estimate similarity coefficients within clusters compared to the other methods.
Additionally, BGWSR-AE and BGWSR provide numerically stable coefficients even in areas with sparse observation locations.
This suggests that the Bayesian Fused Lasso effectively considers the similarity of adjacent locations.
Nonetheless, for the BGWR and GWR, the estimation became unstable near the edge of the cluster.
In areas with sparse observation locations, extremely large or small coefficients were predicted due to the small data size.

\begin{figure}[H]
\centering
\begin{minipage}[b]{0.49\columnwidth}
	\begin{center}
		\includegraphics[width=\linewidth,clip]{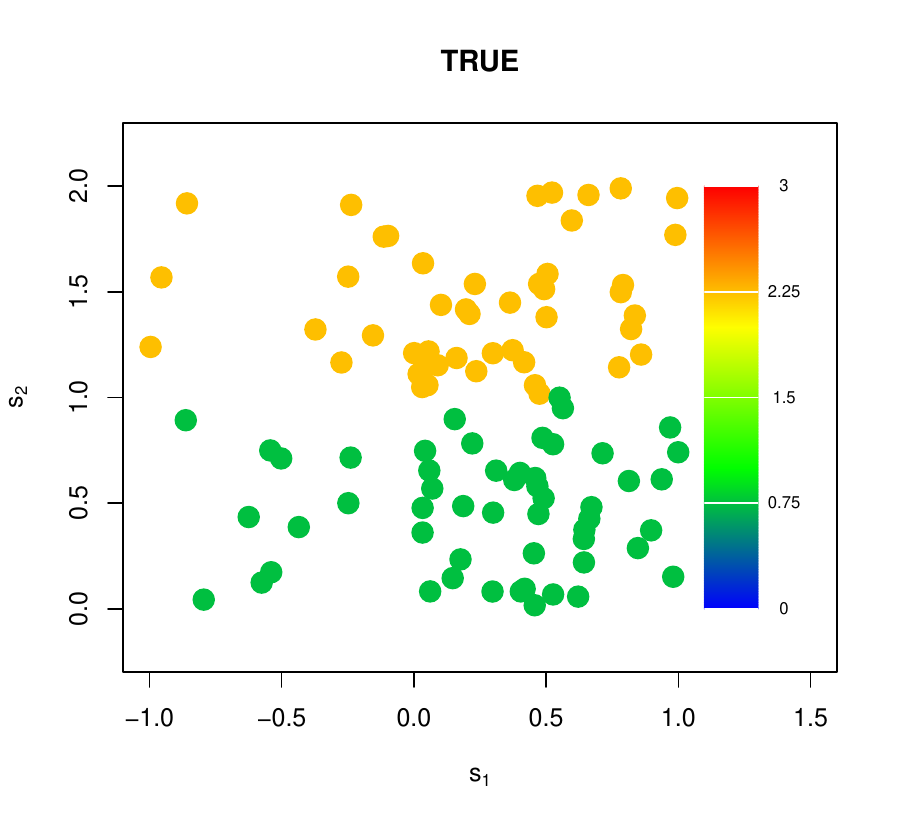}
	\end{center}
\end{minipage}
\\
\begin{minipage}[b]{0.49\columnwidth}
	\begin{center}
		\includegraphics[width=\linewidth,clip]{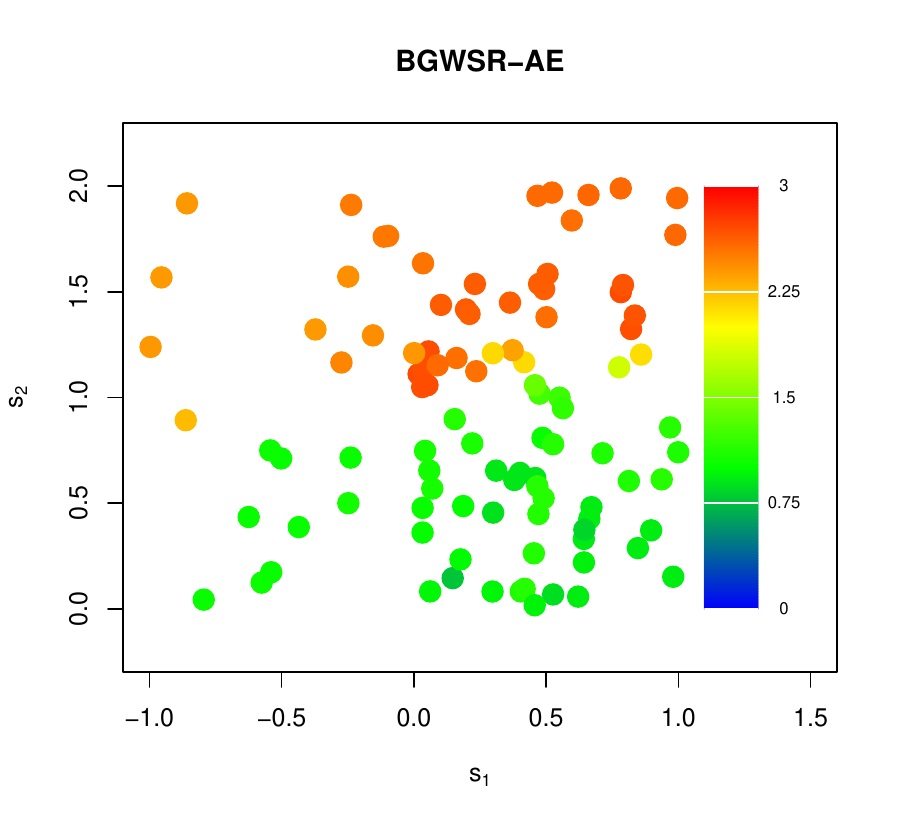}
	\end{center}
\end{minipage}
\begin{minipage}[b]{0.49\columnwidth}
	\begin{center}
		\includegraphics[width=\linewidth,clip]{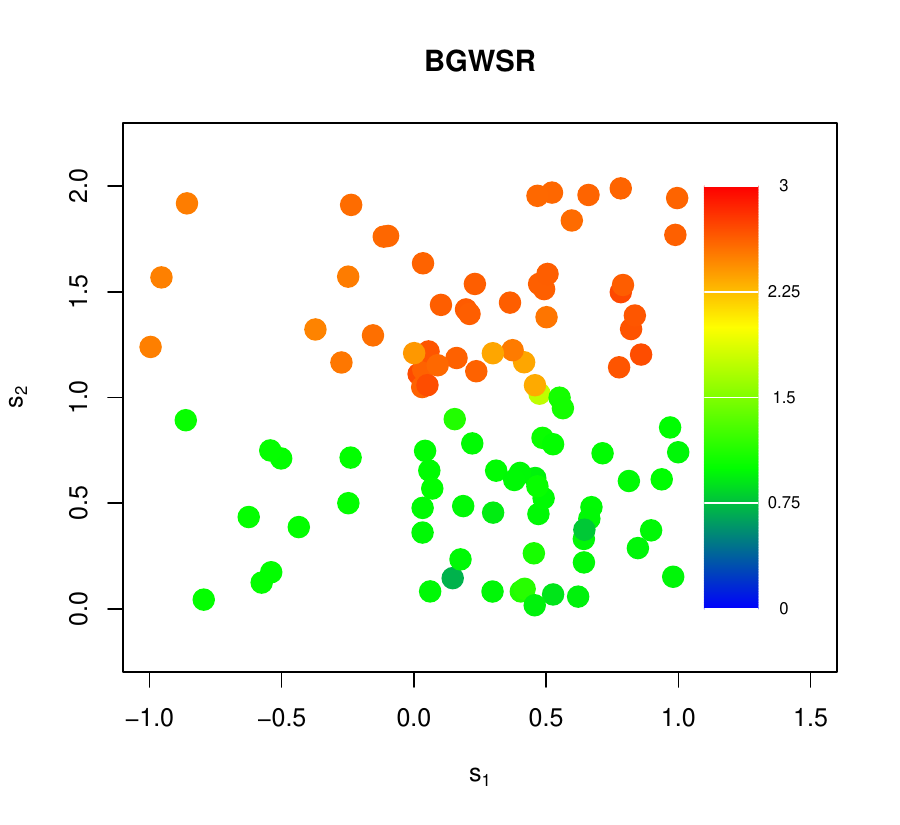}
	\end{center}
\end{minipage}
\\
\begin{minipage}[b]{0.49\columnwidth}
	\begin{center}
		\includegraphics[width=\linewidth,clip]{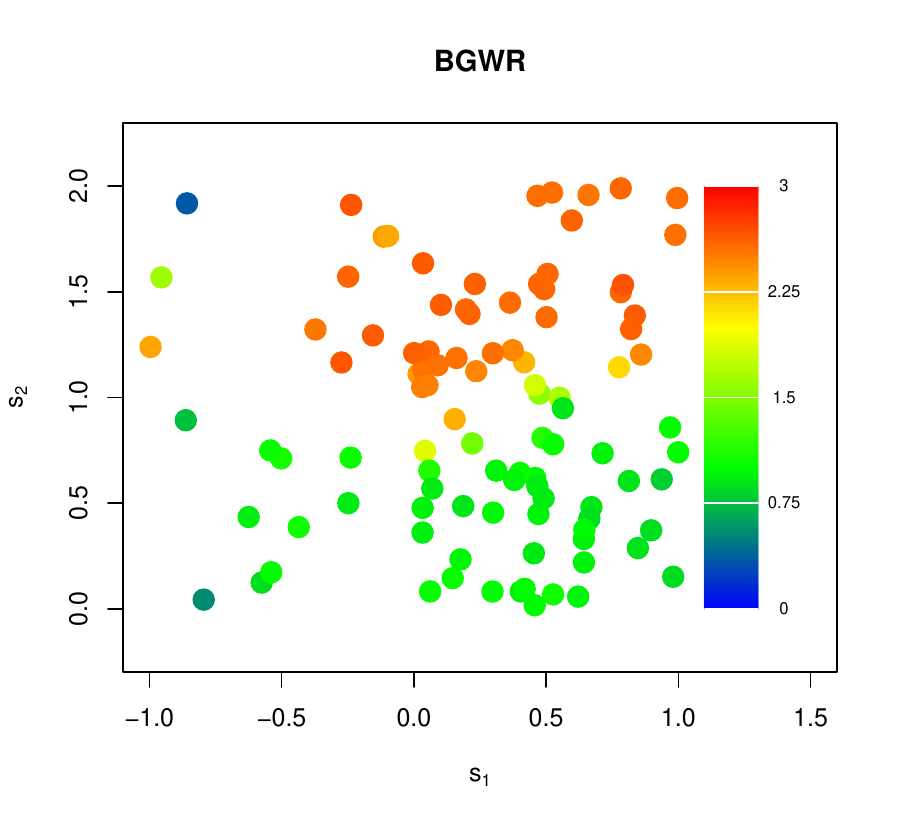}
	\end{center}
\end{minipage}
\begin{minipage}[b]{0.49\columnwidth}
	\begin{center}
		\includegraphics[width=\linewidth,clip]{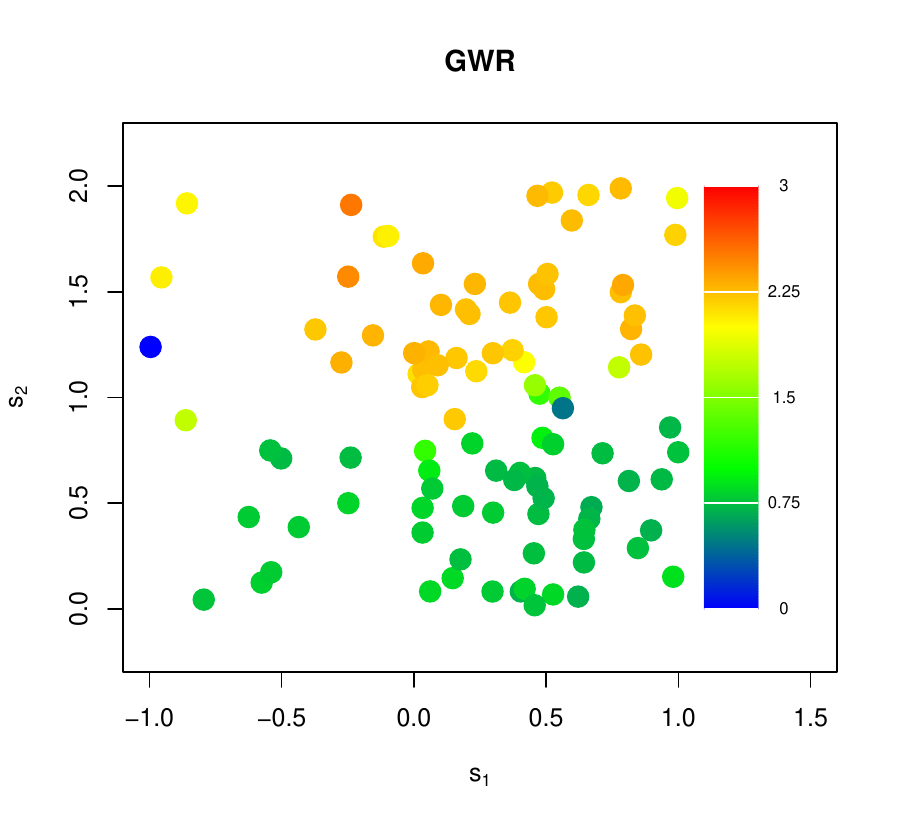}
	\end{center}
\end{minipage}
\caption{True and estimated values of coefficients for observation locations in Scenario 5\label{fig:res5}}
\end{figure}

In Scenario 5, relative to the MSE for the observation locations, BGWSR-AE, and BGWSR had high performance on the objective variables and coefficients at the observation locations.
BGWSR performed the best for the prediction locations.
Meanwhile the GWR yielded the best results on the objective variables, and the BGWSR performed similarly well.
Figure \ref{fig:res5} shows an example of the true values of the coefficients and the estimated values produced by each method in Scenario 5.
The results suggest that BGWSR-AE and BGWSR can estimate coefficients more similarly within clusters than BGWR and GWR.
For BGWSR-AE, BGWSR, and GWR, the prediction of coefficients in the upper clusters yielded larger estimates compared to the true values across clusters due to Bayesian Fused Lasso.
For BGWR and GWR, the coefficient estimates were unstable at the edge of the domain and in areas with sparse observation locations.
\begin{table}[H]
\renewcommand{\arraystretch}{1.8}
\centering
\caption{Median MSE for each method \label{tab:res_numerical}}
\label{tab:res1}
\begin{tabular}{ccccccccccc}
\hline
\multirow{2}{*}{senario} & \multirow{2}{*}{method} & \multicolumn{4}{c}{observation location} &  & \multicolumn{4}{c}{prediction location}\\
\cmidrule(lr){3-6} \cmidrule(lr){8-11}
 &  & $\bm{\beta}_{1}$ & $\bm{\beta}_{2}$ & $\bm{\beta}_{3}$ & $\bm{y}$ &  & $\bm{\beta}_{1}$ & $\bm{\beta}_{2}$ & $\bm{\beta}_{3}$ & $\bm{y}$\\\hline
\multirow[t]{4}{*}{1} & BGWSR-AE & 0.025 & 0.052 & \textcolor{red}{0.027} & 0.037 &  & 0.037 & 0.057 & 0.042 & 0.189\\ 
& BGWSR & \textcolor{red}{0.024} & 0.051 & \textcolor{red}{0.027} & \textcolor{red}{0.036} &  & \textcolor{red}{0.032} & 0.068 & \textcolor{red}{0.033} & \textcolor{red}{0.161}\\
& BGWR & 0.046 & 0.083 & 0.044 & 0.167 &  & 0.041 & 0.075 & 0.037 & 0.187\\
& GWR & 0.033 & \textcolor{red}{0.027} & 0.035 & 0.110 &  & 0.046 & \textcolor{red}{0.035} & 0.046 & 0.164\\
\multirow[t]{4}{*}{2} & BGWSR-AE & \textcolor{red}{0.024} & 0.041 & \textcolor{red}{0.027} & \textcolor{red}{0.032} &  & \textcolor{red}{0.042} & \textcolor{red}{0.057} & 0.059 & 0.172\\
& BGWSR & 0.032 & 0.040 & 0.031 & \textcolor{red}{0.032} &  & 0.046 & 0.062 & \textcolor{red}{0.056} & \textcolor{red}{0.165}\\
& BGWR & 0.046 & 0.054 & 0.034 & 0.156 &  & 0.048 & 0.068 & 0.065 & 0.180\\
& GWR & 0.057 & \textcolor{red}{0.036} & 0.047 & 0.119 &  & 0.152 & 0.064 & 0.219 & 0.323\\
\multirow[t]{4}{*}{3} & BGWSR-AE & 0.045 & 0.031 & \textcolor{red}{0.048} & \textcolor{red}{0.054} &  & 0.070 & 0.083 & 0.080 & 0.223\\
& BGWSR & \textcolor{red}{0.036} & \textcolor{red}{0.028} & 0.052 & 0.055 &  & \textcolor{red}{0.056} & \textcolor{red}{0.049} & \textcolor{red}{0.059} & \textcolor{red}{0.145}\\
& BGWR & 0.077 & 0.075 & 0.089 & 0.199 &  & 0.057 & 0.060 & 0.078 & 0.187\\
& GWR & 0.046 & 0.044 & 0.065 & 0.167 &  & 0.081 & 0.068 & 0.068 & 0.259\\
\multirow[t]{4}{*}{4} & BGWSR-AE & \textcolor{red}{0.034} & \textcolor{red}{0.041} & \textcolor{red}{0.031} & \textcolor{red}{0.052} &  & \textcolor{red}{0.085} & \textcolor{red}{0.086} & 0.088 & \textcolor{red}{0.293}\\
& BGWSR & 0.053 & 0.054 & 0.048 & 0.053 &  & 0.088 & 0.095 & 0.089 & 0.306\\
& BGWR & 0.090 & 0.062 & 0.084 & 0.224 &  & 0.109 & 0.094 & \textcolor{red}{0.078} & 0.345\\
& GWR & 0.052 & 0.071 & 0.049 & 0.210 &  & 0.090 & 0.139 & 0.114 & 0.400\\
\multirow[t]{4}{*}{5} & BGWSR-AE & 0.045 & 0.040 & 0.053 & \textcolor{red}{0.039} &  & 0.078 & 0.086 & 0.104 & 0.278\\
& BGWSR & \textcolor{red}{0.033} & \textcolor{red}{0.036} & \textcolor{red}{0.033} & \textcolor{red}{0.039} &  & \textcolor{red}{0.058} & \textcolor{red}{0.052} & \textcolor{red}{0.064} & 0.164\\
& BGWR & 0.073 & 0.064 & 0.074 & 0.157 &  & 0.075 & 0.069 & 0.068 & 0.196\\
& GWR & 0.049 & 0.047 & 0.078 & 0.142 &  & 0.067 & 0.074 & 0.088 & \textcolor{red}{0.160}\\\hline
\end{tabular}
\end{table}
%
%
%
%
%
\section{Application to land price data\label{chap4:realdata}}
In this section, we evaluate the prediction performance of the proposed method by applying it to real data to demonstrate its usefulness.

\subsection{Data Summary and Applicable Settings}
\begin{figure}[htb]
    \centering
	\begin{center}
		\includegraphics[width=15cm,clip]{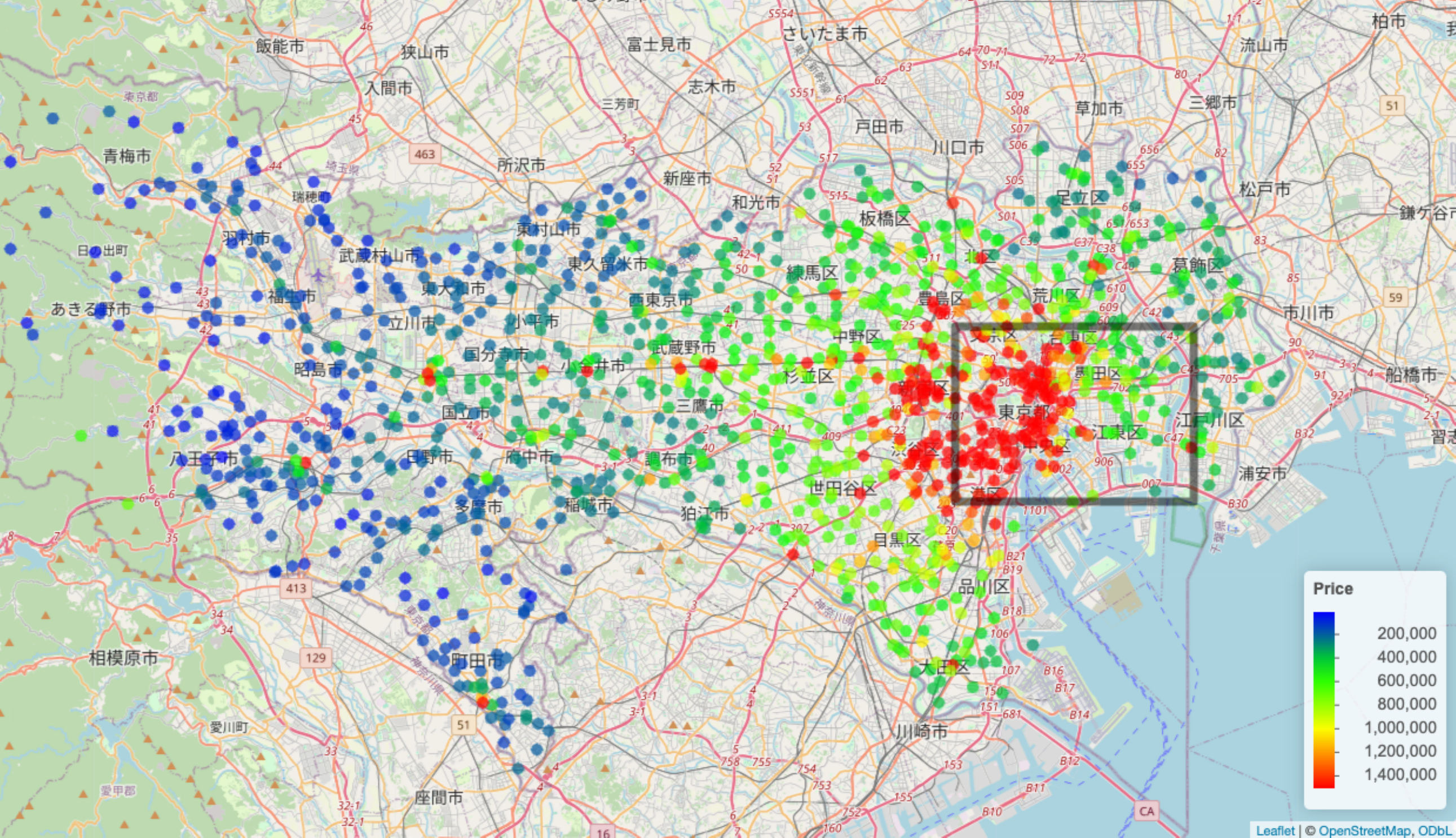}
	\end{center}
	\caption{land price in Tokyo}
\label{fig:land}
\end{figure}

\begin{figure}[H]
\centering
    \begin{minipage}[b]{0.49\columnwidth}
    \begin{center}
        \includegraphics[width=7cm,clip]{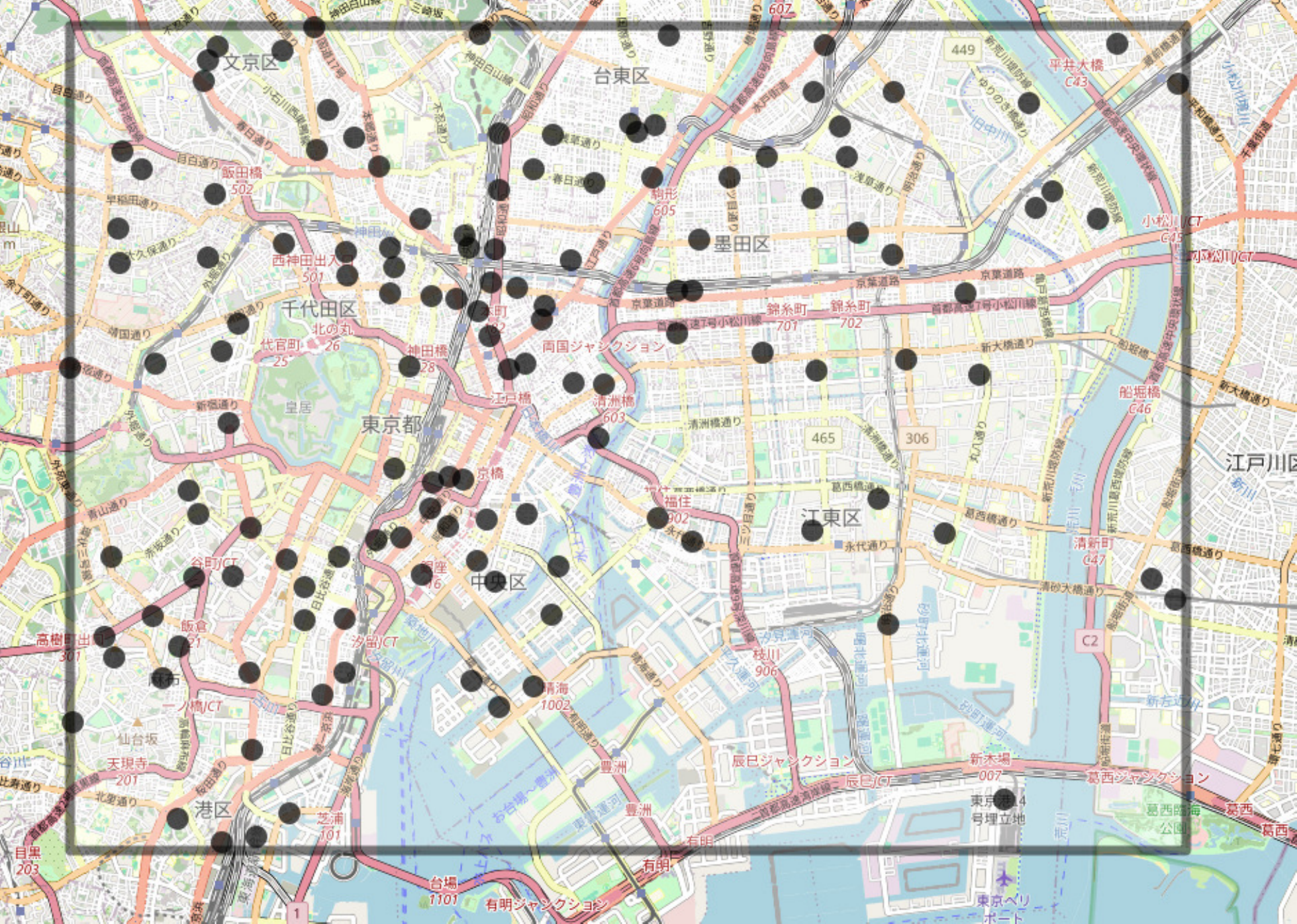}
    \end{center}
    \subcaption{observed location}
    \end{minipage}
    \begin{minipage}[b]{0.49\columnwidth}
    \begin{center}
        \includegraphics[width=7cm,clip]{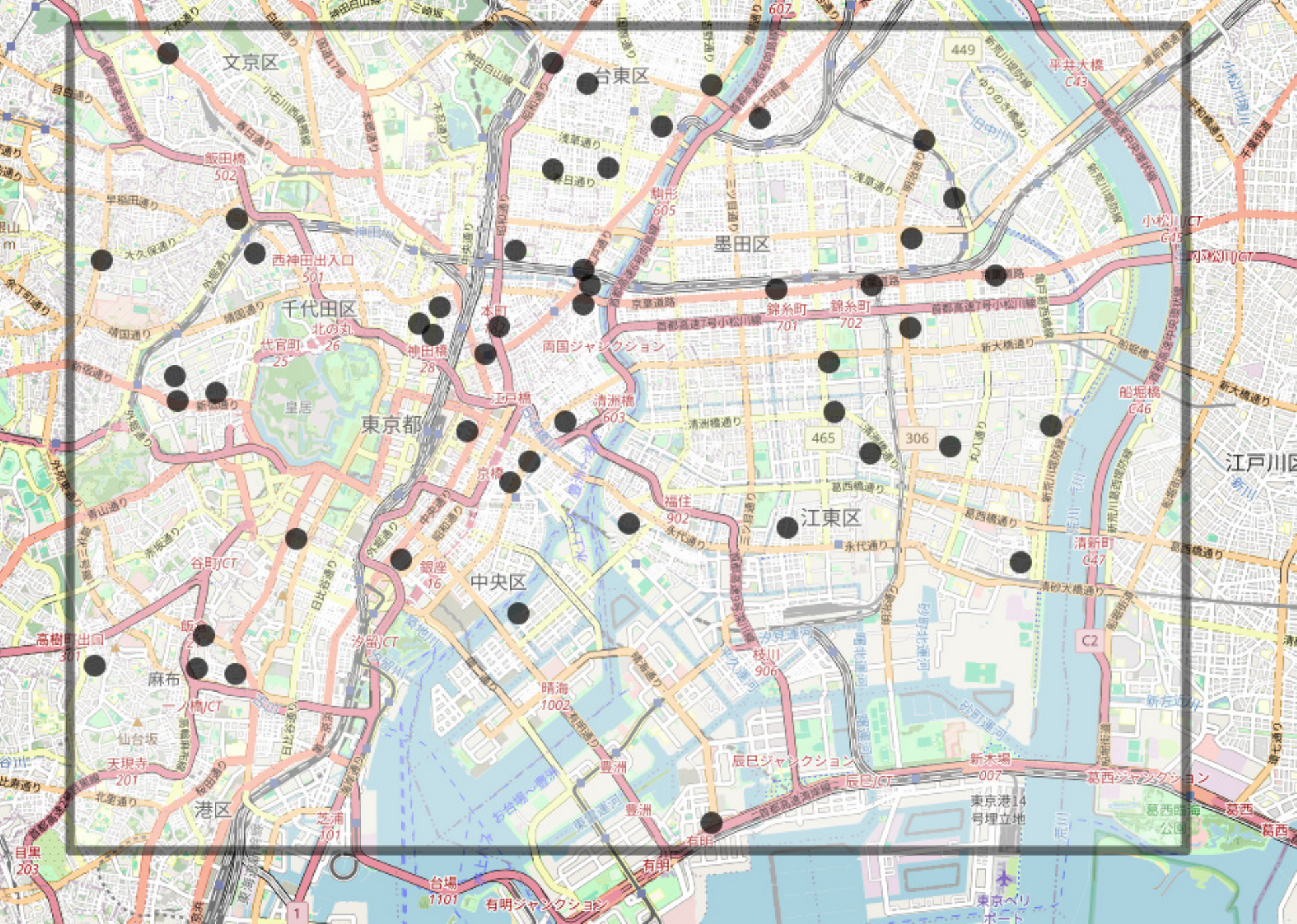}
    \end{center}
    \subcaption{prediction location}
    \end{minipage}
\caption{Locations used in the analysis\label{fig:landdata}}
\end{figure}

The data used are the official land price data for Tokyo for the year $2021$, published by the Ministry of Land, Infrastructure, Transport and Tourism~\citep{landprice}.
The data were obtained for $1219$ locations, and $114$ variables such as land use classification and type of frontage road were obtained based on the location information of latitude and longitude.
Figure \ref{fig:land} plots the obtained data by location information and shows the land prices in 2021 in colors.
In this study, we randomly selected $130$ locations in the area around the city center, which is surrounded by the black square in Figure \ref{fig:land}.
Subsequently, we randomly selected $100$ observation locations and $50$ prediction locations.
Figure \ref{fig:landdata} shows the observation locations and prediction locations.
The observation locations are mostly available in the left half of the city center, while the data are less available in the right half of the city center---the data are coarse-grained in terms of observation locations.
The variables used are shown in table \ref{tab:variabes_land}.
In this application, latitude and longitude were used as location, price as the objective variable, and the volume ratio, land area, and station distance as explanatory variables.
\begin{table}[htb]
\renewcommand{\arraystretch}{1.5}
\centering
\caption{List of variables used in real data example\label{tab:variabes_land}}
\begin{tabular}{c|c|c}
\hline
&variables & details\\
\hline\hline
\multirow[t]{2}{*}{location} & latitude & $Y$-coordinate obtained by World Geodetic System (sec) \\
& longitude & $X$-coordinate obtained by World Geodetic System (sec) \\
objective variable & price & land prices in 2021 (Yen/${\rm{m}}^{2}$, Yen/$10a$ for forest land)  \\
\multirow[t]{3}{*}{covariates} & floor area ratio & Ratio of total building area (total floor space) to site area (\%) \\
& land area & land area (${\rm{m}}^{2}$) \\
& distance & Distance to the nearest station (m) \\ 
\hline
\end{tabular}
\end{table}

We set up the hyperparameters for each method and the method of determining the neighboring locations in the same way as in the numerical studies of ref{chap:numerical study}.
\subsection{Results}
\begin{figure}[H]
\centering
\begin{minipage}[b]{0.49\columnwidth}
    \centering
    \includegraphics[width=7cm,clip]{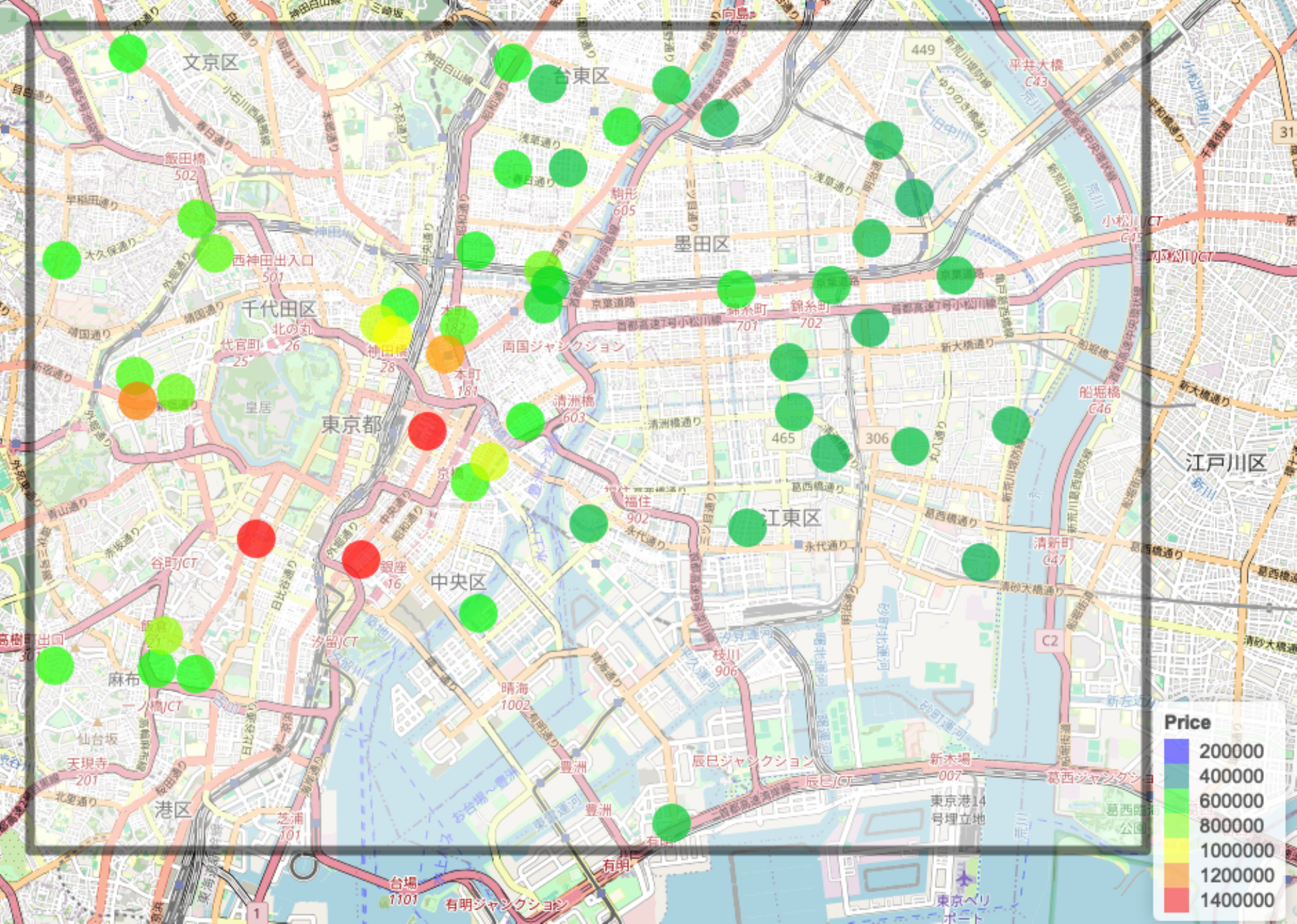}
    \subcaption{TRUE}
\end{minipage}
\\[1ex]
\begin{minipage}[b]{0.49\columnwidth}
	\centering
	\includegraphics[width=7cm,clip]{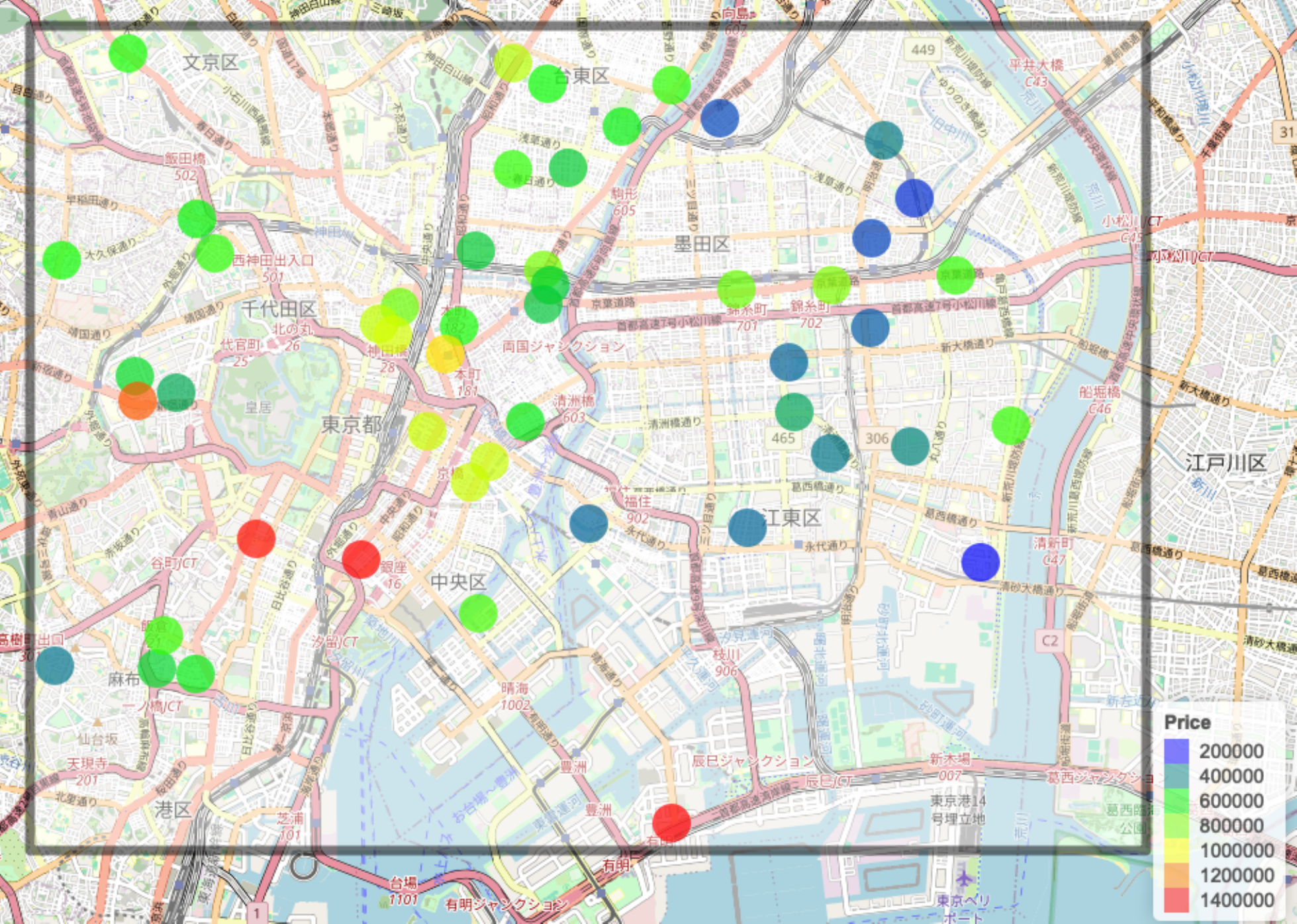}
	\subcaption{BGWSR-AE}
\end{minipage}
\begin{minipage}[b]{0.49\columnwidth}
    \centering
	\includegraphics[width=7cm,clip]{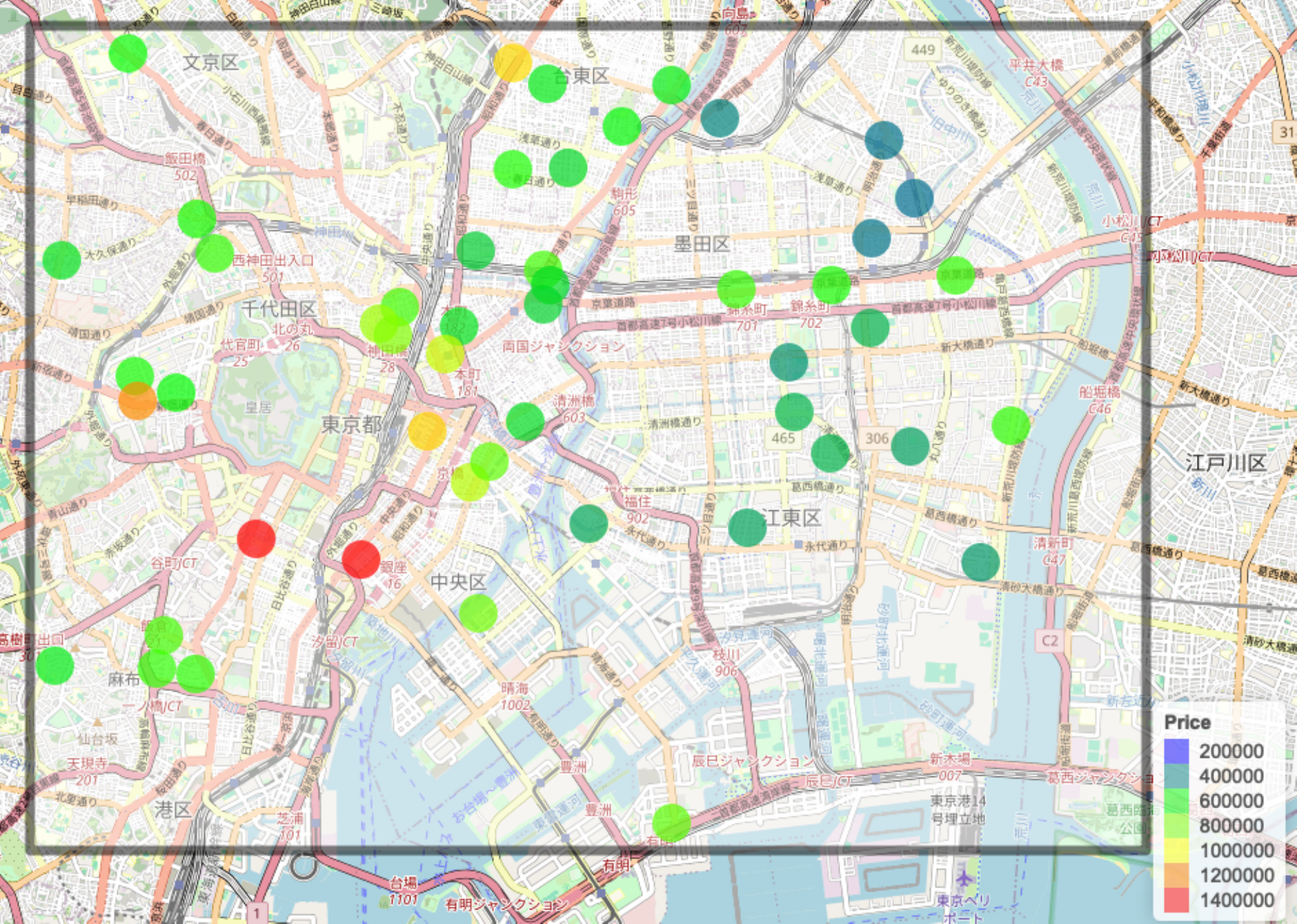}
	\subcaption{BGWSR}
\end{minipage}
\\[1ex] 
\begin{minipage}[b]{0.49\columnwidth}
    \centering
	\includegraphics[width=7cm,clip]{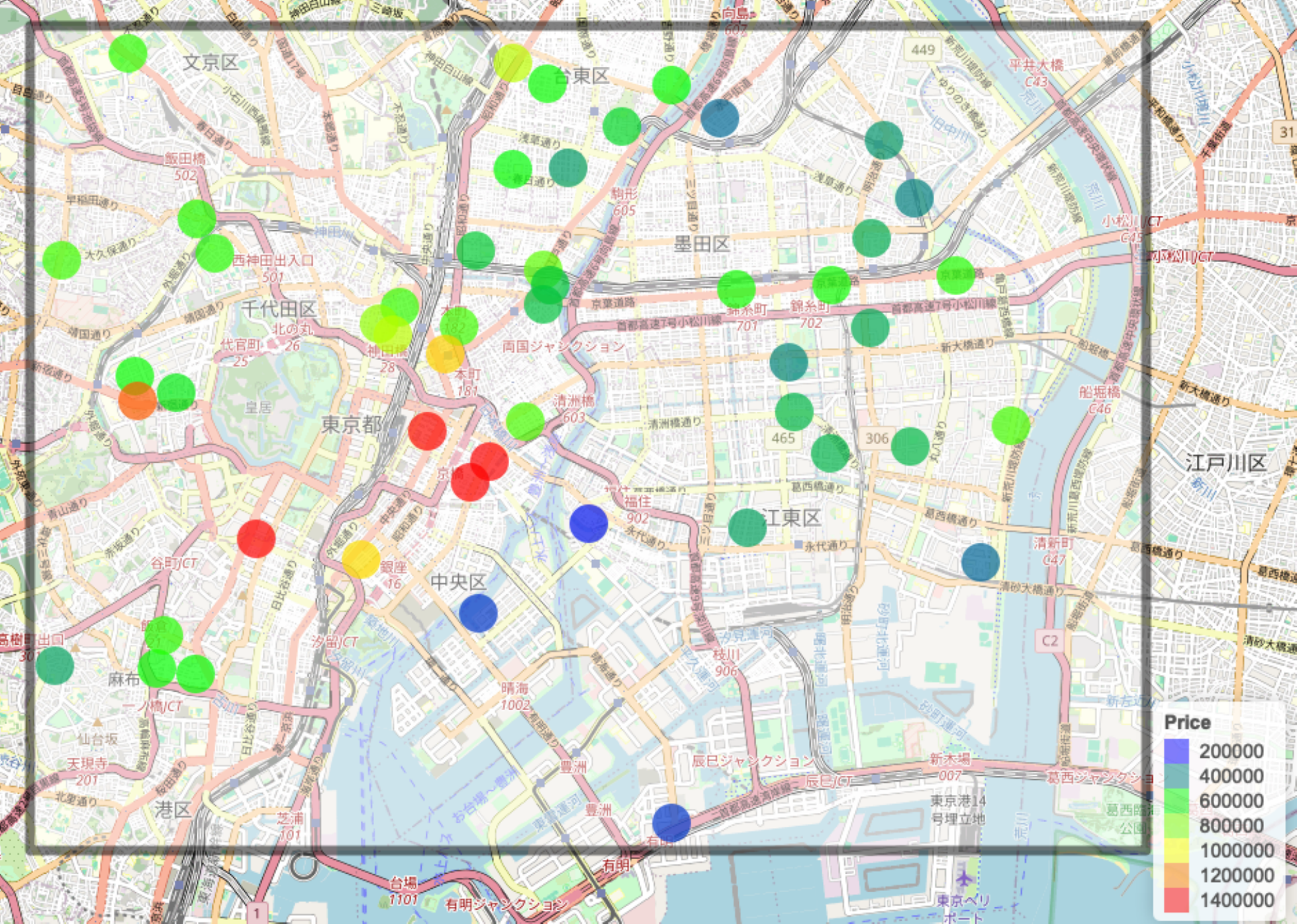}
	\subcaption{BGWR}
\end{minipage}
\begin{minipage}[b]{0.49\columnwidth}
    \centering
	\includegraphics[width=7cm,clip]{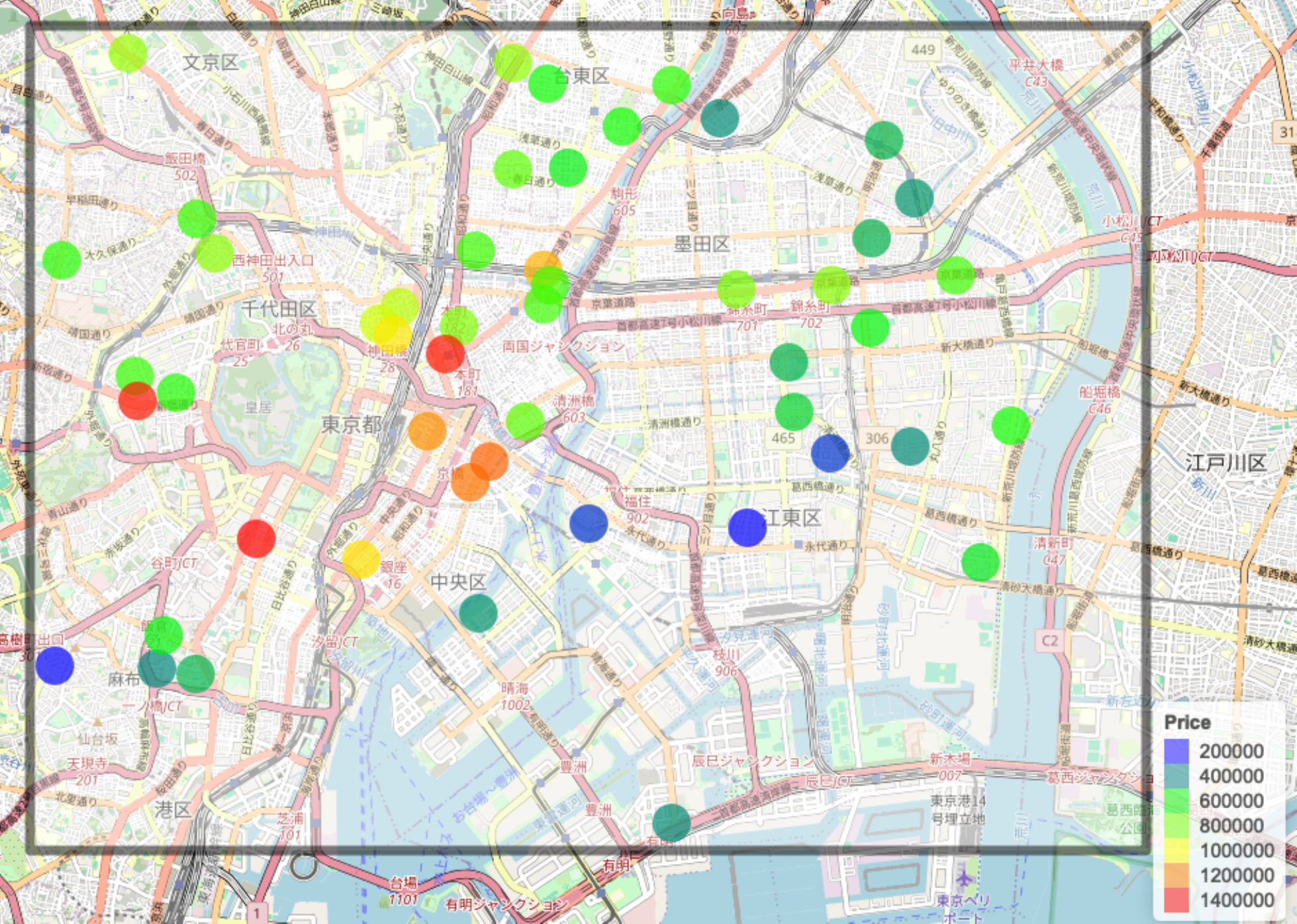}
	\subcaption{GWR}
\end{minipage}
\caption{True and predicted price at the prediction location\label{fig:resland_pred}}
\end{figure}

Table \ref{tab:res_landprice} shows the MSE of the objective variable prediction by each method at the prediction location.
The MSE of BGWSR was the smallest.
It was followed by BGWSR-AE with the smallest MSE and GWR with the largest MSE.
Figure \ref{fig:resland_pred} shows the true value of the price at the prediction location and the predicted value by each method, colored according to the magnitude of the value and visualized on the map.
As shown in Figure \ref{fig:resland_pred}, BGWSR-AE produced the closest prediction to the true value.
Compared to the other methods, BGWSR-AE did not produce extremely small values as predictions, suggesting that the Bayesian Lasso effect stabilizes the estimated values.
However, BGWSR and GWR produced predictions that were extremely large or small compared to the true value in the vicinity of the city center and near the boundaries of the region.

Table \ref{fig:resland_beta_obs} shows the estimated coefficients for each method at the observation location, colored according to the magnitude of the value and visualized on a map.
The $\beta_{1}, \, \beta_{2}, \, \beta_{3}$ are the coefficients for the floor area ratio, land area, and distance.
First, focusing on the estimated coefficients for the floor area ratio, the coefficients are large in the city center and decrease gradually as one moves away from the city center, as is the case for all the methods.
For BGWSR-AE and BGWSR, the coefficients were estimated to vary gradually around the city center.
Nevertheless, for BGWSR and GWR, the estimated coefficients differed between the city center and areas far from it.
Next, focusing on the estimated coefficients for land area, all the methods had a negative value for the coefficient in urban centers.
However, the coefficients of BGWSR, BGWR, and GWR are close to $-1$ in the city center, while those of BGWSR-AE are around $0.4$.
Finally, in the BGWSR-AE, the estimated coefficients for station distance were larger near Chiyoda-ku and smaller away from there.
Nonetheless, for BGWSR, BGWR, and GWR, the coefficients were estimated to be large in the upper right of the region and decreased toward the lower left.

\begin{figure}[H]
\centering
\begin{minipage}[b]{0.49\columnwidth}
	\begin{center}
	\includegraphics[width=7cm,clip]{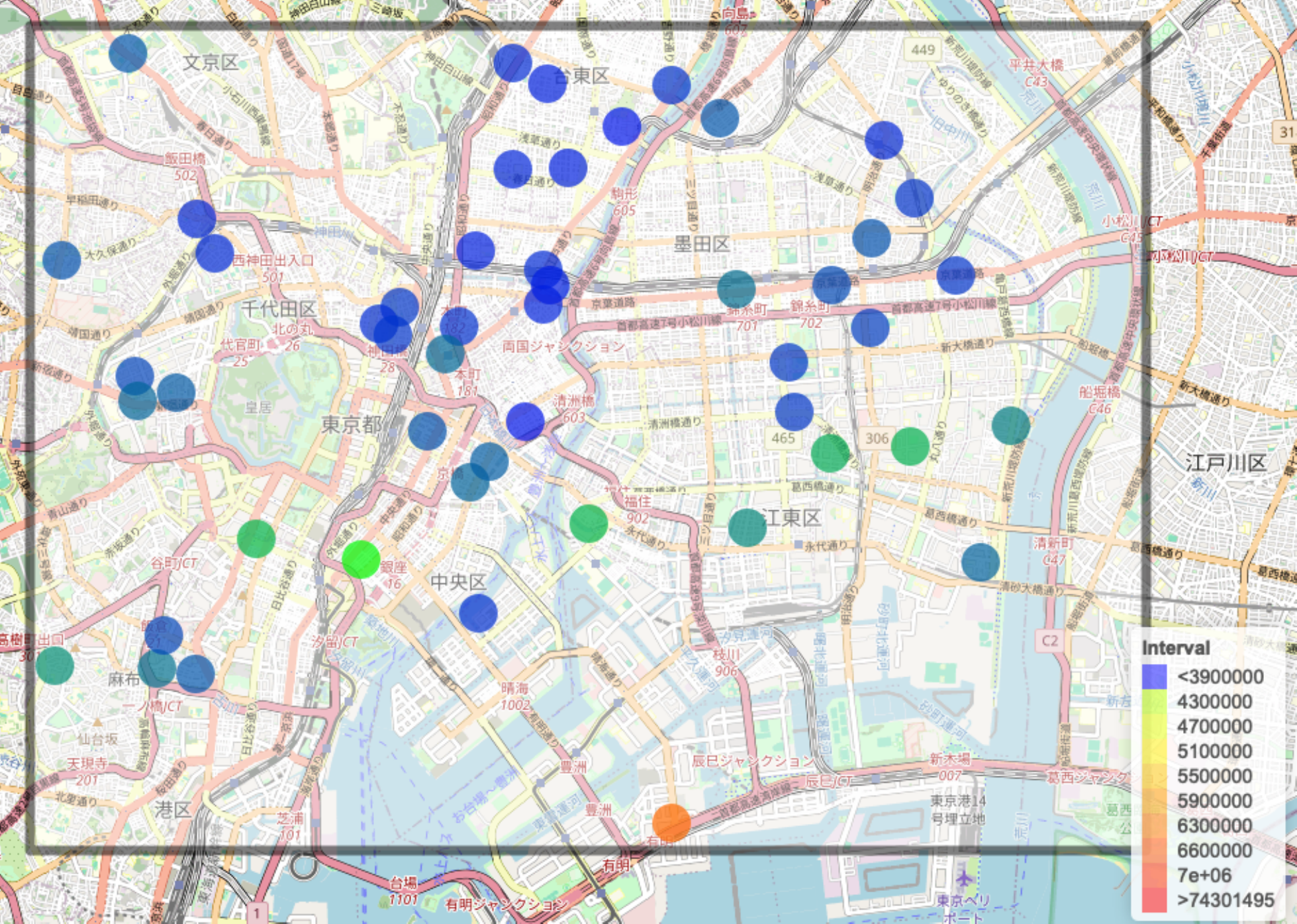}
	\subcaption{BGWSR-AE}
 	\end{center}
\end{minipage}
\begin{minipage}[b]{0.49\columnwidth}
    \begin{center}
	\includegraphics[width=7cm,clip]{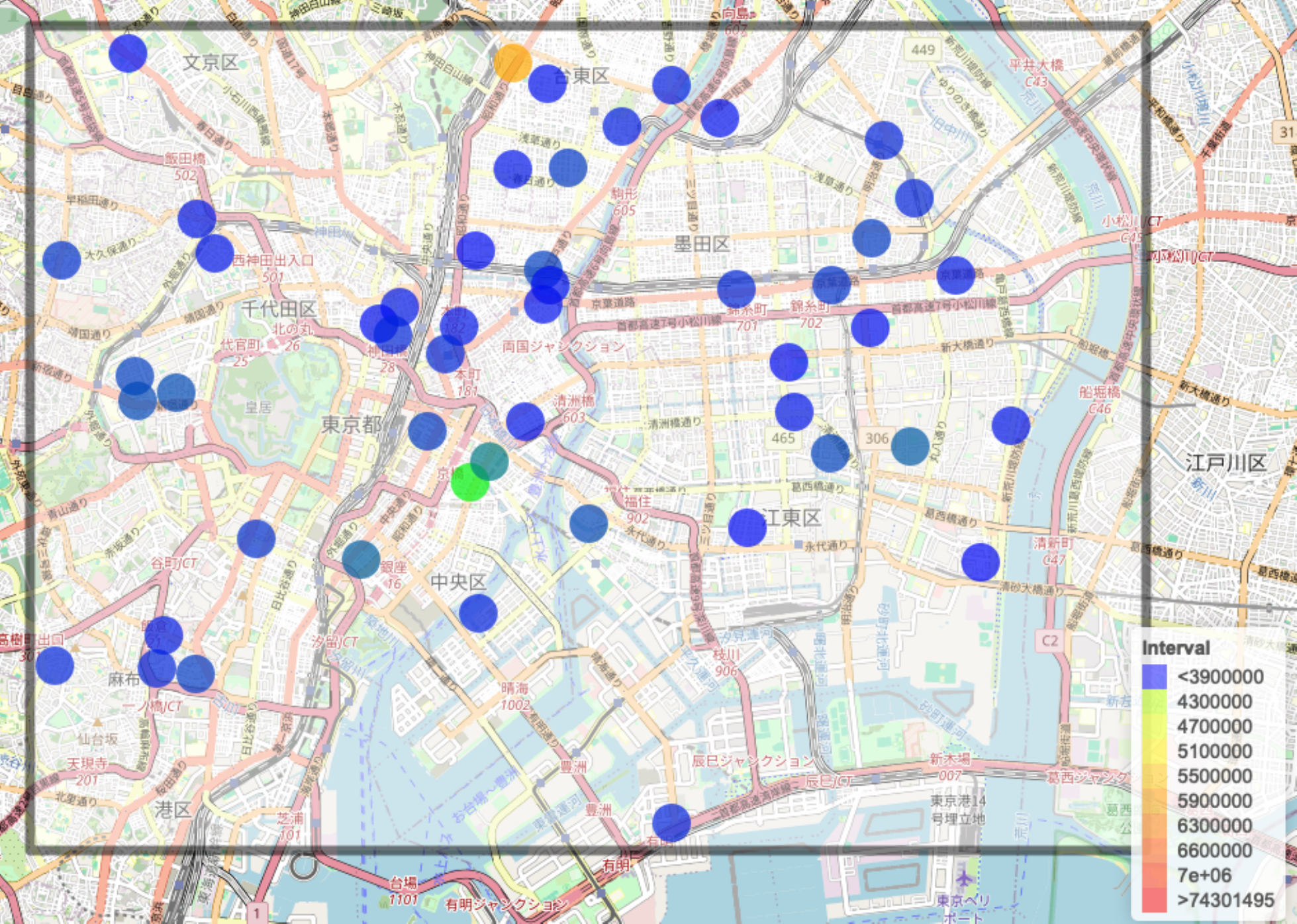}
	\subcaption{BGWSR}
 \end{center}
\end{minipage}
\\[1ex] 
\begin{minipage}[b]{0.49\columnwidth}
    \begin{center}
	\includegraphics[width=7cm,clip]{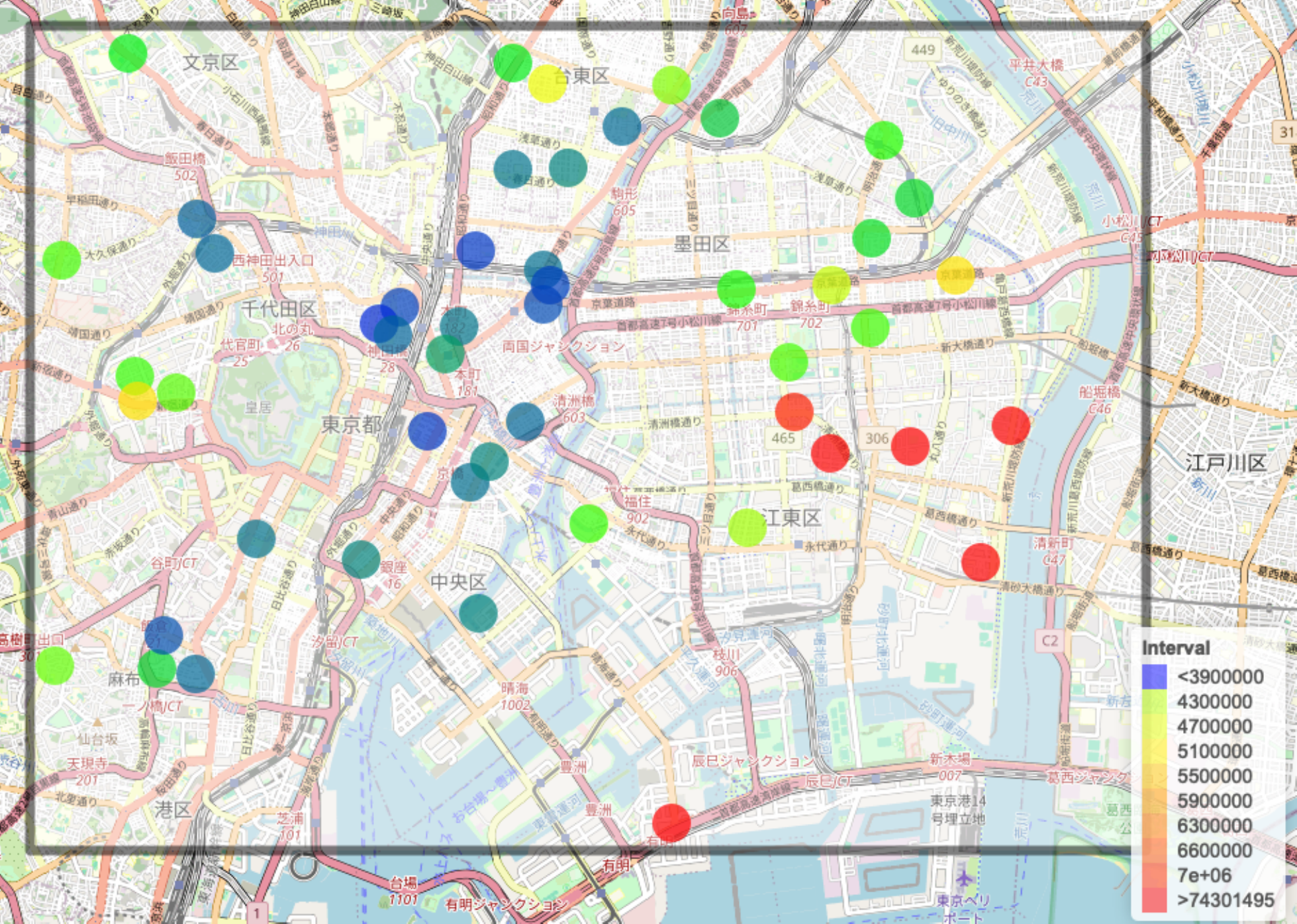}
	\subcaption{BGWR}
 \end{center}
\end{minipage}
\caption{the Range of Credible interval of price at the prediction location\label{fig:resland_pred_interval}}
\end{figure}

Figure \ref{fig:resland_pred_interval} shows a visualization of the predicted values of the objective variable by BGWSR-AE, BGWSR, and BGWR, colored by the size of the range of the $95\%$ credible interval.
From Figure \ref{fig:resland_pred_interval}, BGWSR-AE and BGWSR have more locations with smaller interval widths compared to BGWR.
In particular, the interval widths were smaller for BGWSR-AE and BGWSR even at the edge of the region.
The ratio of true values included in the $95\%$ confidence interval was the same for all methods.
This suggests that BGWSR-AE and BGWSR have less uncertainty than BGWR.
Furthermore, BGWSR-AE and BGWSR are useful because they have smaller MSEs than BGWR and GWR.
%
\begin{table}[H]
\renewcommand{\arraystretch}{1.5}
\centering
\caption{MSE of predicted and true values\label{tab:res_landprice}}
\begin{tabular}{c|c}
\hline
method     & ${\rm{MSE}}_{\bm{y}}$ \\ \hline
BGWSR-AE & \textbf{0.092}                 \\
BGWSR    & \color{red}{\textbf{0.082}}        \\
BGWR     & 0.135                 \\
GWR      & 0.171                 \\ \hline
\end{tabular}
\end{table}
%
\begin{table}[H]
\centering
\begin{tabular}{|p{0.25cm}|c|c|c|}
\hline
& $\beta_{1}$ & $\beta_{2}$ & $\beta_{3}$ \\ \hline
\rotatebox{90}{\quad\quad BGWSR-AE} &
\includegraphics[width=0.285\linewidth]{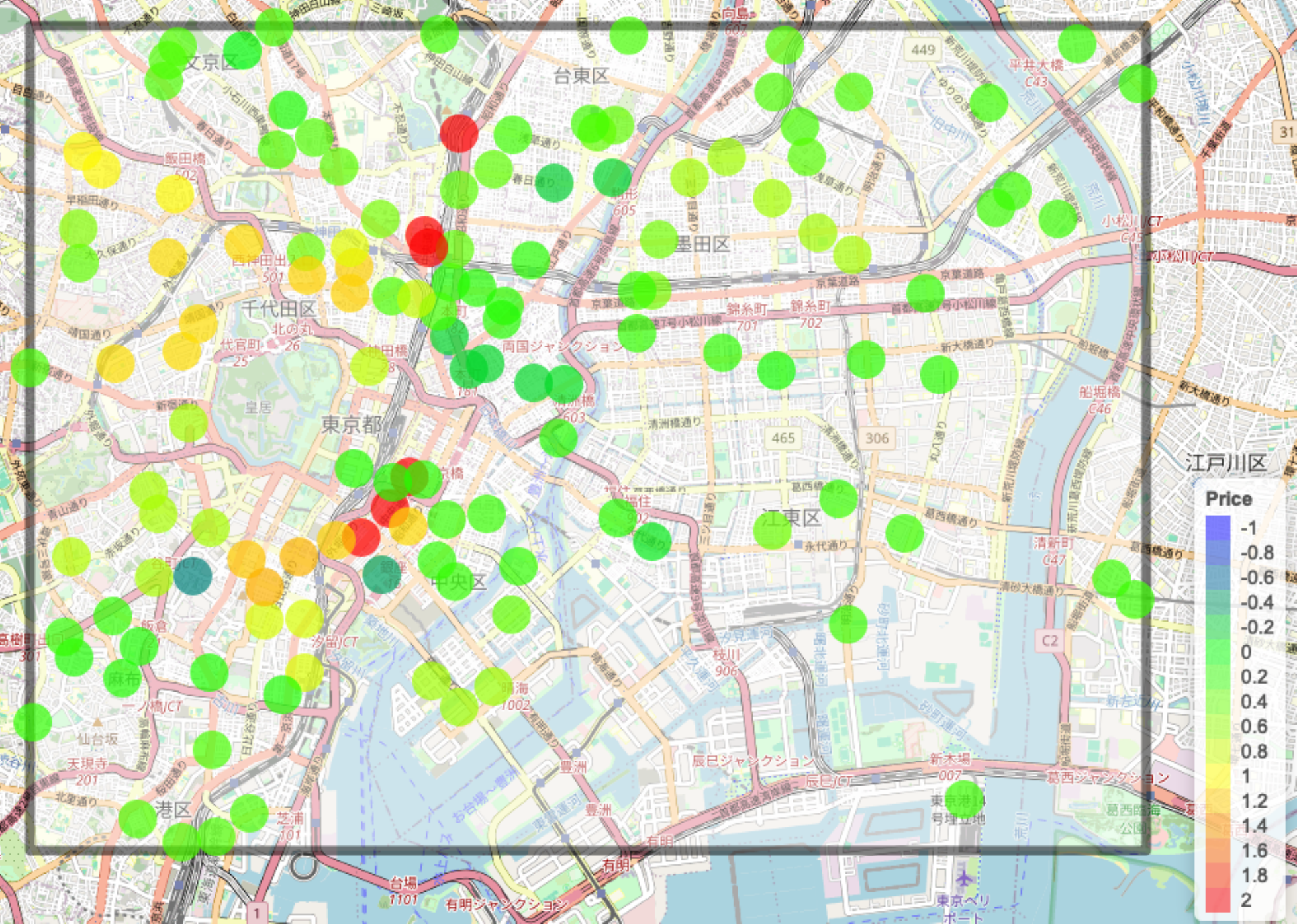} & 
\includegraphics[width=0.285\linewidth]{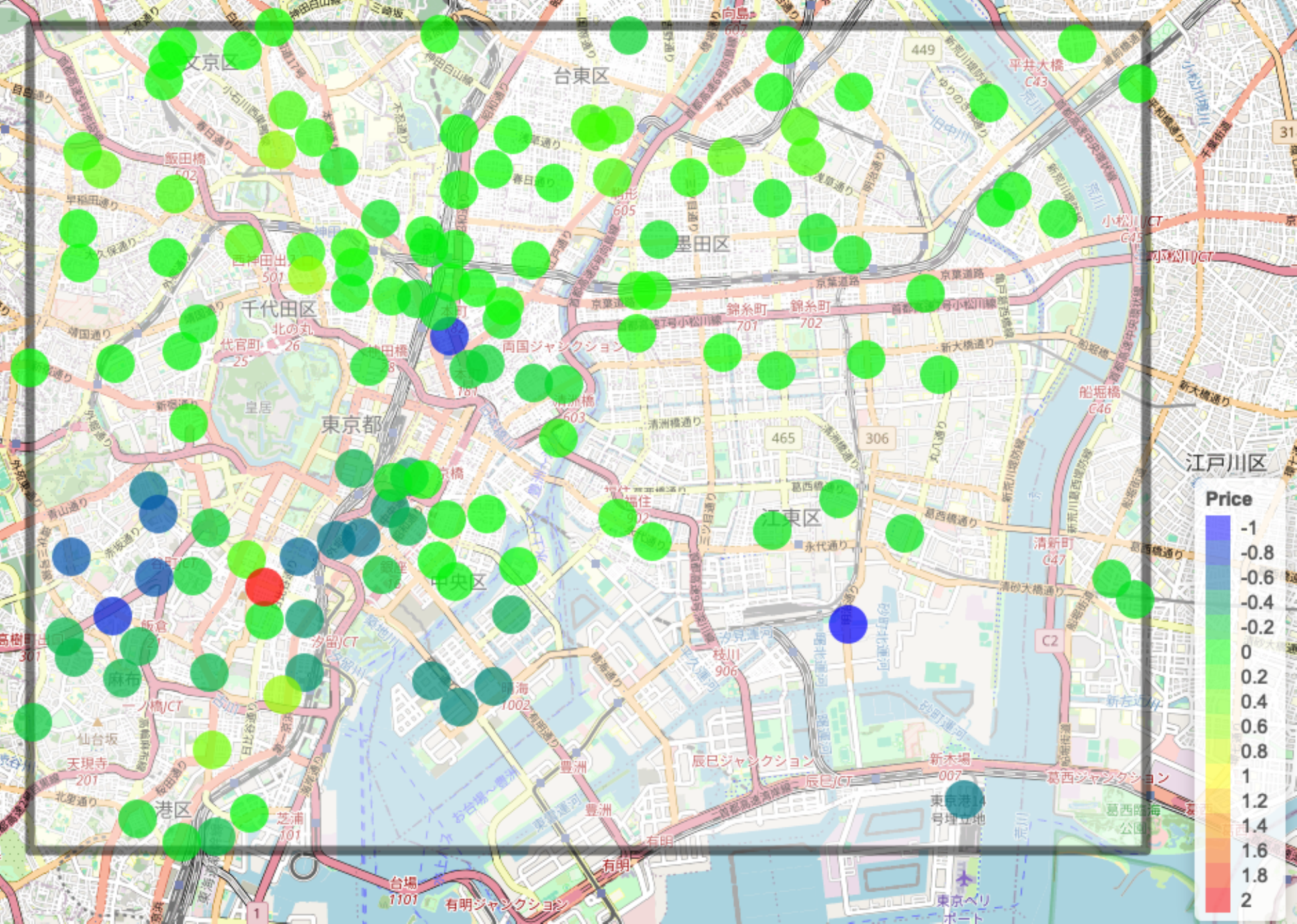} & 
\includegraphics[width=0.285\linewidth]{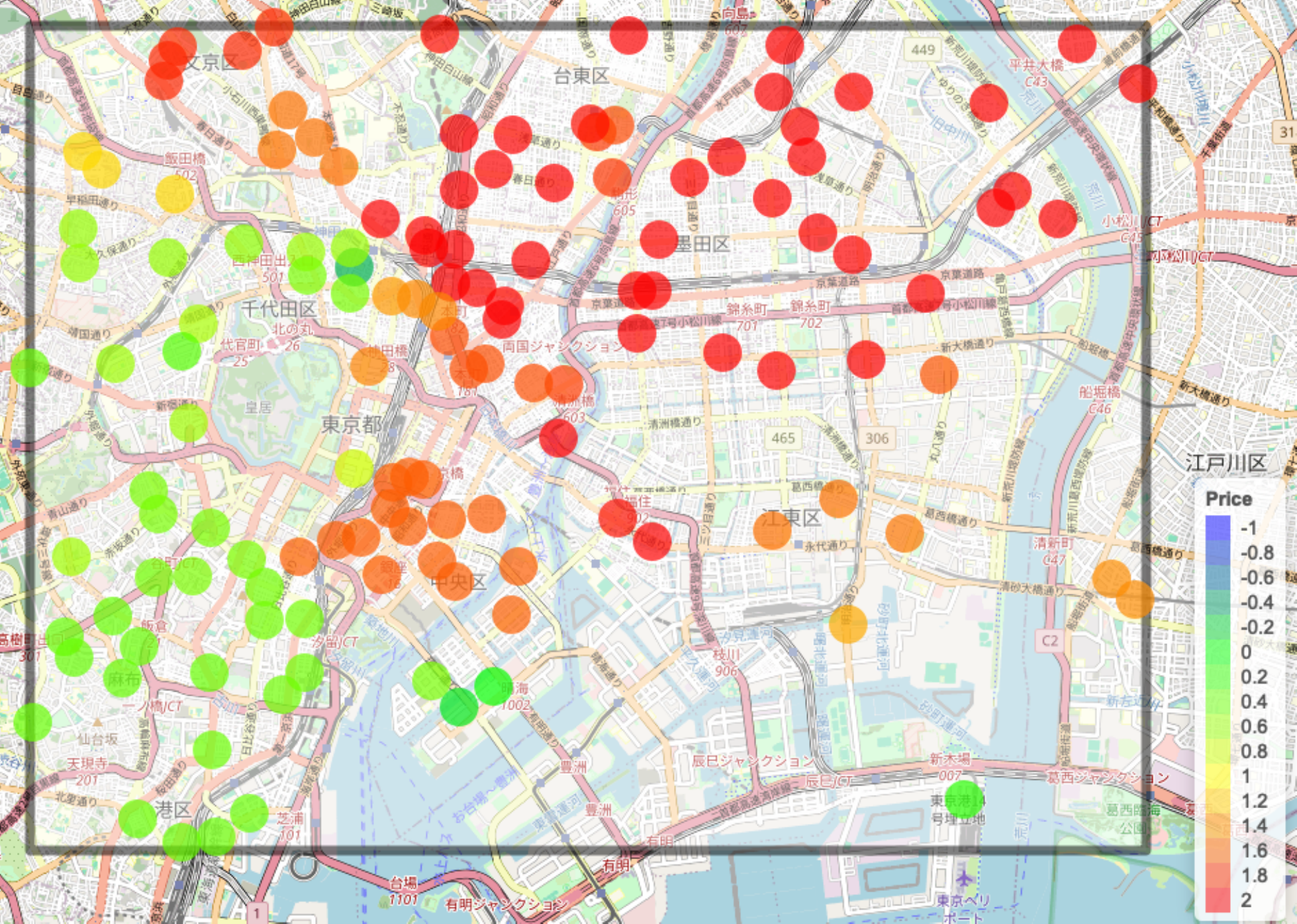} \\ \hline
\rotatebox{90}{\quad \quad BGWSR} &
\includegraphics[width=0.285\linewidth]{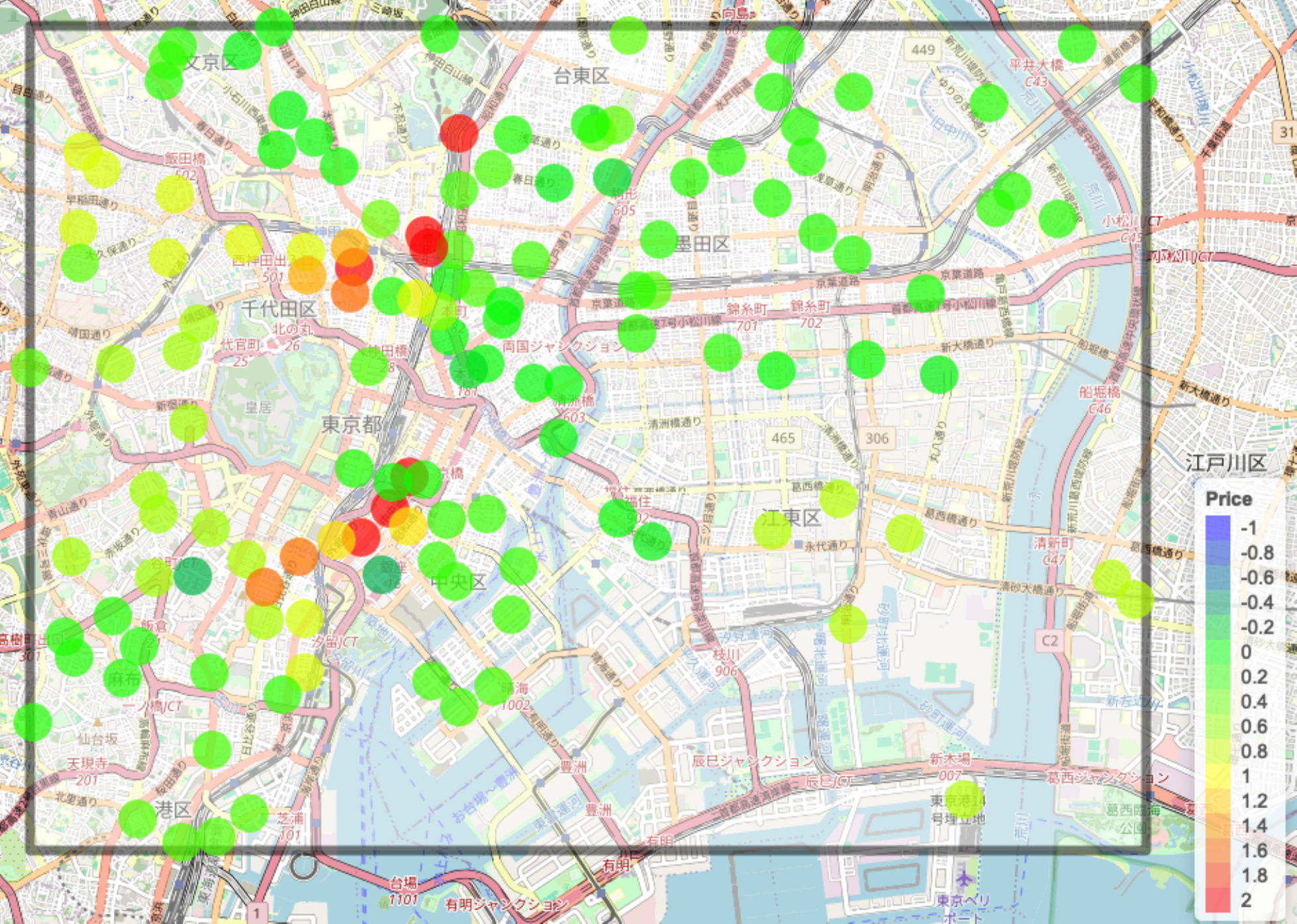} & 
\includegraphics[width=0.285\linewidth]{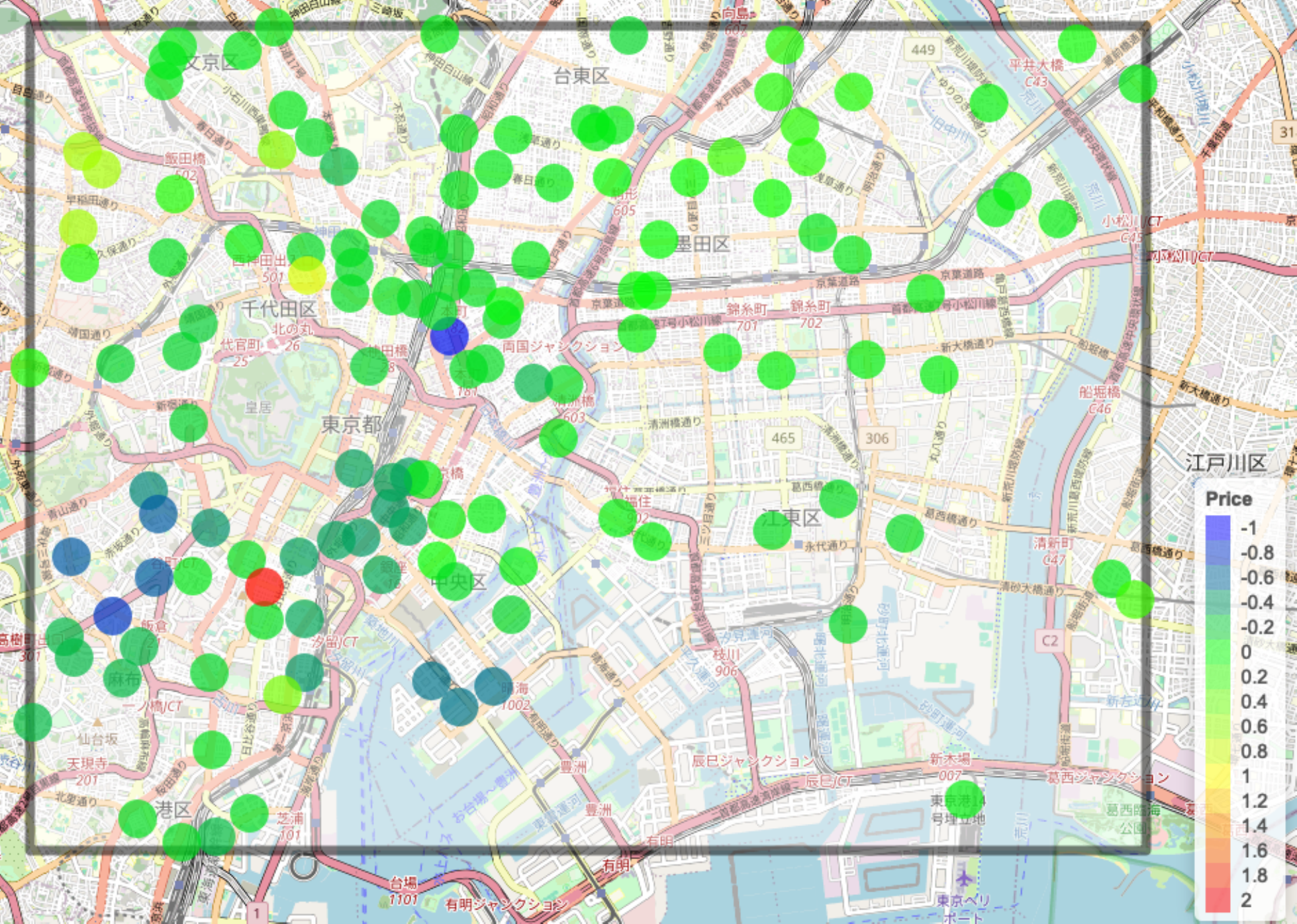} & 
\includegraphics[width=0.285\linewidth]{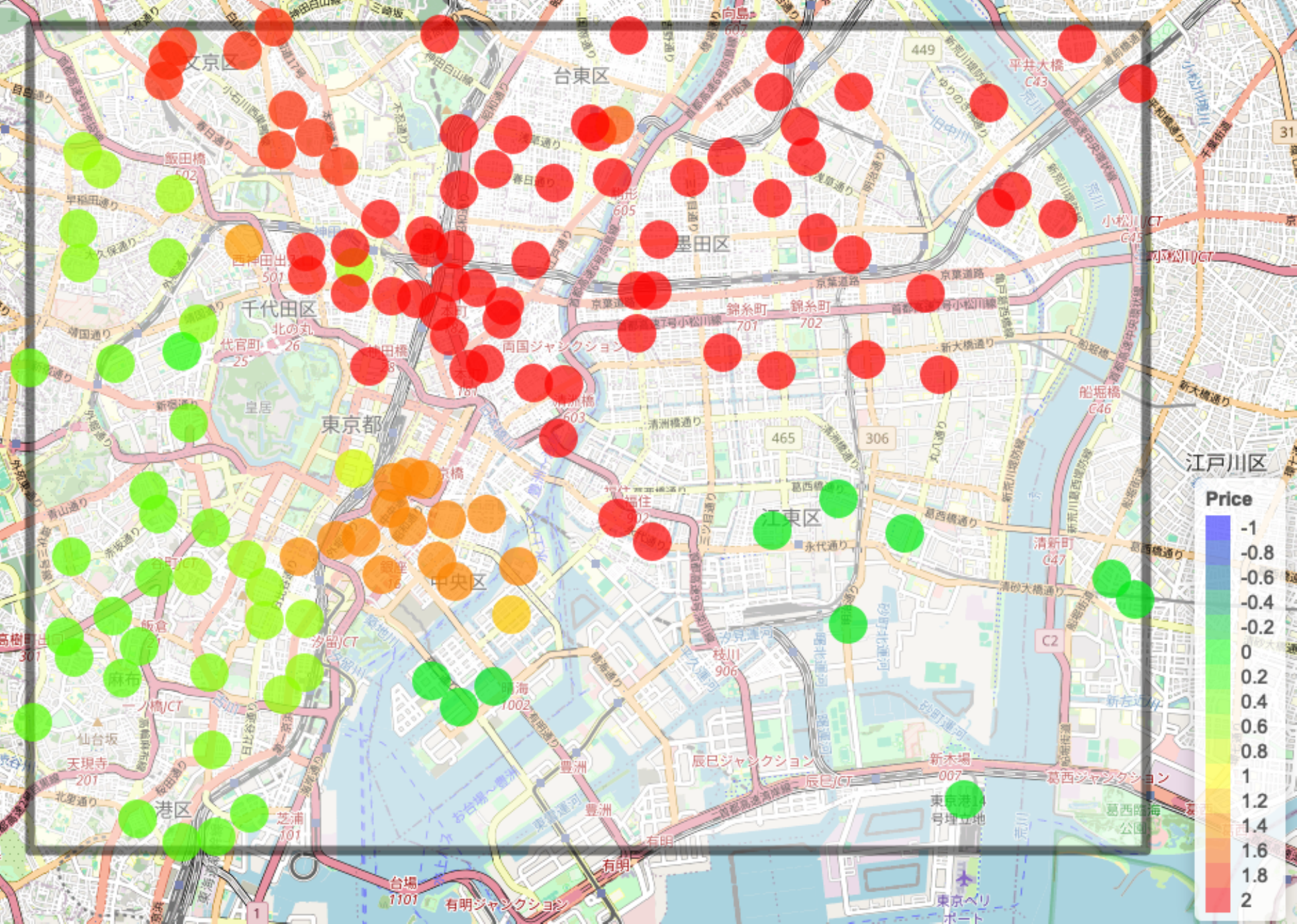} \\ \hline
\rotatebox{90}{\quad \quad \quad BGWR} &
\includegraphics[width=0.285\linewidth]{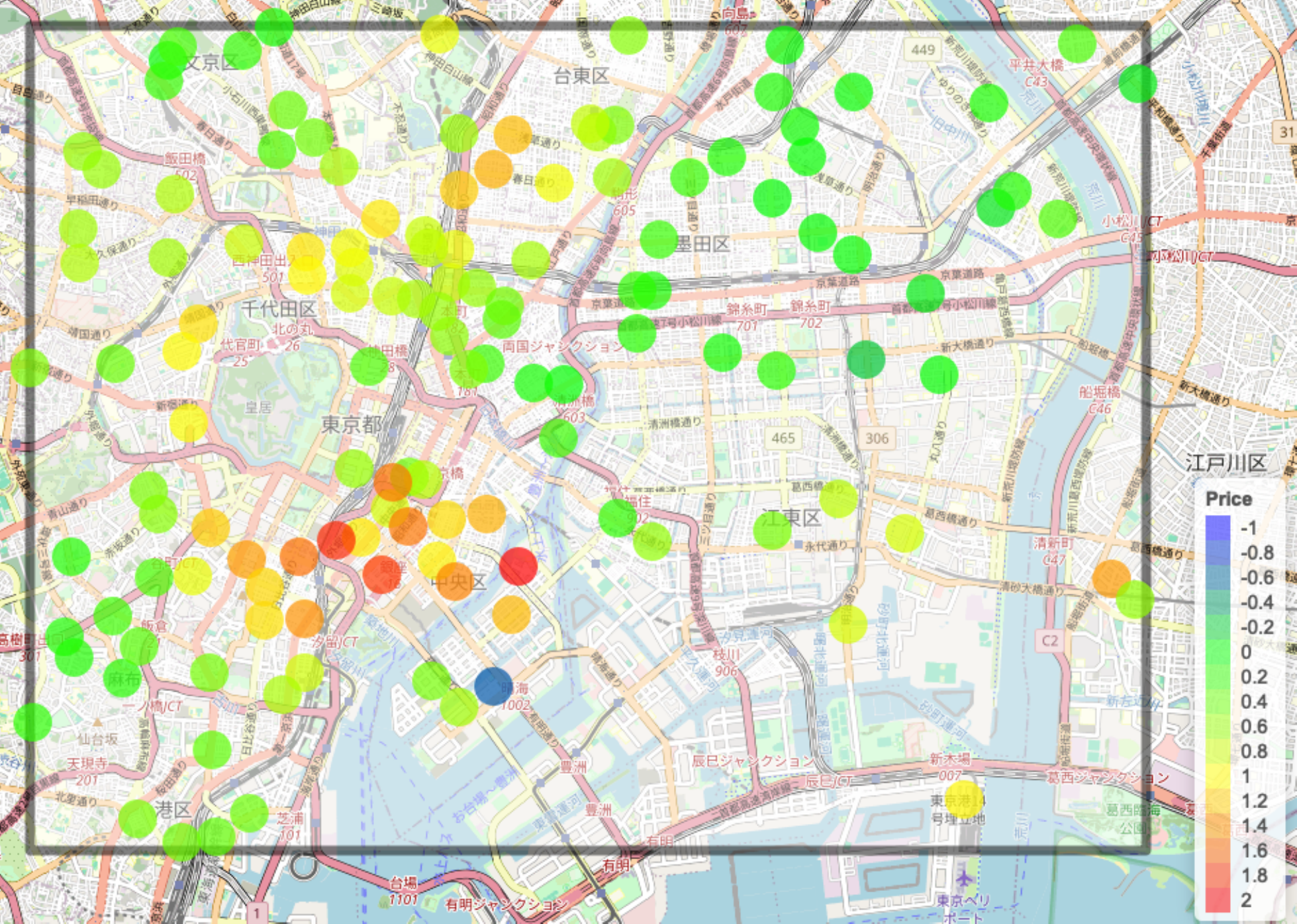} & 
\includegraphics[width=0.285\linewidth]{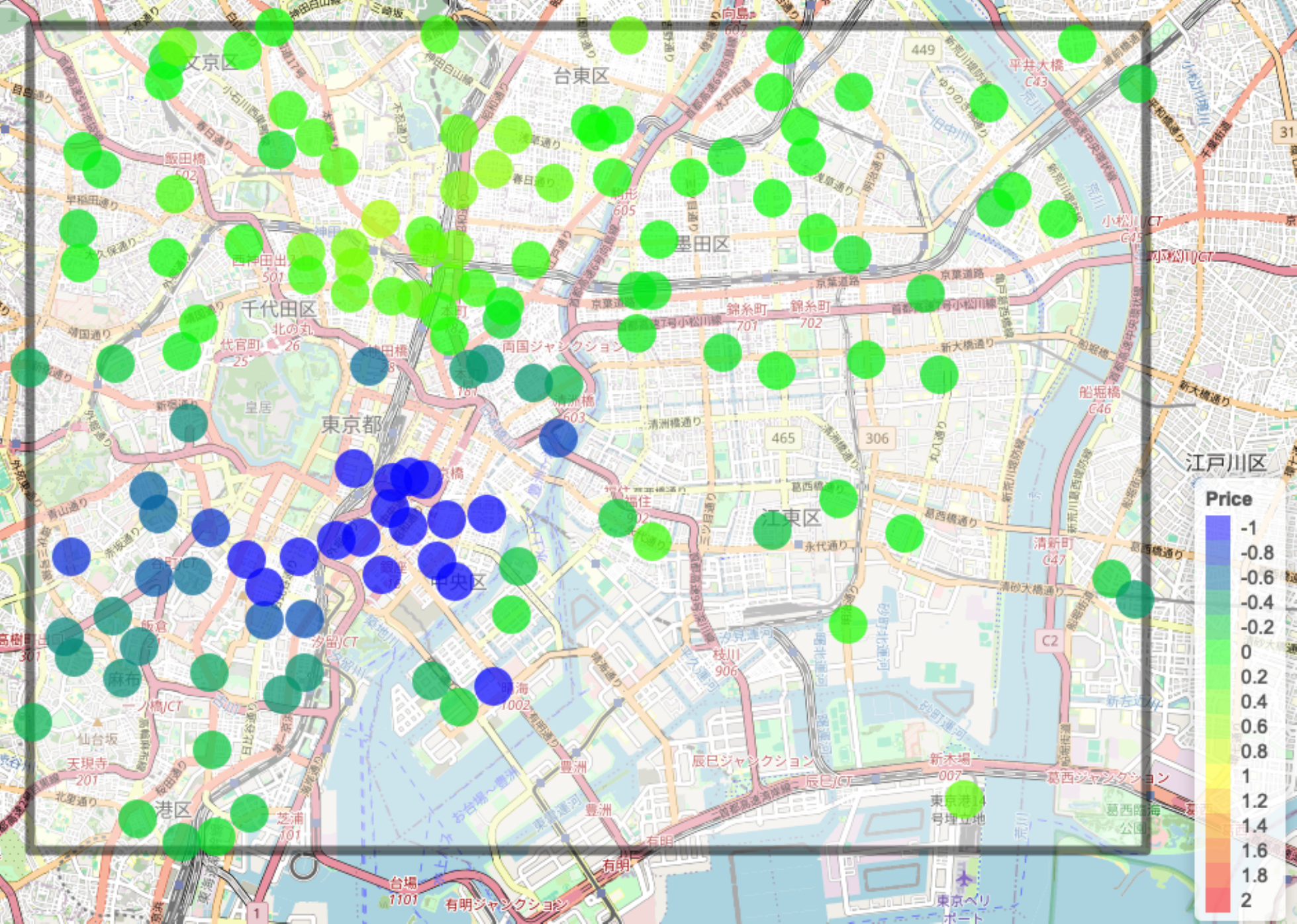} & 
\includegraphics[width=0.285\linewidth]{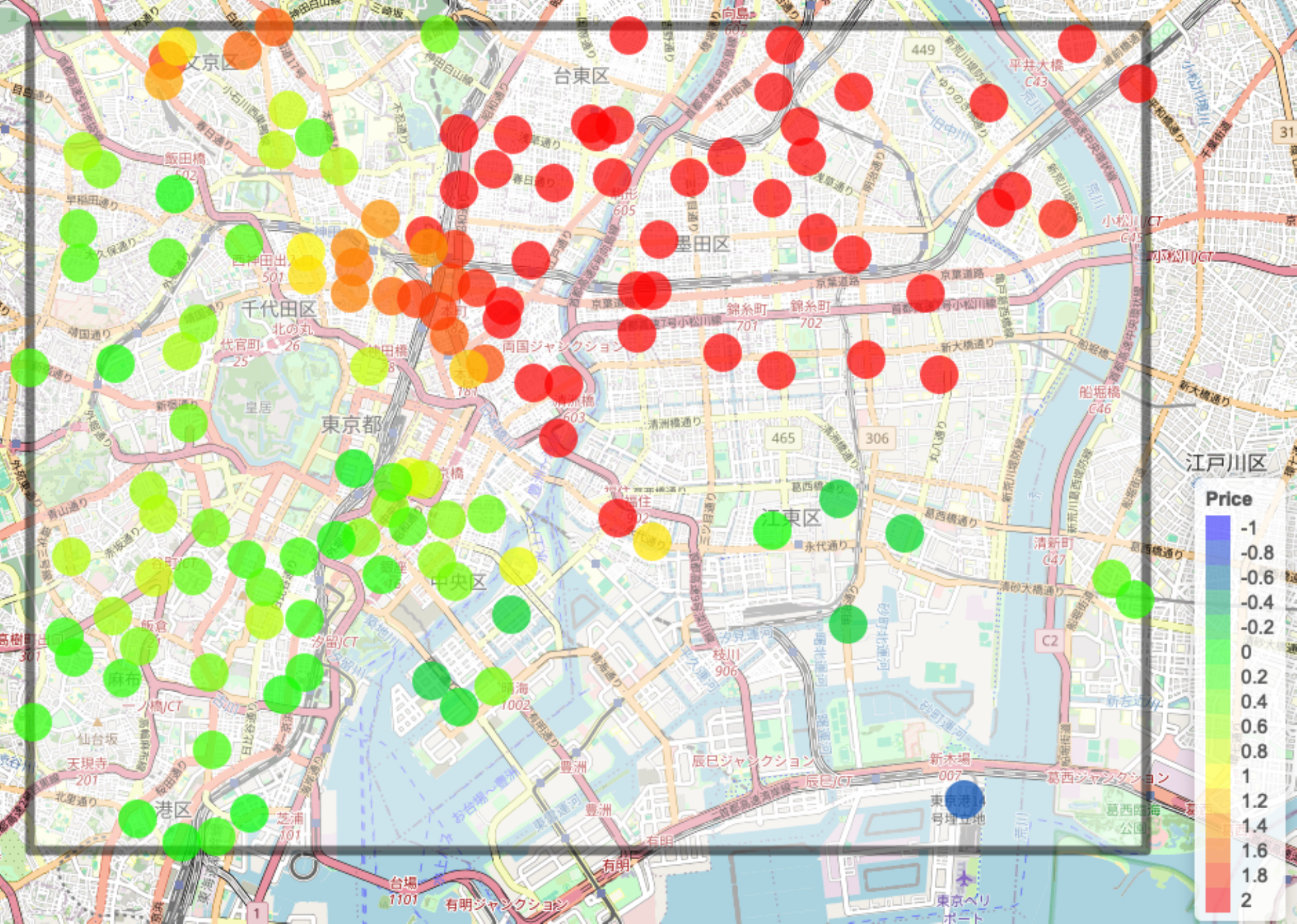} \\ \hline
\rotatebox{90}{\quad \quad \quad \, GWR} &
\includegraphics[width=0.285\linewidth]{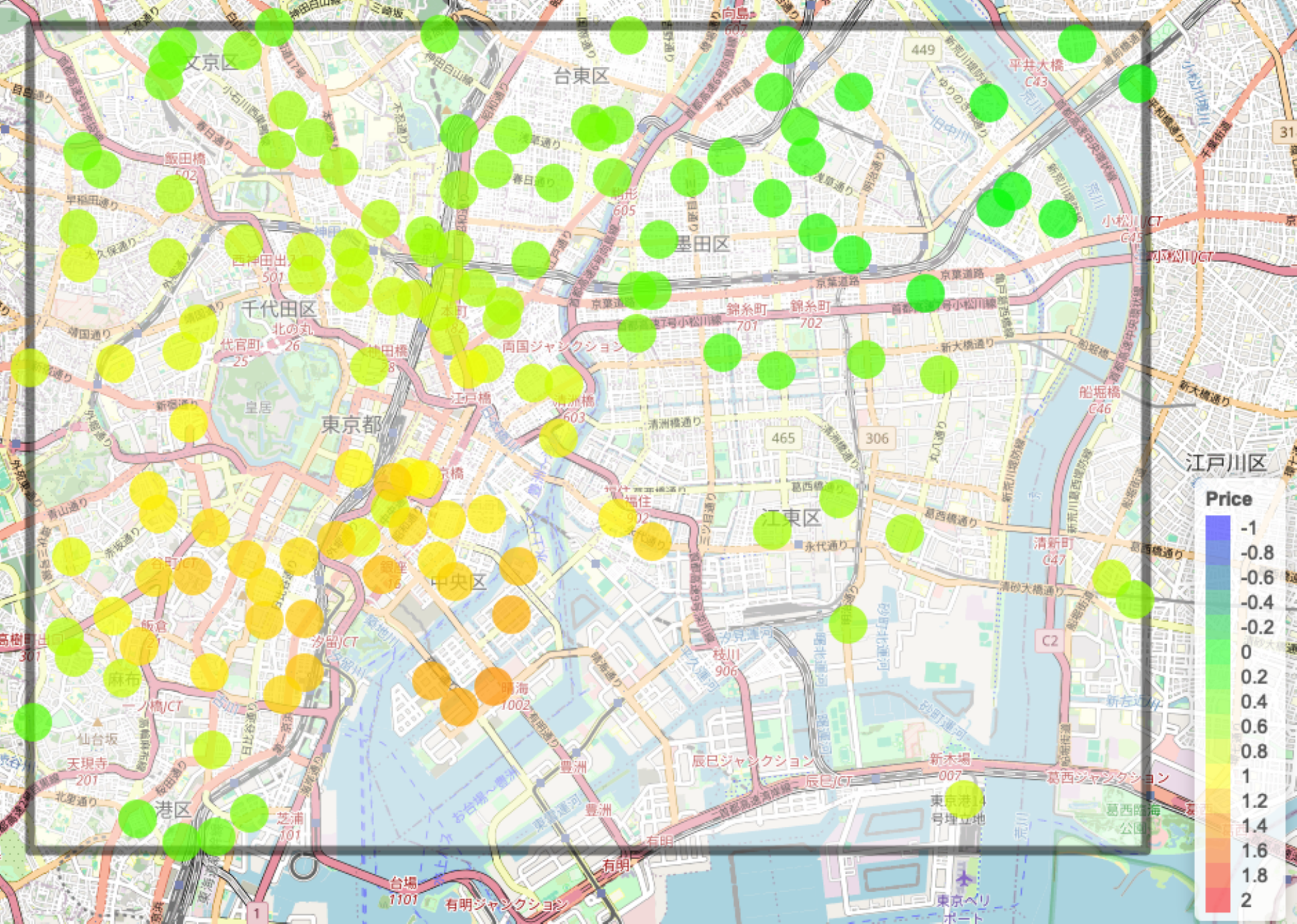} & 
\includegraphics[width=0.285\linewidth]{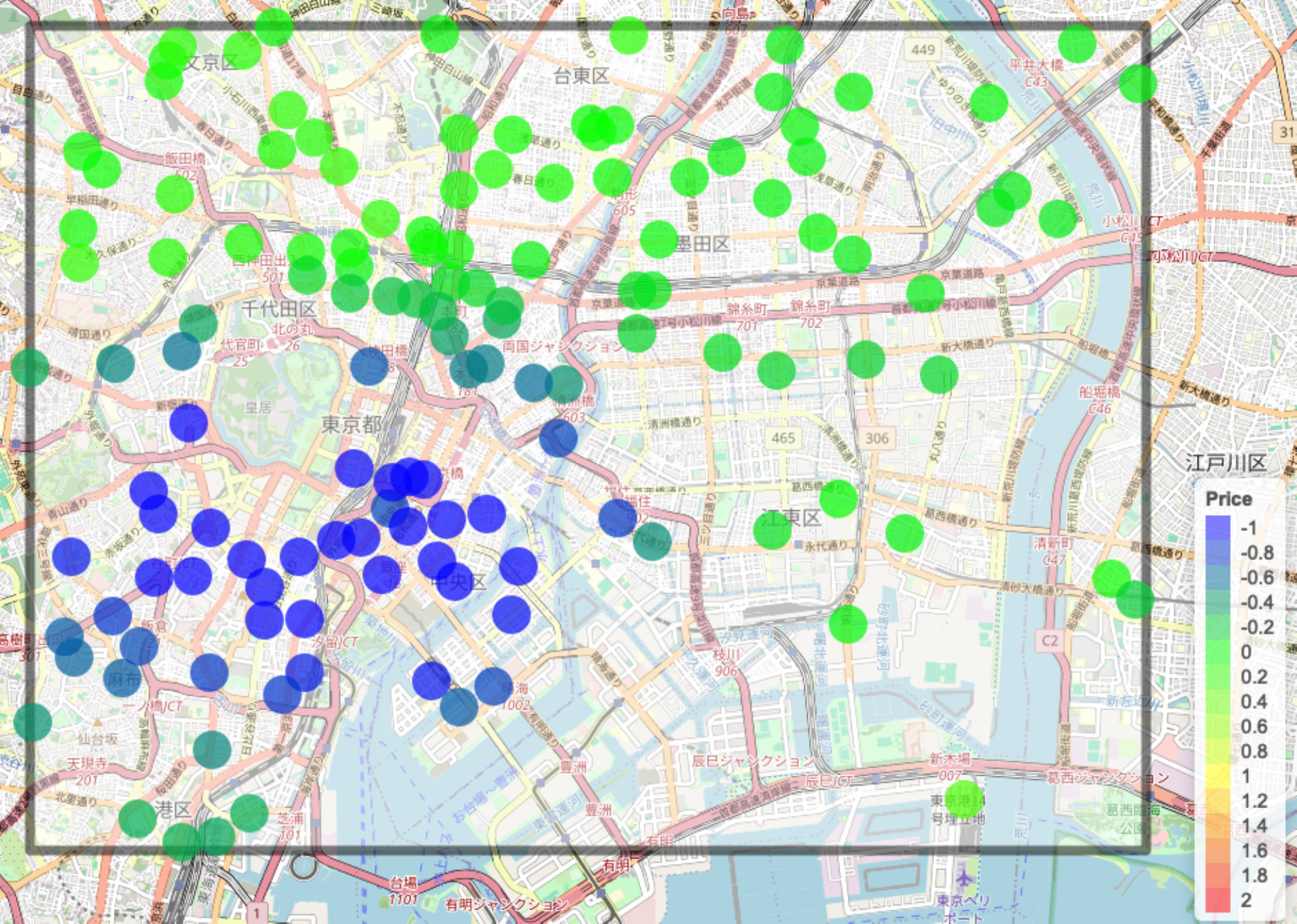} & 
\includegraphics[width=0.285\linewidth]{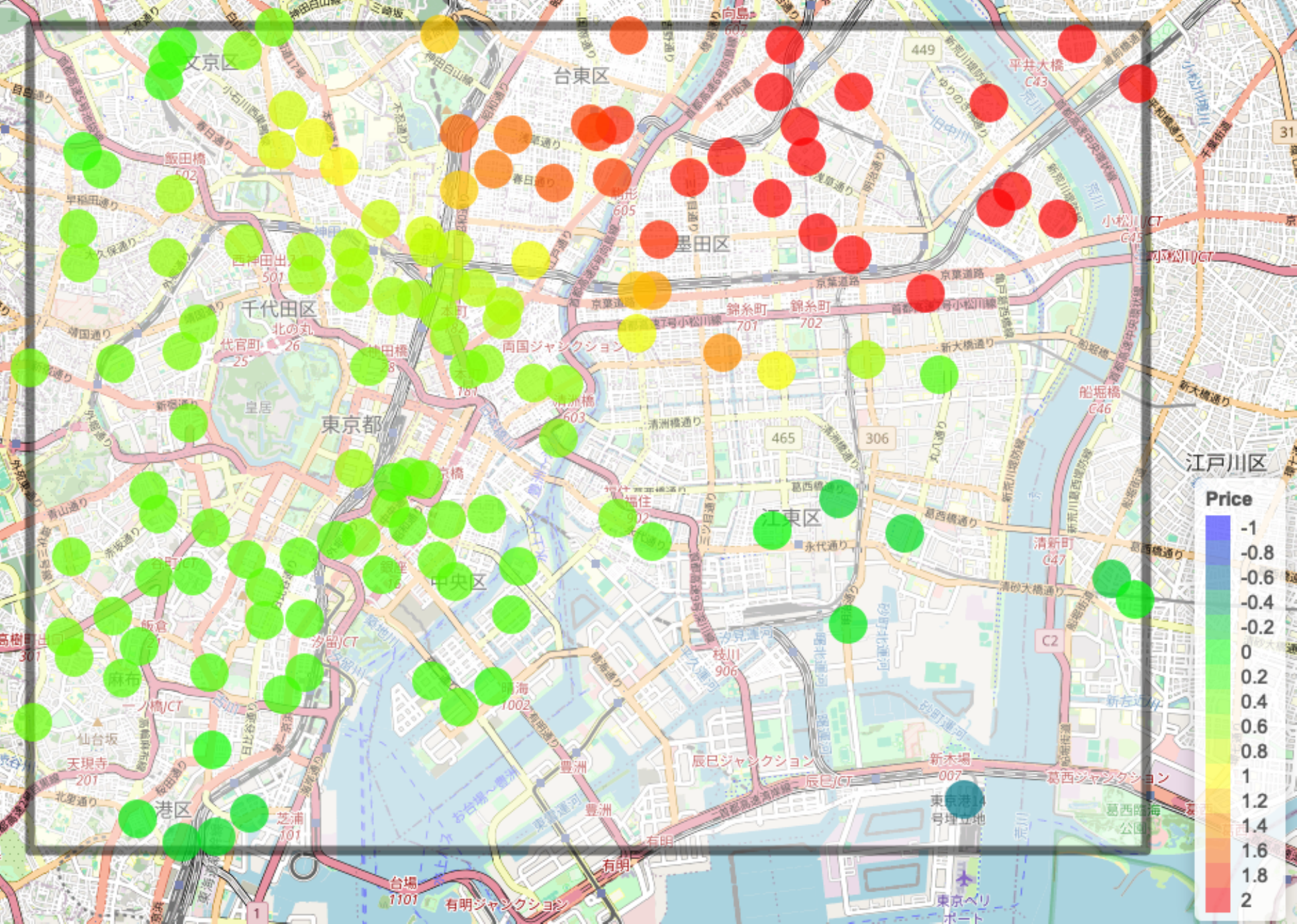} \\ \hline
\end{tabular}
\vspace{10pt}
\caption{Estimated coefficients at observation locations\label{fig:resland_beta_obs}}
\end{table}
\section{Conclusion\label{chap5:summary}}
This paper proposes a BGWSR by combining the idea of Bayesian Fused Lasso and BGWR.
The proposed method enables us to consider spatial correlations even when the number of observation locations is small. 
Moreover, by introducing a method of determining the range of adjacent locations for each location into BGWSR, the prediction performance is effectively maintained even when the observation locations are heterogeneous.

This paper shows the usefulness of the proposed method through numerical studies.
The results show that the proposed method has better prediction performance for the coefficients and the objective variable compared to estimated by GWR and BGWR, where observation locations are obtained heterogeneously in space.
The results suggest that the proposed method effectively considers the similarity of the coefficients between adjacent locations since the estimated values of the coefficients are close among the adjacent locations.
Furthermore, when the locations have a cluster structure, the estimated values of the proposed method are the same coefficients within the clusters.
We also applied the existing and proposed methods to land price data for Tokyo, where the density of observation locations was spatially heterogeneous.
The results showed that the proposed method performed better than the existing methods in predicting the objective variables.
Furthermore, we evaluated the uncertainty of the predicted values of the objective variables.
Consequently, we confirmed that the proposed method has a smaller $95\%$ credible interval than BGWR, even at the edge of the domain and in areas with sparse observation locations, and that most locations include the true value within the interval.
The results suggest that the proposed method has less uncertainty than BGWR.

The proposed method can be widely used for data obtained from heterogeneous observation patterns of location, and when spatial autocorrelation is assumed in the coefficients.
For example, the soil data used in \cite{koh2020} have observation locations where there are sparse and dense densities of locations.
In addition, data for seismic intensity observation locations in earthquake intensity prediction and infection status in epidemiology are often obtained in a spatially heterogeneous pattern.
The proposed method is thus useful for such data.

BGWSR could be potentially improved in several aspects.
First, it is necessary to find a way to determine the function of the weights used to estimate the adjacency locations.
The numerical studies and real data examples presented in this paper use Bi-square, but existing studies have proposed Boxcar, Gaussian, and other weight functions~\citep{fotheringham2003}.
It is necessary to select the appropriate approach among these weighting function based on the characteristics of the data to be applied.
For example, when using Boxcar, the predicted values of the coefficients tend to be similar across the whole set. If the purpose of the estimation is to emphasize the mean structure of the coefficients, the use of Boxcar therefore may be appropriate.

The second point of potential improvement is enhancing interpretability for high-dimensional data.
BGWSR uses a Laplace distribution for the prior distribution of the coefficients.
Hence, the estimates of the coefficients are not exactly $0$ in the methods that use the posterior mean or posterior median as the estimator of the coefficient.
This may reduce interpretability. In the case of multivariate data analysis, variable selection methods are often used and known to improve interpretability~\citep{tibshirani1996}.
For achieving this issue, one of the directions is the use of Horseshoe prior for the prior distribution~\citep{carvalho2010}, and Spike and Slab prior~\citep{rovckova2018}.
This allows to drive the estimation of some coefficient estimates to $0$ and is expected to improve interpretability.
For BGWSR, similarly changing the prior distribution is expected to improve interpretability through variable selection when applied to high-dimensional data.
%
\bibliographystyle{apa}
\bibliography{references}
\end{document}